\documentclass[11pt,a4paper]{article}
\usepackage{amssymb,amsmath,amsfonts,amsthm,amstext,amscd,array,amsopn}
\usepackage{mathrsfs}
\usepackage{hyperref}
\usepackage{bbm}
\usepackage{bm}
\usepackage[arrow,matrix,curve]{xy}
\usepackage{bbding}
\usepackage{wasysym}
\usepackage{color}
\usepackage{soul}

\def\nn{\nonumber }

 1

\newcommand{\eq}{\begin{equation}}
\newcommand{\eqa}{\begin{eqnarray}}
\newcommand{\en}{\end{equation}}
\newcommand{\ena}{\end{eqnarray}}
\newcommand{\enn}{\nonumber \end{equation}}

\def\st {\star}

\def\of{{\overline{{\rm{f}}\,}}}

\def\D/h{\widehat{\fmslash D}}

\def\om{\omega}

\def\al{\alpha}

\def\be{\beta}
\def\ga{\gamma}

\def\5bar{{\overline 5}}

\def\ee{{\hat e}}

\def\RRR{{\mathscr R}}

\def\R{{\rm R}}
\def\oR{{\overline{\R}}}

\def\DD{{\Xi}}
\def\FF{\mathcal F}

\def\bd{{{{\,}\scriptstyle{{\displaystyle\bar{_{}}}}}} \!\!\:\!\delta}

\def\s'O{\stackrel{_{{\displaystyle\st \footnotesize '}}}{_{^{^{\displaystyle\otimes}}}}}

\def\D{\Delta}
\def\1s{{1_\st }}
\def\3s{{3_\st }}
\def\2s{{2_\st }}
\def\ef1{{1_\FF}}
\def\ef2{{3_\FF}}
\def\ef3{{2_\FF}}

\def\le{\langle}
\def\re{\rangle}

\def\dd{{\triangledown}}

\def\g{\mathsf{g}}

\def\r4{\mathbb{R}^4}

\def\nn{\nonumber}

\newcommand{\mbf}[1]{{\boldsymbol {#1} }}
\def\ii{{\,{\rm i}\,}}
\def\dd{{\rm d}}
\def\DD{{\rm D}}

\def\Id{{\rm id}}

\def\G{{\sf G}}
\def\W{{\sf W}}

\def\LC{{\sf LC}}

\newcommand{\unit}{\mathbbm{1}}   			% identity map/matrix

\def\mcR{{\mathcal R}}

\def\Lie{{\mathcal L}}

\newcommand{\eqend}{\end{equation}}
\newcommand{\nonueqa}{\begin{eqnarray*}}
\newcommand{\eqaend}{\end{eqnarray}}
\newcommand{\nonueqaend}{\end{eqnarray*}}

\newcommand{\bma}[1]{\begin{array}{#1}}
\newcommand{\ema}{\end{array}}
\newcommand{\bc}{\begin{center}}
\newcommand{\ec}{\end{center}}

\newcommand{\vect}{{\sf Vec}}

\renewcommand{\thefootnote}{\fnsymbol{footnote}}
\newcommand{\newsection}{\setcounter{equation}{0}\section}

\newcommand{\real}{{\mathbb R}} %% real numbers

\def\Mcal{{\mathcal M}}

\def\ot{{\, \otimes\, }}

\newif\ifold             \oldtrue

\def\nn{\nonumber}

\def\e{{\,\rm e}\,}

\def\be{\begin{equation}}
\def\ee{\end{equation}}
\def\bea{\begin{eqnarray}}
\def\eea{\end{eqnarray}}
\def\bd{\begin{displaymath}}
\def\ed{\end{displaymath}}

\def\s{\sigma}

\newcommand{\beq}{\begin{eqnarray}}
\newcommand{\eeq}{\end{eqnarray}}

\makeatletter
\newdimen\normalarrayskip              % skip between lines
\newdimen\minarrayskip                 % minimal skip between lines
\normalarrayskip\baselineskip
\minarrayskip\jot
\newif\ifold             \oldtrue            
\def\arraymode{\ifold\relax\else\displaystyle\fi} % mode of array entries
\def\@arrayskip{\ifold\baselineskip\z@\lineskip\z@
     \else
     \baselineskip\minarrayskip\lineskip2\minarrayskip\fi}
\def\@arrayclassz{\ifcase \@lastchclass \@acolampacol \or
\@ampacol \or \or \or \@addamp \or
   \@acolampacol \or \@firstampfalse \@acol \fi
\edef\@preamble{\@preamble
  \ifcase \@chnum
     \hfil$\relax\arraymode\@sharp$\hfil
     \or $\relax\arraymode\@sharp$\hfil
     \or \hfil$\relax\arraymode\@sharp$\fi}}
\def\@array[#1]#2{\setbox\@arstrutbox=\hbox{\vrule
     height\arraystretch \ht\strutbox
     depth\arraystretch \dp\strutbox
     width\z@}\@mkpream{#2}\edef\@preamble{\halign \noexpand\@halignto
\bgroup \tabskip\z@ \@arstrut \@preamble \tabskip\z@ \cr}%
\let\@startpbox\@@startpbox \let\@endpbox\@@endpbox
  \if #1t\vtop \else \if#1b\vbox \else \vcenter \fi\fi
  \bgroup \let\par\relax
  \let\@sharp##\let\protect\relax
  \@arrayskip\@preamble}
\makeatother

\def\FF{{\cal F}}
\def\al{\alpha}
\def\be{\beta}

\theoremstyle{definition}

\usepackage{mathtools}

\usepackage{hyperref}
\hypersetup{colorlinks,%
citecolor=black,%
filecolor=black,%
linkcolor=black,%
urlcolor=black}

\begin{document}

\begin{titlepage}

\begin{flushright}
\small
EMPG--17--16
\end{flushright}

\begin{center}

\vspace{1cm}

\baselineskip=24pt

{\Large\bf Nonassociative differential geometry \\ and gravity with non-geometric fluxes}

\baselineskip=14pt

\vspace{1cm}

{\bf Paolo Aschieri}${}^{1}$,  \ {\bf Marija Dimitrijevi\'c \'Ciri\'c}${}^2$ \ and \ {\bf Richard
  J. Szabo}${}^{3}$
\\[5mm]
\noindent  ${}^1$ {\it Dipartimento di Scienze e Innovazione
  Tecnologica, Universit\`a del Piemonte Orientale,\\ Viale T. Michel 11, 15121 Alessandria, Italy}\\
\noindent {\it {\rm and}  INFN, Sezione di Torino, via P. Giuria 1, 10125
Torino, Italy}\\
and {\it Arnold-Regge Center, via P. Giuria 1, 10125, Torino, Italy}\\
Email: \ {\tt
    aschieri@to.infn.it}
\\[3mm]
\noindent  ${}^2$ {\it Faculty of Physics, University of
  Belgrade}\\ {\it Studentski trg 12, 11000 Beograd, Serbia}\\
Email: \ {\tt
    dmarija@ipb.ac.rs}
\\[3mm]
\noindent  ${}^3$ {\it Department of Mathematics, Heriot-Watt University\\ Colin Maclaurin Building,
  Riccarton, Edinburgh EH14 4AS, U.K.}\\ and {\it Maxwell Institute for
Mathematical Sciences, Edinburgh, U.K.} \\ and {\it The Higgs Centre
for Theoretical Physics, Edinburgh, U.K.}\\
Email: \ {\tt R.J.Szabo@hw.ac.uk}
\\[30mm]

\end{center}

\begin{abstract}

\baselineskip=12pt

\noindent
We systematically develop the metric aspects of nonassociative
differential geometry tailored to the parabolic phase space model of constant
locally non-geometric closed string vacua, and use it to construct 
preliminary steps towards a
nonassociative theory of gravity on spacetime. We obtain explicit
expressions for the torsion, curvature, Ricci tensor and Levi-Civita
connection in nonassociative Riemannian geometry on phase space, and
write down Einstein field equations. We apply this
formalism to construct $R$-flux corrections to the Ricci tensor on
spacetime, and comment on the potential implications of these structures in
non-geometric string theory and double field theory.

\end{abstract}

\end{titlepage}
\setcounter{page}{2}

\newpage

{\baselineskip=12pt
\tableofcontents
}

\newpage

\renewcommand{\thefootnote}{\arabic{footnote}}
\setcounter{footnote}{0}

\newsection{Introduction\label{sec:intro}}

Deformations of spacetime geometry through compactifications of
string theory may help elucidate the precise mechanism by which closed
strings provide a framework for a quantum theory of gravity. This has
been the hope in some recent investigations surrounding non-geometric string
theory, in which noncommutative and nonassociative deformations of target space geometry
have been purported to be probed by closed strings in non-geometric
flux compactifications~\cite{Blumenhagen2010,Lust2010,Blumenhagen2011,Condeescu2012,Andriot2013,Blumenhagen2013,Blair2014,Bakas2015}. In
particular, in locally non-geometric backgrounds one aims to find a
low-energy limit of closed string theory which is described by an effective
nonassociative theory of gravity on spacetime.

In this paper we focus on the parabolic phase space model for strings
propagating in locally non-geometric constant $R$-flux
backgrounds in $d$ dimensions~\cite{Lust2010}. In this framework the canoncial commutation
relations of phase space are deformed by a trivector $R$-flux on
target space to the
quasi-Poisson coordinate algebra
\bea
[x^\mu,x^\nu] = \mbox{$\frac{\ii\ell_s^3}{3\hbar}$} \, R^{\mu\nu\rho}\, p_\rho \ , \qquad
[x^\mu,p_\nu] = \ii\hbar\, \delta^\mu{}_\nu \qquad \mbox{and}
\qquad [p_\mu,p_\nu] = 0 \ ,
\eea
with the nonassociativity of spacetime captured by the non-vanishing
Jacobiators
\bea
[x^\mu,x^\nu,x^\rho] := [x^\mu,[x^\nu,x^\rho]]+[x^\nu,[x^\rho,x^\mu]]
+ [x^\rho,[x^\mu,x^\nu]] = \ell_s^3\, R^{\mu\nu\rho} \ .
\eea
On fields these deformations
are described by nonassociative phase space
star-products~\cite{Mylonas2012,Bakas2013,Kupriyanov2015}, and using
them the formal
mathematical development of nonassociative differential geometry of
phase space has been pursued
in~\cite{Mylonas2013,Barnes2014,Aschieri2015,Barnes2015}; a more
pedestrian approach to these developments is given in~\cite{Barnes2016,Blumenhagen2016}. The purpose
of the present paper is two-fold.

Firstly, we present a self-contained construction of nonassociative
differential geometry based on the constant
parabolic $R$-flux background which is rooted in two guiding 
principles: equivariance under the twist deformed quasi-Hopf algebra 
of infinitesimal diffeomorphisms on the one hand (including invariance
of the operations of multiplication, inner derivation and exterior
derivation of fields, and covariance of
tensor fields) and, on the other hand, the equivalent
descriptions of tensor fields as sections of vector bundles and as
maps between vector bundles.
These constructions are compatible with the category theory formalism
of~\cite{Barnes2014,Barnes2015}; indeed, Sections~\ref{sec:algebraic}--\ref{sec:connections}
of the present paper can also be
regarded as unravelling that general construction for the
specific cochain twist deformation provided by the constant parabolic $R$-flux
model in phase space. This unravelling 
complements that undertaken in~\cite{Barnes2016}. However, as pursued
by~\cite{Blumenhagen2016}, the main viewpoint here is to avoid the use of
category theory altogether and yet, in contrast to
the more pedestrian approach of~\cite{Blumenhagen2016}, 
to provide a
self-consistent and mathematically rigorous construction of nonassociative
differential geometry. 
This leads to a notion of torsion that coincides with that introduced in~\cite{Blumenhagen2016} (see \eqref{Torsion}). On the
other hand, it leads to key new results, including a simple definition of curvature tensor as
the square of the covariant derivative, its equivalent
description as an operator on vector fields (the second Cartan
structure equation, see \eqref{eq:Curvexpl}), and a well-defined
Ricci tensor (see \eqref{defRic} and \eqref{eq:RicBC}).

Secondly, we use this framework to systematically study the
metric aspects of nonassociative differential geometry.
This is a nonassociative generalization of the noncommutative
Riemannian geometry developed in \cite{Aschieri2004,Aschieri2005}. One
of our main achievements is the construction of the analog of the Levi-Civita connection (see \eqref{Gamma3}),
wherein we describe how to circumvent the problems encountered in~\cite{Blumenhagen2016}.
We thus obtain a  metric formulation of nonassociative gravity on
phase space. A complementary
vielbein or first order formalism for nonassociative gravity has been
considered in~\cite{Barnes2015,Barnes2016}. Here we have chosen to
develop the metric aspects of nonassociative gravity, because it also
represents the most direct way to explore the potential relevance of
nonassociative gravity to string theory, in that in closed string
theory the fundamental field is the metric tensor rather than the vielbein.

Although it is interesting in its own right to be able to implement general
covariance under the quasi-Hopf algebra of deformed diffeomorphisms
and to formulate Einstein equations in nonassociative
space, in order to arrive at a theory that can be potentially
considered as providing a low-energy effective action for closed strings in the
presence of non-geometric fluxes, it is necessary to project nonassociative
gravity from phase space to spacetime.
To this aim we further develop the approach 
of~\cite{Aschieri2015}, which demonstrated how to extract results of
closed string scattering amplitudes in non-geometric
backgrounds~\cite{Blumenhagen2011} from the nonassociative deformation
of phase space. We conclude in particular that, in the constant
parabolic $R$-flux model, the curvature of spacetime is deformed in a
non-trivial way by locally non-geometric fluxes. Our main result for the Ricci tensor of nonassociative gravity
is presented in \eqref{eq:Ricpolarised} and reproduced here:
\bea
{\sf Ric}^\circ_{\mu\nu} &=& {\sf Ric}_{\mu\nu}^{\LC} +
\mbox{$\frac{\ell_s^3}{12}$} \, R^{\alpha\beta\gamma}\,\Big(
\partial_\rho\big(\partial_\alpha\g^{\rho\sigma}\, (\partial_\beta
\g_{\sigma\tau})\, \partial_\gamma \Gamma_{\mu\nu}^{\LC\, \tau} \big) 
- \partial_\nu\big(\partial_\alpha\g^{\rho\sigma}\, (\partial_\beta
\g_{\sigma\tau})\, \partial_\gamma \Gamma_{\mu\rho}^{\LC\, \tau}\big)
\nn\\ && \qquad \qquad \qquad \qquad +\,  \partial_\gamma \g_{\tau\omega}\big( \partial_\alpha
(\g^{\sigma\tau}\, \Gamma_{\sigma\nu}^{\LC\,\rho})\, \partial_\beta
\Gamma_{\mu\rho}^{\LC\,\omega} - \partial_\alpha
(\g^{\sigma\tau}\, \Gamma_{\sigma\rho}^{\LC\,\rho})\,\partial_\beta
\Gamma_{\mu\nu}^{\LC\,\omega}
\nn\\ && \qquad\qquad \qquad \qquad +\, (\Gamma_{\mu\rho}^{\LC\,\sigma}\, \partial_\alpha
\g^{\rho\tau} - \partial_\alpha \Gamma_{\mu\rho}^{\LC\,\sigma} \,\g^{\rho\tau})\,           
\partial_\beta \Gamma_{\sigma\nu}^{\LC\,\omega}
\nn\\ && \qquad\qquad \qquad \qquad - \, (\Gamma_{\mu\nu}^{\LC\,\sigma}\, \partial_\alpha
\g^{\rho\tau} -\partial_\alpha \Gamma_{\mu\nu}^{\LC\,\sigma}\,\g^{\rho\tau})\,              
\partial_\beta \Gamma_{\sigma\rho}^{\LC\,\omega} \big) \Big) \ ,
\label{eq:Ricciintro}\eea
where ${\sf Ric}_{\mu\nu}^{\LC}$ is the usual Ricci tensor of the classical
Levi-Civita connection $\Gamma_{\mu\nu}^{\LC\, \rho}$ of a metric
tensor $\g_{\mu\nu}$ on spacetime. This expression is valid to linear
order in the $R$-flux, which is the order at which the corresponding
conformal field theory calculations are
reliable~\cite{Blumenhagen2011}. It represents the first non-trivial
starting point for understanding how to define
a nonassociative theory of gravity describing the low-energy effective
dynamics of closed strings in non-geometric backgrounds, although in this paper
we do not address in detail the implications of this structure on string theory
or double field theory~\cite{Blumenhagen2013};
see~\cite{Blumenhagen2016} for some discussion of points which should
be addressed in this latter context.

A somewhat perplexing aspect about the development of the geometry of
the phase space model for the $R$-flux background concerns the precise
meaning
of Riemannian geometry of phase space. Superficially, our approach is
reminescent of recent discussions of Born geometry in string
theory~\cite{Freidel2013}, wherein it is argued that the fundamental
string symmetries should contain diffeomorphisms of phase space, in
accord with Born's original proposal to unify quantum theory with
general relativity by treating spacetime and momentum space on equal
footing. It is precisely the phase space formulation that is
responsible for nonassociativity in the parabolic $R$-flux
model~\cite{Mylonas2012,Mylonas2013,Aschieri2015}, and it would be
interesting to find explicit connections between our constructions and
the proposal of~\cite{Freidel2013}.

The outline of the remainder of this paper is as follows. In
Section~\ref{sec:algebraic} we describe some preliminary Hopf
algebraic ingredients and define the quasi-Hopf algebra of
infinitesimal diffeomorphisms. This is the symmetry algebra that
leads us into the nonassociative deformations of differential
geometry, and in this preliminary section we
follow~\cite{Mylonas2013,Barnes2014,Aschieri2015}, where further
details can be found; we also comment on how
our constructions fit into frameworks suitable for a double field theory
formulation of all our developments. In Section~\ref{sec:functions}
we use these ingredients to fully develop nonassociative tensor
calculus on phase space, and apply it in
Section~\ref{sec:connections} to construct a nonassociative theory of
connections, obtaining new definitions of curvature and Ricci
tensors, together with the main results of the Cartan structure equations
for curvature and torsion. This section builds and expands on the
nonassociative differential geometry machinery developed
by~\cite{Barnes2015}, and on the noncommutative geometry techniques and results of~\cite{Aschieri2005,Aschieri2017}.
In Section~\ref{sec:Riemannian} we introduce metric tensors
and develop the Riemannian aspects
of nonassociative differential geometry, including the extension to
the nonassociative setting of the noncommutative metric compatibility condition~\cite{Aschieri2004,Aschieri2017}, the explicit construction of the Levi-Civita
connection, the corresponding Ricci tensor and vacuum Einstein
equations on nonassociative phase space, and the induced corrections
to the spacetime Ricci tensor given in~\eqref{eq:Ricciintro}. Finally, in Section~\ref{sec:Conclusions} we conclude by summarising our main findings and highlighting key open issues for further investigation.

\newsection{Algebraic structure of non-geometric flux deformations\label{sec:algebraic}}

\subsection{$R$-flux induced cochain twist and quasi-Hopf algebra}

The presence of a constant non-geometric $R$-flux on $M=\real^d$ has been
proposed to be captured by a certain nonassociative deformation of the
geometry of
phase space $\Mcal=T^*M=\real^d\times(\real^d)^*$. Coordinates on $\Mcal$
will be denoted $x^A=(x^\mu,\tilde x_\mu)$ with $A=1,\dots, 2d$, where
$x^\mu$ are spacetime coordinates on $M$ while $\tilde
x_{{\mu}}=p_\mu$ are momentum coordinates for $\mu=1,\dots,
d$.
Derivatives are denoted in a similar way:
$\partial_A=\big(\frac{\partial}{\partial x^\mu}=\partial_\mu,
\frac{\partial}{\partial p_\mu}=\tilde\partial^{{\mu}} \big)$. In string theory applications, the metric on $M$ is usually taken to have Euclidean signature, but in what follows our results do not depend on the signature of the spacetime metric tensor which can be either Euclidean or Lorentzian.

The geometry of phase space $\Mcal$ is deformed by using a particular
cochain twist element ${\cal F}$ in the universal enveloping Hopf algebra
${U}\vect(\Mcal)$ of the Lie
algebra of vector fields $\vect(\Mcal)$ on $\Mcal$. It is defined by
\begin{eqnarray}
{\cal F}= \exp\big(-\mbox{$\frac{\ii\hbar}{2}$}\, (\partial_\mu\otimes \tilde\partial^{{\mu}} -
\tilde\partial^{{\mu}}\otimes\partial_\mu)
-\mbox{$\frac{\ii\ell_s^3}{12\hbar}$} \, R^{\mu\nu\rho}\, (p_\nu\, \partial_\rho\otimes \partial_\mu -
\partial_\mu\otimes p_\nu\, \partial_\rho) \big) \ , \label{twist}
\end{eqnarray}
where $R^{\mu\nu\rho}$ are the totally antisymmetric constant $R$-flux
components, and implicit summation over repeated upper and lower
indices is always understood. We will write
\begin{eqnarray}
{\cal F} =:{\rm f}\,^\al\otimes {\rm f}\,_\al = 1\otimes 1 +O(\hbar,\ell_s^3) \ ,
\end{eqnarray}
where ${\rm f}\,^\al, {\rm f}\,_\al$ are elements in ${U}\vect(\Mcal)$ and
summation on $\alpha$ is understood; for the inverse of the twist we write ${\cal
  F}^{-1} =:\of\,^\al\otimes \of\,_\al$. 
Following~\cite{Mylonas2013,Aschieri2015}, it will
sometimes be convenient to regard the twist element \eqref{twist} as
the result of applying successively two commuting abelian cocycle twists
\bea
{\cal F}=F\, F_R = F_R\, F \ ,
\label{eq:calFFR}\eea
where the Hopf 2-cocycle
\bea\label{eq:twistF}
F =\exp\big(-\mbox{$\frac{\ii\hbar}{2}$}\, (\partial_\mu\otimes \tilde\partial^{{\mu}} -
\tilde\partial^{{\mu}}\otimes\partial_\mu)\big) =:{\tt
  f}\,^\al\otimes{\tt f}\,_\al = 1\otimes1 +O(\hbar)
\eea
implements the
standard Moyal-Weyl deformation of canonical phase space, while the 2-cocycle
\bea\label{eq:twistFR}
F_R=\exp\big(-\mbox{$\frac{\ii\kappa}{2}$} \, R^{\mu\nu\rho}\,(p_\nu \, \partial_\rho\otimes \partial_\mu -
\partial_\mu\otimes p_\nu \, \partial_\rho) \big)=:{\tt
  f}_R^{\,\al}\otimes {\tt f}_{R\,\al} = 1\otimes1 + O(\kappa) \ ,
\eea
with
\bea
\kappa:= \frac{\ell_s^3}{6\hbar} \ ,
\label{eq:kappadef}\eea
implements the
deformation by the $R$-flux; in the following we shall sometimes treat
$\hbar$ and $\kappa$ as independent (small) deformation parameters.

The Hopf algebra ${U}\vect(\Mcal)$ has coproduct $\Delta$ defined as $\Delta(1)=1\otimes1$,
$\Delta(\partial_A)=1\otimes\partial_A+\partial_A\otimes 1$, counit
$\epsilon$ defined as
$\epsilon(1)=1$, $\epsilon(\partial_A)=0$, and
antipode $S$ defined as $S(1)=1$, $S(\partial_A)=-\partial_A$, with
$\Delta$ and $\epsilon$ extended to all of ${U}\vect(\Mcal)$ as algebra
homomorphisms,
and $S$ extended as an algebra antihomomorphism (linear and anti-multiplicative).
With the twist ${\cal F}$, following \cite{Mylonas2013} we deform the
Hopf algebra ${U}\vect(\Mcal)$ (considered to be extended with power series in
$\hbar$ and $\kappa$) to the quasi-Hopf algebra 
$U\vect^\FF(\Mcal)$. It has the same algebra structure as $U\vect(\Mcal)$ and coproduct $\Delta_\FF=\FF\,\Delta\, \FF^{-1}$; explicitly,
on the basis vector fields we have
\begin{eqnarray}
\Delta_\FF(\partial_\mu) &=& 1\otimes \partial_\mu + \partial_\mu\otimes 1
                         \ ,\nonumber\label{DeltaPartial1}\\[4pt]
\Delta_\FF(\tilde\partial^{{\mu}}) &=& 1\otimes \tilde\partial^{{\mu}} + \tilde\partial^{{\mu}}\otimes
1 +\ii\kappa \, R^{\mu\nu\rho}\, \partial_\nu\otimes\partial_\rho \ . \label{DeltaPartial2} 
\end{eqnarray}
The quasi-antipode is $S_\FF=S$, where the quasi-antipode elements $\al_\FF$ and $\beta_\FF$ are
the identity in the case of the twist ${\cal F}$, because $\al=\be=1$ in ${U}\vect(\Mcal)$
 and
\beq
{\tt f}\,^\al\, S({\tt f}\,_\al)=  {\tt  f}_R^{\,\al}\, S({\tt
  f}_{R\,\al})=\overline{\tt f}\,^\al\, S(\,\overline{\tt f}\,_\al)=
\overline{\tt  f}_R^{\,\al}\, S(\,\overline{\tt
  f}_{R\,\al})=1 \ , 
\label{eq:fS1}\eeq
see \cite[Section~2.2]{Mylonas2013}. The counit is
$\epsilon_\FF=\epsilon$.
The twist ${\cal F}$ does not fulfill the 2-cocycle condition, and instead one obtains
\begin{equation}
\Phi \, (\FF\otimes 1)\, (\Delta\otimes \Id)\FF=(1\otimes
\FF)\, (\Id\otimes \Delta)\FF \ , \label{cocycle}
\end{equation}
where the element $\Phi$, called the associator, is the Hopf 3-cocycle
\beq
\Phi=\exp\big(\, \mbox{$\frac{\ell_s^3}{6}$}\,
R^{\mu\nu\rho}\, \partial_\mu\otimes\partial_\nu\otimes\partial_\rho\, \big)
=:\phi_1\otimes \phi_2\otimes \phi_3 = 1\otimes1\otimes1 + O(\ell_s^3)
\label{eq:associator}\eeq
with summation understood in the second expression (as in e.g.
$\FF={\rm f}\,^\al \otimes {\rm f}\,_\al$). The inverse 
associator is denoted
$\Phi^{-1}=:\bar\phi_1\otimes\bar\phi_2\otimes\bar\phi_3$. The failure
of the 2-cocycle condition implies that the twisted coproduct $\Delta_\FF$
is no longer coassociative, as one sees from the quasi-coassociativity relation
\bea
\Phi \ (\Delta_\FF\otimes\Id)\,\Delta_\FF(\xi) =
(\Id\otimes\Delta_\FF)\,\Delta_\FF(\xi) \: \Phi
\label{eq:DeltaPhi}\eea
for  all $\xi\in U\vect(\Mcal)$.

The sextuple $(U\vect(\Mcal), \, \cdot \, ,\Delta_\FF,\Phi, S,\epsilon)$
defines on the vector space $U\vect(\Mcal)$ the structure of a
quasi-Hopf algebra $U\vect^\FF(\Mcal)$~\cite{Drinfeld}. In
$U\vect^\FF(\Mcal)$ the only relaxation of the Hopf algebra structure is the presence of a
non-trivial associator $\Phi$ for the coproduct $\Delta_\FF$. 
The quasi-Hopf algebra $U\vect^\FF(\Mcal)$ will play the role of the symmetry
algebra of infinitesimal diffeomorphisms of the nonassociative
deformation of phase
space ${\cal M}$.

For later use, we rewrite the relation (\ref {eq:DeltaPhi}), which
expresses the failure of coassociativity of $\Delta_\FF$, in the form
\bea
\big(\xi_{(1)_{(1)}}\otimes\xi_{(1)_{(2)}}\otimes\xi_{(2)}\big)\,
\Phi^{-1} = \Phi^{-1}\, \big(\xi_{(1)}\otimes\xi_{(2)_{(1)}}\otimes
  \xi_{(2)_{(2)}} \big) \ .
\label{eq:DeltaPhirewrite}\eea
Here we introduced the Sweedler notation
\bea
\Delta_\FF(\xi)=:\xi_{(1)}\otimes\xi_{(2)}
\eea
for the coproduct (with
implicit summation) and its iterations, for example
\bea
(\Delta_\FF\otimes\Id)\Delta_\FF(\xi)=: \xi_{(1)_{(1)}}\otimes\xi_{(1)_{(2)}}\otimes
\xi_{(2)} \ . 
\eea
Recalling that the quasi-antipode is just the
undeformed antipode $S$, we also observe that its compatibility with the 
coproduct $\Delta_\FF$, 
\bea\label{eq:Sepsilon}
\xi_{(1)}\, S(\xi_{(2)})= \epsilon(\xi) = S(\xi_{(1)})\, \xi_{(2)}~,
\eea
for all $\xi\in U\vect^\FF(\Mcal)$, follows from the equalities \eqref{eq:fS1}.

A further relevant property of the quasi-Hopf algebra
$U\vect^\FF(\Mcal)$ is its triangularity.  We denote by
$\Delta_\FF^{\rm op}$ the opposite coproduct, obtained by
flipping the two legs of $\Delta_\FF$, i.e.,  if
$\Delta_\FF(\xi)=\xi_{(1)}\otimes\xi_{(2)}$ then $\Delta_\FF^{\rm
  op}(\xi):=\xi_{(2)}\otimes\xi_{(1)}$, for all $\xi\in
U\vect^\FF(\Mcal)$.
Since the coproduct of $U\vect(\Mcal)$ satisfies $\Delta=\Delta^{\rm op}$,
it follows immediately that the coproduct $\Delta_\FF$ of
$U\vect^\FF(\Mcal)$ satisfies the property $\Delta_\FF ^{\rm op}(\xi)={\cal
  R}\, \Delta_\FF(\xi)\, {\cal R}^{-1}$,  for all
$\xi\in U\vect^\FF(\Mcal)$, with the $\cal R$-matrix given by 
\beq
{\cal R} =
{\cal F}^{-2}=: {\rm R}\,^\al\otimes {\rm R}\,_\al \ .
\eeq
Its inverse ${\cal R}^{-1}={\cal F}^2=:\oR\,^\al\otimes\oR\,_\al$ satisfies
\bea\label{eq:triangular}
\oR\,^\al\otimes \oR\,_\al = {\rm R}\,_\al\otimes{\rm R}\,^\al
\eea
so that the ${\cal R}$-matrix is triangular. The quasi-Hopf algebra
$U\vect^\FF(\Mcal)$ with this ${\cal R}$-matrix is a
triangular quasi-Hopf algebra \cite{Kassel, Barnes2014}.
The coproduct of the inverse of the ${\cal R}$-matrix can be explicitly
computed and reads
\bea
(\Delta_\FF\otimes\Id){\cal R}^{-1}&=&
\Phi^{-3} \ \oR\,^\beta\otimes\oR\,^\al \otimes\oR\,_\al\,\oR\,_\beta \
, \label{eq:DeltaRPhia} \\[4pt]
(\Id\otimes\Delta_\FF){\cal R}^{-1}&=&
\Phi^3 \ \oR\,^\al\,\oR\,^\beta\otimes \oR\,_\al\otimes\oR\,_\beta \ .
\label{eq:DeltaRPhib}\eea
These equalities are a simplified version of the compatibility conditions between
the coproduct $\Delta_\FF$ and the ${\cal R}$-matrix that is due to
the antisymmetry of the trivector $R^{\mu\nu\rho}$ (see
(\ref{Rantsymm}) below).

\subsection{Associator identities}

There are various noteworthy identities for the associator that arise for
the particular cochain twist induced by the constant $R$-flux
background, which we summarise here as they will be used extensively in
our calculations throughout this paper. 

A main simplification is that the legs of the
associator commute among themselves, $\phi_a\,\phi_b=\phi_b\,\phi_a$,
and also with the legs of the twist, $\phi_a\,{\rm f}\,^\al={\rm
  f}\,^\al\, \phi_a$ and $\phi_a\,{\rm f}\,_\al={\rm f}\,_\al\,
\phi_a$.
Moreover,
by antisymmetry of the trivector $R$, we have
\begin{eqnarray}\label{Rantsymm}
\phi_a\otimes\phi_b\otimes\phi_c = \bar\phi_a\otimes\bar\phi_c\otimes\bar\phi_b
\end{eqnarray}
and  $\phi_a\otimes\phi_b\,\phi_c=1\otimes1$, where here and in the
following $(a,b,c)$ denotes a permutation of $(1,2,3)$.
Furthermore, since the antipode is an antihomomorphism
we have
\bea
(\Id\otimes\Id\otimes S)\Phi = \exp\big(\, \mbox{$\frac{\ell_s^3}{6}$}\,
R^{\mu\nu\rho}\, \partial_\mu\otimes\partial_\nu\otimes
S(\partial_\rho) \,\big) = \exp\big(-\mbox{$\frac{\ell_s^3}{6}$}\,
R^{\mu\nu\rho}\, \partial_\mu\otimes\partial_\nu\otimes \partial_\rho
\big) 
\eea
which leads to
\begin{eqnarray}
\phi_a\otimes\phi_b \otimes S(\phi_c)=\bar\phi_a\otimes\bar\phi_b\otimes\bar\phi_c \ .
\label{eq:antipodeid}\end{eqnarray}

Since the coproduct $\Delta: U\vect(\Mcal)\to
U\vect(\Mcal)\otimes U\vect(\Mcal) $ is an algebra homomorphism, we have
\bea
(\Id\otimes\Id\otimes\Delta)\Phi &=& \exp\big(\, \mbox{$\frac{\ell_s^3}{6}$}\,
R^{\mu\nu\rho}\, \partial_\mu\otimes\partial_\nu\otimes\Delta(\partial_\rho)
\, \big) \nn \\[4pt]
&=& \exp\big(\, \mbox{$\frac{\ell_s^3}{6}$}\,
R^{\mu\nu\rho}\,
(\partial_\mu\otimes\partial_\nu\otimes \partial_\rho\otimes 1
+ \partial_\mu\otimes \partial_\nu\otimes 1\otimes \partial_\rho)\,
\big) \\[4pt]
&=&  \exp\big(\, \mbox{$\frac{\ell_s^3}{6}$}\,
R^{\mu\nu\rho}\, \partial_\mu\otimes\partial_\nu\otimes\partial_\rho\otimes
1\, \big)\,\exp\big(\, \mbox{$\frac{\ell_s^3}{6}$}\,
R^{\mu\nu\rho}\, \partial_\mu\otimes\partial_\nu\otimes1\otimes \partial_\rho\,
\big) \nonumber\\[4pt]
&=&\phi_1\,
\varphi_1\otimes\phi_2\,\varphi_2\otimes\phi_{3}\otimes\varphi_{3} ~
\nn \ ,
\eea
where here and in the following we use different
symbols for multiple associator insertions in order to avoid
confusion.
We further have
$(\Id\otimes\Id\otimes\Delta)\Phi=(\Id\otimes\Id\otimes\Delta_\FF)\Phi$
because $\Delta_\FF(\xi)=\FF\, \Delta(\xi)\, \FF^{-1}$ and $\FF$ commutes
with the legs of the associator. Hence we have 
\begin{eqnarray}
\phi_a\otimes\phi_b\otimes\phi_{c_{(1)}}\otimes\phi_{c_{(2)}}= \phi_a\, \varphi_a\otimes\phi_b\,\varphi_b\otimes\phi_{c}\otimes\varphi_{c} \ .
\label{eq:phicoprodid}\end{eqnarray}

Finally, we also rewrite the identity $\Phi\,
\Phi^{-1}=\Id$ as 
\bea
\phi_a\,\bar\varphi_a\otimes\phi_b\,\bar\varphi_b\otimes\phi_{c}\,\bar\varphi_c=1\otimes1\otimes1 \ .
\eea

\subsection{Double field theory formulation\label{sec:DFT}}

Before deforming the geometry of ${\cal M}$ into a nonassociative differential geometry with the cochain twist $\FF$  and
the associated quasi-Hopf algebra $U\vect^\FF(\Mcal)$,
let us describe the general extent to which our results will be
applicable, particularly from the perspective of double field theory, as they will mostly be suppressed in the following in
order to streamline our presentation and formulas.

Firstly, if globally non-geometric $Q$-flux is also present~\cite{Lust2010,Condeescu2012,Andriot2013,Blumenhagen2013},
then it has the effect of modifying the twist element \eqref{eq:calFFR} to
\bea
\FF= F\, F_R\, F_Q \ ,
\label{eq:FFRFQ}\eea
where
\bea
F_Q=\exp\big(-\mbox{$\frac{\ii\kappa}{2}$} \,
Q^{\mu\nu}{}_{\rho} \, ( w^\rho\, \partial_\mu\otimes \partial_\nu -
\partial_\nu\otimes w^\rho\, \partial_\mu) \big)
\eea
with $w^\mu$ closed string winding
coordinates which may be regarded as momenta $\hat p^\mu$ conjugate to coordinates
$\hat x_\mu$ that are T-dual to the spacetime variables $x^\mu$. The
twist $F_Q$ is an abelian 2-cocycle, and the vector fields
$w^\rho\, \partial_\mu$ commute with the other vector fields
$\partial_A$ and 
$p_\mu\, \partial_\nu$ generating the twists $F$ and $F_R$, so the
Hopf coboundary of \eqref{eq:FFRFQ} is
still the associator \eqref{eq:associator}; indeed, unlike the
$R$-flux, the $Q$-flux only sources noncommutativity. In this setting the $Q$-flux and $R$-flux are independent deformation parameters, while coexistent constant $Q$-flux and $R$-flux in string theory are constrained in the absence of geometric fluxes by the Bianchi identities
\bea
R^{\mu[\nu\rho}\,Q^{\lambda\tau]}{}_\mu=0 \qquad \mbox{and} \qquad Q^{[\nu\rho}{}_\mu\,Q^{\lambda]\mu}{}_\tau=0
\eea
which are easily imposed as additional constraints on the twist element \eqref{eq:FFRFQ}.

In fact,
analogously to~\cite{Bakas2013} one can extend
the twist element \eqref{twist} to the full phase space
$\Mcal\times\hat\Mcal$ of double field theory in the $R$-flux frame as
\bea
\hat\FF= \FF \, \hat F \ ,
\label{eq:DFTtwist}\eea
where 
\bea
\hat F =\exp\big(-\mbox{$\frac{\ii\hbar}{2}$} \, (\hat\partial^\mu\otimes \tilde{\hat\partial}_{{\mu}} -
\tilde{\hat\partial}_{{\mu}}\otimes \hat\partial^\mu)\big)
\eea
implements
the deformation of the canonical T-dual phase space $\hat\Mcal$ with
coordinates $\hat x_\mu,\hat p^\mu=w^\mu$. The cochain twist
\eqref{eq:DFTtwist} is $O(2d,2d)$-invariant so that it can be rotated to any other
T-duality frame by using an $O(2d,2d)$ transformation on
$\Mcal\times\hat\Mcal$, and in this way one can write down a
nonassociative theory which is manifestly invariant under $O(2d,2d)$
rotations. In particular, by restricting to rotations which also
preserve the canonical symplectic 2-form of double phase space, we
obtain a formulation that is invariant under a subgroup
\bea
O(d,d) \ \subseteq \ O(2d,2d)\cap Sp(2d) \ .
\eea
As the inclusion of $\hat F$ affects neither the
commutation nor the association relations on the original phase space
$\Mcal$, while yielding the standard Moyal-Weyl deformation of the T-dual
phase space $\hat\Mcal$, we will regard the phase space $\hat\Mcal$ as implicitly hidden in
the background in all of our subsequent treatments, with the
understanding that all of our formalism can be rotated to any
T-duality frame by including a dependence on the T-dual coordinates of
$\hat\Mcal$ and suitably inserting $\hat F$ in formulas. In this way we obtain
a manifestly $O(d,d)$-invariant formulation of the gravity theory
which follows.

From this perspective, there are also natural modifications of our
formalism, analogous to those of Moyal-Weyl spaces~\cite{AC}, which fit nicely into the flux formulation of double field
theory appropriate to curved backgrounds~\cite{Blumenhagen2013}. The vector fields
\bea
X_\mu=\partial_\mu \ , \qquad \tilde X^\mu=\tilde\partial^\mu \qquad
\mbox{and} \qquad X_{\mu\nu}=p_\mu\,\partial_\nu-p_\nu\,\partial_\mu
\label{eq:Xmunu}\eea
on $\Mcal$, defining the
twist deformation of flat space $M=\real^d$, represent a nilpotent
subalgebra $\mathfrak{k}$ of
the Lie algebra $\mathfrak{iso}(2d)$ with the nonvanishing Lie
brackets
\bea
\big[\tilde X^\mu,X_{\nu\rho}\big] = \delta^\mu{}_\nu\, X_\rho-
\delta^\mu{}_\rho\, X_\nu \ .
\eea
For any collection of vector fields
$\{X_\mu,\tilde X^\mu,X_{\mu\nu}\,,\,\mu,\nu=1,\dots,d\}$ satisfying these Lie
bracket relations on an \emph{arbitrary} manifold ${\cal M}$, the
cochain twist element
\bea
\FF_{\rm c} = \exp\big(-\mbox{$\frac{\ii\hbar}{2}$}\, (X_\mu\otimes
\tilde X^{{\mu}} -
\tilde X^{{\mu}}\otimes X_\mu)
-\mbox{$\frac{\ii\kappa}{4}$}\, R^{\mu\nu\rho}\, (X_{\nu\rho}\otimes X_\mu -
X_\mu \otimes X_{\nu\rho}) \big)
\eea
of the Hopf algebra $U \mathfrak{iso}(2d)\subset U\vect(\Mcal)$ provides a nonassociative deformation of $\Mcal$, all of whose features
fit into the framework we develop in the following. 

For example, in the cases considered in the present paper we will see that
there is a preferred basis $\partial_A, \dd x^A$ of vector fields  and
1-forms on $\Mcal$ that is invariant under the action of the associator and which greatly
simplifies calculations. 
In particular, the Cartan structure equations expressing torsion and
curvature as operators on vector fields will be established by 
checking that these operators define tensor fields, and by showing
that in the preferred basis they coincide with the torsion and curvature coefficients.
These simplifications can be carried out as well
for the more general twist $\FF_{\rm c}$, by considering as basis the commuting
vector fields $X_\mu, \tilde X^\mu$ and their dual 1-forms; more generally,
if the vector fields $X_\mu, \tilde X^\mu$  do not form
a basis (e.g. they become degenerate), this can be achieved by completing them to a basis
that is still invariant under the action of the associator with the methods
described in~\cite[Section 4]{AC}.

\newsection{Nonassociative deformation of tensor calculus\label{sec:functions}}

\subsection{Principles of twist deformation}

The tensor algebra on $\cal M$ is covariant under the action of the
universal enveloping algebra of infinitesimal diffeomorphisms
$U\vect(\Mcal)$. 
We have seen how the $R$-flux induces a twist deformation of
the Hopf algebra $U\vect(\Mcal)$ into the quasi-Hopf algebra
$U\vect^\FF(\Mcal)$.
We construct a nonassociative differential geometry on $\cal M$ by
requiring it to be covariant with respect to the quasi-Hopf algebra
$U\vect^\FF(\Mcal)$. 

Every time we
have an algebra $\cal A$ that carries a representation of the Hopf algebra
$U\vect(\Mcal)$, and where vector fields $u\in \vect(\Mcal)$ act on
$\cal A$ as derivations: $u(a\, b)=u(a)\, b+a \, u(b)$, i.e., every time we
have a $U\vect(\Mcal)$-module algebra $\cal A$, then deforming the multiplication in
$\cal A$ into the star-multiplication
\eq
a\star b=\of\,^\al(a)\, \of\,_\al(b)
\en
yields a noncommutative and nonassociative algebra ${\cal A}_\st$
that carries a representation of the quasi-Hopf algebra
$U\vect^\FF(\Mcal)$, where
\eq
\xi(a\star b)=\xi_{(1)}(a)\star \xi_{(2)}(b) \ ,
\label{starA}\en
for all $\xi\in U\vect^\FF(\Mcal)$ and
$a,b\in \cal A$; in particular, the vector fields $\partial_\mu$ and $\tilde\partial^\mu$ act as deformed
derivations according to the Leibniz rule implied by the coproduct (\ref{DeltaPartial2}).
We say that ${\cal A}_\st$ is a $U\vect^\FF(\Mcal)$-module algebra
because of the compatibility  (\ref{starA}) of the action of $U\vect^\FF(\Mcal)$ with the
product in  ${\cal A}_\st$.
For later use we recall the proof of the key property (\ref{starA}):
\eqa\label{proofstarA}
\xi(a\star b) &=& \xi\big(\,\of\,^\al(a)\, \of\,_\al(b) \big) \nn\\[4pt]
&=&
\xi_{(1_0)}\big(\,\of\,^\al(a)\big)\, \xi_{(2_0)}\big(\,\of\,_\al(b)\big)\nn\\[4pt]
&=&
\of\,^\al\big(\xi_{(1)}(a)\big) \, \of\,_\al\big(\xi_{(2)}(b)\big)
\nn\\[4pt]
&=& \xi_{(1)}(a)\star \xi_{(2)}(b) \ ,
\ena
where we used the notation $\Delta(\xi)=\xi_{(1_0)}\otimes
\xi_{(2_0)}$ for the undeformed coproduct together with
$\Delta(\xi)\, \FF^{-1}=\FF^{-1}\, \Delta_\FF(\xi)$.

If the algebra $\cal A$ is commutative then the
noncommutativity of ${\cal A}_\st$ is controlled by the
${\cal R}$-matrix as
\beq\label{Rmab}
a\star b=\oR\,^\al(b)\star\oR\,_\al(a)=: {}^\al b\star {}_\al a \ ,
\eeq
where in the last equality we used the notation ${}^{\al\!} a:=\oR\,^\al(a)$ and ${}_\al
a:=\oR\,_\al(a)$; the expression (\ref{Rmab}) is easily proven by recalling
that ${\cal R}=\FF^{-2}$ and that $a\star b=\of\,^\al(a)\, \of\,_\al(b)=\of\,_\al(b)\, \of\,^\al(a)$.
If the algebra $\cal A$ is associative then the nonassociativity of
${\cal A}_\st$ is controlled by the associator $\Phi$ as
\begin{equation}
(a\star b)\star c = {}^{\phi_1}a\star ({}^{\phi_2} b\star
{}^{\phi_3}c) \ ,\label{fgh} 
\end{equation}
where we denote ${}^{\phi_1}a:= \phi_1(a)$; an explicit proof can be found in \cite[Section 4.2]{Aschieri2015}.

In the following we deform the algebra of functions, the exterior
algebra of differential forms and the algebra of tensor fields on
$\Mcal$ according to this prescription.

\subsection{Functions}

The action of a vector field $ u\in\vect(\Mcal)$ on a function $f\in
C^\infty(\Mcal)$ is via the Lie derivative $\Lie_u(f)= u(f)$, which is
indeed a derivation. The action of the Lie
algebra $\vect(\Mcal)$ on functions is extended to an action of the
universal enveloping algebra $U\vect(\Mcal)$ by defining the
Lie derivative on products of vector fields as $\Lie_{u_1\, u_2 \,\cdots\, u_n}:=\Lie_{u_1}\circ \Lie_{u_2}\circ
\cdots\circ\Lie_{u_n}$ and by linearity. 
The $U\vect(\Mcal)$-module algebra
$C^\infty(\Mcal)$ (extended with power series in $\hbar$ and $\kappa$) is then deformed to the 
$U\vect^\FF(\Mcal)$-module algebra $A_\st:=C^\infty(\Mcal)_\st$,
which as a vector space is the same as $C^\infty(\Mcal)$ but with
multiplication given by the star-product
\begin{eqnarray}
f\star g &=& \of\,^\al(f)\cdot \of\,_\al(g) \label{fstarg}\\[4pt]
&=& f\cdot g + \mbox{$\frac{\ii\hbar}{2}$}\,\big(\partial_\mu f\cdot \tilde\partial^{{\mu}}g \,-\,
\tilde\partial^{{\mu}}f\cdot \partial_\mu g\big)
    +\ii\kappa \, R^{\mu\nu\rho}\, p_\nu\,
    \partial_\rho f\cdot \partial_\mu g \,+\, \cdots \ , \nn 
\end{eqnarray}
where the ellipses denote terms of higher order in $\hbar$ and $\kappa$. Noncommutativity is
controlled by the $\cal R$-matrix as
$f\star g=\oR\,^\al(g)\star \oR\,_\al(f)=: {}^\al g\star {}_\al f,
$
and nonassociativity by the associator $\Phi$ as $
(f\star g)\star h = {}^{\phi_1}f\star ({}^{\phi_2} g\star
{}^{\phi_3}h) $. The constant function $1$ on $\Mcal$ is also the unit
of $A_\st$ because $f\st 1=f=1\st f$.

Denoting the star-commutator of functions by
$[f,g]_\star:= f\st g-g\st f$, we reproduce in this way the defining
phase space quasi-Poisson coordinate algebra
\bea
[x^\mu,x^\nu]_\st = 2\ii\kappa\, R^{\mu\nu\rho}\, p_\rho \ , \qquad
[x^\mu,p_\nu]_\st = \ii\hbar\, \delta^\mu{}_\nu \qquad \mbox{and}
\qquad [p_\mu,p_\nu]_\st = 0
\eea
of the parabolic $R$-flux background, with the non-vanishing
Jacobiators
\bea
[x^\mu,x^\nu,x^\rho]_\st = \ell_s^3\, R^{\mu\nu\rho} \ .
\eea

\subsection{Forms}

Similarly to \eqref{fstarg}, we can deform the exterior algebra of
differential forms $\Omega^\sharp(\Mcal)$ by introducing the star-exterior product
\beq
\omega\wedge_\star\eta = \of\,^\al(\omega)\wedge \of\,_\al(\eta) \ .
\eeq
The algebra of differential forms with the nonassociative product
$\wedge_\star$ is denoted $\Omega_\star^\sharp$, with $\Omega_\star^0=A_\star$.
Here too the vector fields in the twist  act on differential forms via the Lie
derivative; in particular, for the basis 1-forms we find
\begin{eqnarray}\
{\cal L}_{\partial_A} (\mathrm{d}x^B) = 0 
  \label{LieTwistForms1} 
\end{eqnarray}
along with (recalling that ${\tilde x}_\mu:=p_\mu$)
\begin{eqnarray}
{\cal L}_{R^{\mu\nu\rho}\, X_{\nu\rho}} (\mathrm{d}x^\sigma)
  = 2\,R^{\mu\nu\sigma}\, \mathrm{d}\tilde x_{{\nu}} \qquad \mbox{and} \qquad 
{\cal L}_{R^{\mu\nu\rho}\, X_{\nu\rho}}(\mathrm{d}\tilde
  x_{{\sigma}}) =0 \ , \label{LieTwistForms3}
\end{eqnarray}
where the vector fields $X_{\mu\nu}$ are defined in \eqref{eq:Xmunu}
and we used the fact that the Lie derivative commutes with the exterior derivative.
Iterating the commutativity of the exterior derivative $\dd:\Omega_\st^\sharp\to \Omega_\st^{\sharp+1}$ with the Lie
derivative along vector fields implies $\dd\, \of\,^\al(\omega)=\of\,^\al(\dd \omega)$
and $\dd\, \of\,_\al(\omega)=\of\,_\al(\dd \omega)$, giving the undeformed
Leibniz rule
\bea
\dd(\omega\wedge_\st\eta)=\dd\omega\wedge_\st\eta+(-1)^{|\omega|}\,
\omega\wedge_\star\dd \eta \ 
\eea
where $\omega$ is a homogeneous form of degree $|\omega|$.

The star-exterior product of 1-forms $\dd x^A$ reduces to the usual
antisymmetric associative exterior product:
Using  (\ref{LieTwistForms1})  we have
\begin{eqnarray}
\mathrm{d}x^A \wedge_\star \mathrm{d}x^B = \mathrm{d}x^A \wedge 
\mathrm{d}x^B  = - \mathrm{d}x^B \wedge \mathrm{d}x^A = -\mathrm{d}
x^B \wedge_\star \mathrm{d}x^A \ . \label{goodbasis1} 
\end{eqnarray}
In particular, the volume element is undeformed.
For this, we note that the action of the associator \eqref{eq:associator}
trivializes on the exterior products of basis 1-forms: In the case of three basis
1-forms we obtain
\begin{eqnarray}
(\mathrm{d}x^A \wedge_\star \mathrm{d}x^B)\wedge_\star \mathrm{d}x^C &=&
{}^{\phi_1} (\mathrm{d}x^A) \wedge_\star \big({}^{\phi_2} (\mathrm{d}x^B)\wedge_\star
{}^{\phi_3} (\mathrm{d}x^C) \big)\nn\\[4pt]
&=& \mathrm{d}x^A \wedge_\star (\mathrm{d}x^B\wedge_\star
\mathrm{d}x^C) \nn\\[4pt] &=& \mathrm{d}x^A \wedge \mathrm{d}x^B\wedge
\mathrm{d}x^C \ ,
\end{eqnarray}
where $\phi_{a}$ act via Lie derivatives on forms and we used (\ref{LieTwistForms1}). 

The exterior product between 0-forms (functions) and 1-forms gives
the space of 1-forms the structure of a $C^\infty(\Mcal)$-bimodule. Similarly,
restricting the star-exterior product of differential forms to functions and 1-forms defines
the $A_\st$-bimodule structure of the space of 1-forms $\Omega_\st^1$.  In particular,
the star-exterior product of functions and basis 1-forms is given by
\begin{eqnarray}
f\star \mathrm{d}\tilde x_{{\mu}} &=&  f\cdot \mathrm{d}\tilde
                                      x_{{\mu}}  \ \ = \ \ 
                                      \mathrm{d}\tilde x_{{\mu}}\star
                                      f \ , \nn\\[4pt]
f\star \mathrm{d}x^\mu &=& f\cdot \mathrm{d}x^\mu 
-\mbox{$\frac{\ii\kappa}{2}$}\, R^{\mu\nu\rho}\,
                           \partial_\nu f\cdot \mathrm{d}\tilde
                           x_{{\rho}} \ \ = \ \ 
\mathrm{d}x^\mu\star f -\dd\tilde x_\rho\star \ii\kappa \,
                           R^{\mu\nu\rho}\, \partial_\nu
f  \ . \label{fstardx}
\end{eqnarray}
Similarly to~\cite{Blumenhagen2016}, it is convenient to package the
relations in \eqref{fstardx} into a single relation by defining an antisymmetric
tensor $\RRR^{AB}{}_C$ on $\Mcal$ whose only non-vanishing components are $\RRR^{x^\mu,x^\nu}{}_{\tilde x_\rho}= R^{\mu\nu\rho}$, so that
\bea
f\star \dd x^A=\dd
x^C\star\big(\delta^A{}_C\, f -\ii\kappa \, \RRR^{AB}{}_C\,\partial_Bf\big) \ .
\label{fstardxA}\eea
As a useful special case, this implies
\bea
\dd f=\partial_Af \, \dd x^A = \partial_Af\star \dd x^A= {\rm d}x^A \star 
\partial_A f
\eea
by antisymmetry of $R^{\mu\nu\rho}$ (and hence of $\RRR^{AB}{}_C$).

\subsection{Tensors}\label{sectTens}

The usual tensor product $\otimes_{C^\infty(\Mcal)}$ over $C^\infty(\Mcal)$ is deformed to the star-tensor product $\otimes_\star$ over $A_\star$ defined by
\bea
T\otimes_\star U=\of\,^\al(T)\otimes_{C^\infty(\Mcal)} \of\,_\al(U) \ ,
\eea
where the action of the twist on the tensor fields $T$ and $U$  is via the Lie derivative. 
Due to nonassociativity, for $f\in A_\star$ one has
\bea\label{eq:quotientrel}
(T\star f)\otimes_\star U = {}^{\phi_1}T\otimes_\star ({}^{\phi_2}f\star {}^{\phi_3}U) \ .
\eea
Here the star-tensor product $\otimes_\star$ between functions and tensor fields
is denoted $\st$. In particular, it gives the space of vector fields $\vect({\cal M})$ an
$A_\st$-bimodule structure. We denote $\vect({\cal M})$ with this
$A_\st$-bimodule structure by $\vect_\st$.  In order to explicitly write the star-product
between functions and  the basis vector fields
$\partial_\mu, \tilde\partial^\mu$, we first compute the Lie derivative
action of the vector fields in the twist on the basis vector fields (i.e., the Lie brackets):
\begin{eqnarray}
{\cal L}_{\partial_A} (\partial_B) = 0 
  \label{LieTwistVecs1} 
\end{eqnarray}
together with
\begin{eqnarray}
{\cal L}_{R^{\mu\nu\rho}\, X_{\nu\rho}} (\partial_\sigma)
  = 0 \qquad \mbox{and} \qquad 
{\cal L}_{R^{\mu\nu\rho}\, X_{\nu\rho}}(\tilde\partial^{{\sigma}}) =-2\,R^{\mu\nu\sigma}\, \partial_{{\nu}} \ . \label{LieTwistVecs3}
\end{eqnarray}
Then we have
\bea
f\star \partial_{{\mu}} &=&  f\cdot \partial_{{\mu}}  \ \ = \ \ 
                                      \partial_{{\mu}}\star
                                      f \ , \nn\\[4pt]
f\star \tilde\partial^\mu &=& f\cdot \tilde\partial^\mu 
-\mbox{$\frac{\ii\kappa}{2}$}\, R^{\mu\nu\rho}\,
                           \partial_\nu f\cdot \partial_{{\rho}} \ \ = \ \ 
\tilde\partial^\mu\star f + \partial_\nu\star \ii\kappa \,
                           R^{\mu\nu\rho}\, \partial_\rho
f \ ,
\eea
where here $\partial_\mu\star f$ denotes the right $A_\st$-action on
$\vect_\st$ (and not the action of $\partial_\mu$ on the function
$f$). Again we can write these relations collectively in the form
\bea\label{RRRpartial}
f\star\partial_A = \partial_C\star\big( \delta^C{}_A\,
f+\ii\kappa\, \RRR^{CB}{}_A\,\partial_Bf\big) \ .
\eea

Using the star-tensor product, we can extend the $A_\star$-bimodule
$\vect_\star$ of vector fields to the $\Omega_\star^\sharp$-bimodule
$\vect_\star^\sharp= \vect_\star\otimes_\star\Omega_\star^\sharp$: The
left and right actions of the exterior algebra $\Omega_\st^\sharp$ on
$\vect_\star^\sharp$ are given by
\bea
(u\otimes_\star\omega)\wedge_\star\eta&=&{}^{\phi_1}u\otimes_\star({}^{\phi_2}
\omega\wedge_\star {}^{\phi_3} \eta) \ , \nn \\[4pt]
\eta\wedge_\star(u\otimes_\star\omega)&=&
{}^{\al}\big({}^{\bar{\phi}_1}({}_\be u)\big)
\otimes_\star\big({}_{\al_{(1)}}({}^{\bar{\phi}_2} \eta)
\wedge_\star{}^\beta({}_{\al_{(2)}}({}^{\bar{\phi}_3}\omega)) \big) \ ,
\label{eq:Omegavectactions}\eea
where ${}^{\al_{(a)}} (T):={\oR\,^\al}_{\!\!{}_{}(a)}(T)$ and ${}_{\al_{(a)}}
T:={\oR\,_{\al_{}}}_{(a)}(T)$ for $T$ a tensor or a form; the left action in \eqref{eq:Omegavectactions}
follows from
\bea
\eta\wedge_\star(u\otimes_\star\omega)=
\eta\wedge_\st({}^\be\omega\otimes_\st {}_\be u) =
({}^{\bar\phi_1}\eta\wedge_\st{}^{\bar\phi_2\,\be}\omega) \otimes_\st
{}^{\bar\phi_3}{}_\be u = {}^{\al}({}^{\bar\phi_3}{}_\be u)\otimes_\st
{}_\al({}^{\bar\phi_1}\eta \wedge_\st{}^{\bar\phi_2\,\be}\omega)
\eea
where in the first equality we used the fact that the tensor product between
contravariant and covariant
tensors is commutative, in particular  $u\otimes_{C^\infty(\Mcal)} \omega=\omega
\otimes_{C^\infty(\Mcal)}  u$, and similarly in the
last equality.

Analogously, we can extend the $A_\star$-bimodule of 1-forms
$\Omega_\star^1$ to an $\Omega_\star^\sharp$-bimodule
$\Omega^\sharp_\star\otimes_\star\Omega_\star^1$. 

\subsection{Duality}\label{Duality}

The three star-multiplications $\star$, $\wedge_\st$ and $\otimes_\st$ thus far constructed
are compatible with the $U\vect^\FF(\Mcal)$-action according to (\ref{starA}). This compatibility can be regarded as 
equivariance of these products under the $U\vect^\FF(\Mcal)$-action: There is no action of $\xi$ on the star-multiplication in
(\ref{starA}), only on (the functions, forms or tensors) $a$
and $b$. This notion of equivariance under the universal enveloping
algebra of diffeomorphisms $U\vect^\FF(\Mcal)$ (invariance and covariance in physics parlance) is the guiding principle in constructing a
noncommutative and nonassociative differential geometry on ${\cal M}$.
The recipe thus far considered, which consists in deforming a multiplication
$m$ to the star-multiplication $\star$ defined by composing the classical
product with the inverse twist, $\star:=m\circ \FF^{-1}$, extends more
generally to any bilinear map that is equivariant under infinitesimal
diffeomorphisms, i.e., under $U\vect(\Mcal)$.

For example, the pairing between 1-forms and vectors $\langle ~\,,\,~\rangle:
\Omega^1(\Mcal) \times \vect({\cal M})\to C^\infty(\Mcal)$ is deformed to the star-pairing
\eq
\langle ~\,,\,~\rangle_\st:=\langle ~\,,\,~\rangle\circ \FF^{-1}\,: \,
\Omega_\st^1\times \vect_\st \ \longrightarrow \ C^\infty(\Mcal)~,
\en
which is explicitly given by
\begin{equation}
\le~ \omega\,,\, u ~\re_\st = \big\le~\of\,^\al(\omega)\,,\,\of\,_\al(u)~\big\re~ . \label{pairing}
\end{equation}
Equivariance of the star-pairing under the quasi-Hopf algebra
$U\vect^\FF(\Mcal)$,
\begin{equation}
{}^\xi\le~ \omega\,,\, u ~\re_\st = \le~ {}^{\xi_{(1)}}\omega\,,\,
{}^{\xi_{(2)}} u ~\re_\st~,
\end{equation}
follows from equivariance of the undeformed pairing under the Hopf algebra
$U\vect(\Mcal)$, with the proof being analogous to (\ref{proofstarA}).

Since the usual pairing is $C^\infty(\Mcal)$-linear: 
$\le~ \omega\cdot f\,,\, u~\re = \le~ \omega\,,\, f\cdot u~\re$,
$\le~ f\cdot \omega\,,\, u~\re = f\cdot\le~ \omega\,,\, u~\re$, and
$\le~ \omega\,,\, u \cdot f~\re =\le~ \omega\,,\, u~\re \cdot f
$,
it follows that the star-pairing is $A_\st$-linear:
\begin{eqnarray}
\le~ \omega\star f\,,\, u~\re_\st &=& \le~
                                      {}^{\phi_1}\omega\,,\,
                                      {}^{\phi_2} f\star {}^{\phi_3} u~\re_\st  \ ,\nn\\[4pt]
\le~ f\star \omega\,,\, u~\re_\st &=& {}^{\phi_1}f\star\le~ {}^{\phi_2}\omega\,,\, {}^{\phi_3}u~\re_\st 
 \ ,\nn\\[4pt]
\le~ \omega\,,\, u \star f~\re_\st &=& \le~ {}^{\bar{\phi}_1}\omega\,,\, {}^{\bar{\phi}_2}u~\re_\st \star
{}^{\bar{\phi}_3}f \ , \label{l-rpairing}
\end{eqnarray}
with the proof being analogous to that of quasi-associativity \eqref{eq:quotientrel} of the star-tensor
product. The first equality in \eqref{l-rpairing} shows that the star-pairing is a well-defined map
\begin{equation}
\langle \quad \rangle_\star \,:\, \Omega^1_\star\otimes_\star \vect_\star \ \longrightarrow \ A_\star ~.
\end{equation}
At zeroth order in the deformation parameters $\hbar$ and $\kappa$, this is just the canonical undeformed pairing $\langle\quad\rangle$ of 1-forms with vector fields which is nondegenerate, and hence the star-pairing $\langle \quad \rangle_\star$ is nondegenerate as well.
Because of (\ref{LieTwistForms1}) and (\ref{LieTwistVecs1}), the star-pairing between basis vector fields $\partial_A$ and basis 1-forms ${\rm d}x^A$ is
undeformed: $\le~ \dd x^A\,,\, \partial_B ~\re_\st = \delta^A{}_B$. 

This pairing can be extended to star-tensor products
\begin{equation}
\langle \quad \rangle_\star \,:\, \big(\Omega^1_\star\otimes_\star \Omega^1_\star\big) \otimes_\star \big(\vect_\star\otimes_\star \vect_\star\big) \ \longrightarrow \ A_\star 
\end{equation}
in the following way. Firstly, for $\omega,\eta\in\Omega^1_\star$ and $u\in\vect_\star$ we define the 1-form
\begin{equation}\label{eq:pair1forms}
\le~ (\omega\otimes_\star\eta) \,,\, u ~\re_\st := {}^{\phi_1}\omega \star\le~ {}^{\phi_2}\eta\,,\, {}^{\phi_3}u ~\re_\st~ .
\end{equation}
This definition is compatible with equivariance under the quasi-Hopf algebra action, since for $\xi\in U\vect^\FF(\Mcal)$ we have
\bea
{}^\xi\le~ (\omega\otimes_\star\eta) \,,\, u ~\re_\st &=& 
{}^\xi\big({}^{\phi_1}\omega \star\le~ {}^{\phi_2}\eta\,,\, {}^{\phi_3}u ~\re_\st\big) \nonumber\\[4pt]
&=& {}^{\xi_{(1)}\, \phi_1}\omega \star\, {}^{\xi_{(2)}} \le~ 
  {}^{\phi_2}\eta\,,\,  {}^{\phi_3}u ~\re_\st
\nonumber\\[4pt]
&=& {}^{\xi_{(1)}\, \phi_1}\omega \star\le~ {}^{\xi_{(2)_{(1)}}\,
  \phi_2}\eta\,,\, {}^{\xi_{(2)_{(2)}}\, \phi_3}u ~\re_\st
\nonumber\\[4pt]
&=& {}^{\phi_1\,\xi_{(1)_{(1)}}}\omega \star\le~ {}^{\phi_2\,\xi_{(1)_{(2)}}}\eta\,,\, {}^{\phi_3\,\xi_{(2)}}u ~\re_\st \nonumber\\[4pt]
&=&\le~ {}^{\xi_{(1)}}(\omega\otimes_\star\eta) \,,\, {}^{\xi_{(2)}}u ~\re_\st 
\eea
where in the second line we used the equivariance of the star-product, in
the third line the equivariance of the star-pairing, and in the fourth line the quasi-coassociativity property
\eqref{eq:DeltaPhirewrite}. We then define the pairing
\bea
\le~ (\omega\otimes_\star\eta) \,,\, (u\otimes_\star v) ~\re_\st &:=&
\le~ \le~ {}^{\bar\phi_1}(\omega\otimes_\star\eta) \,,\,
{}^{\bar\phi_2}u ~\re_\st \,,\, {}^{\bar\phi_3}v ~\re_\star \nn
\\[4pt]
&=& \le~ \le~ {}^{\bar\phi_1}\omega\otimes_\st{}^{\bar\varphi_1}\eta \,,\,
{}^{\bar\phi_2\, \bar\varphi_2}u ~\re_\st \,,\, {}^{\bar\phi_3\,
  \bar\varphi_3}v ~\re_\st \nn \\[4pt]
&=& \le~ {}^{\zeta_1\, \bar\phi_1}\omega\st \le~ {}^{\zeta_2\,
  \bar\varphi_1}\eta \,,\, {}^{\zeta_3\, \bar\phi_2}u ~\re_\st \,,\,
{}^{\bar\phi_3\, \bar\varphi_3}v ~\re_\st \nn \\[4pt]
&=& {}^{\rho_1\, \al}\le~ {}^{\zeta_2\, \bar\varphi_1}\eta \,,\,
{}^{\zeta_3\, \bar\phi_2}u ~\re_\st \st \le~ {}^{\rho_2\, \zeta_1\,
  \bar\phi_1}{}_\al\omega \,,\, {}^{\rho_3\, \bar\phi_3\, \bar\varphi_3}v
~\re_\st \ ,
\label{pairingtensor}\eea
and one again checks that it is equivariant under $U\vect^\FF(\Mcal)$ by using quasi-coassociativity \eqref{eq:DeltaPhirewrite}. This definition can be straightforwardly iterated to arbitrary star-tensor products.

\subsection{Module homomorphisms\label{sec:modulehom}}

Tensors can be regarded either as sections of vector bundles or as maps between sections of
vector bundles. In Section~\ref{sectTens} we have taken
the first point of view and deformed the product of sections to the star-tensor product.
Thanks to the pairing $\le ~ \,,\, ~\re_\st$, we can also consider the second
perspective; for example, for any 1-form $\omega$ the object $\le~ \om \,,\,~ \re_\st$ is
a right $A_\st$-linear map from the $A_\st$-bimodule $\vect_\st$ to
$A_\st$. More generally, given $A_\star$-bimodules $V_\star$ and $W_\star$,
we can consider the space of module homomorphisms (linear maps) 
${\rm hom}(V_\star,W_\star)$.
This space carries the adjoint action of the Hopf algebra, which is given by
\bea\label{adjact}
{}^\xi L(v) := ({}^\xi L)(v) = \xi_{(1)} \big(L (S(\xi_{(2)})(v)) \big) \ ,
\eea
for $\xi\in U\vect^\FF(\Mcal)$, $L\in{\rm hom}(V_\star,W_\star)$ and $v\in
V_\star$. It is straightforward to check equivariance of the
evaluation of $L$ on $v$:
\bea
{}^\xi\big(L(v)\big) =
{}^{\xi_{(1)}}L \big({}^{\xi_{(2)}}v\big) \ .
\label{eq:evalequiv}\eea
Indeed the right-hand side can be written as
\bea
{}^{\xi_{(1)}}L \big({}^{\xi_{(2)}}v\big) &=&
{}^{\xi_{(1)_{(1)}}}\big(L \big({}^{S(\xi_{(1)_{(2)}})\,{\xi_{(2)}}}v\big)\big)
\nonumber\\[4pt]&=& 
{}^{\bar\phi_1\,\xi_{(1)}\,\varphi_1}\big(L\big({}^{S(\bar\phi_2\,\xi_{(2)_{(1)}}\,\varphi_2)\,
\bar\phi_3\,\xi_{(2)_{(2)}}\, \varphi_3}v\big)\big) \nonumber\\[4pt]
&=& 
{}^{\xi_1\,\varphi_1}\big(L\big({}^{\varphi_2\,S(\xi_{(2)_{(1)}})\,
  \xi_{(2)_{(2)}}\, \varphi_3} v\big)\big) \nonumber\\[4pt]
&=& {}^\xi\big(L(v)\big) \ ,
\label{eq:evalequivproof}\eea
where we used \eqref{eq:DeltaPhirewrite}, antimultiplicativity of the
antipode $S$, the compatibility \eqref{eq:Sepsilon} and $\phi_a\otimes\phi_b\,
\phi_c=1\otimes1$.
Since the vector fields comprising the associator commute with those
of the twisting cochain $\FF$, using \eqref{eq:antipodeid} and
$\phi_a\otimes\phi_b\, \phi_c=1\otimes1$ we obtain the following
identities that will be frequently used:
\begin{eqnarray}\label{properties1}
\phi_a\otimes {}^{\phi_b}(v\star {}^{\phi_c}f) &=& \phi_a\otimes ({}^{\phi_b}v\star {}^{\phi_c} f) \ , \\[4pt]
{}^{\phi_a}L({}^{\phi_b}v)\otimes \phi_c &=& {}^{\phi_a}\big(L({}^{\phi_b}v)\big)\otimes \phi_c \
,\label{properties2}
\end{eqnarray}
for $v\in V_\star$, $f\in A_\star$ and $L\in\hom(V_\st,W_\st)$. 

We define the composition of homomorphisms by
\begin{eqnarray}
(L_1\bullet L_2)(v):= {}^{\phi_1}L_1\big({}^{\phi_2}L_2({}^{\phi_3}v) \big)
\end{eqnarray}
for $L_1\in {\rm hom}(W_\star,X_\star)$, $L_2\in{\rm
  hom}(V_\star,W_\star)$ and $v\in V_\star$. One can readily check
equivariance of this composition, i.e., compatibility with the $U\vect^\FF(\Mcal)$-action:
\bea
{}^\xi(L_1\bullet L_2)={}^{\xi_{(1)}}L_1\bullet {}^{\xi_{(2)}}L_2 \ ,
\eea
with the proof being similar to \eqref{eq:evalequivproof}, see also~\cite{Barnes2014}.
In particular, with this composition the $U\vect^\FF(\Mcal)$-module 
${\rm end}(V_\st)$ of linear maps on $V_\st$ is a
quasi-associative algebra:
\bea
(L_1\bullet L_2)\bullet L_3 = {}^{\phi_1}L_1\bullet
\big({}^{\phi_2}L_2\bullet {}^{\phi_3}L_3 \big) \ ,
\eea
for all $L_1,L_2,L_3\in {\rm end}(V_\st)$.
We define the twisted commutator of endomorphisms $L_1,L_2\in {\rm
  end}(V_\st)$ through
\beq
[L_1,L_2]_{\bullet}=L_1\bullet L_2-{}^\al L_2\bullet
{}_\al L_1 \ ,
\label{eq:braidedcomm}\eeq
where the braiding with the $\cal R$-matrix ensures equivariance of
$[~,~]_\bullet$ under the $U\vect^\FF(\Mcal)$-action:
$
{}^\xi[L_1,L_2]_\bullet =
\big[{}^{\xi_{(1)}}L_1,{}^{\xi_{(2)}}L_2\big]_\bullet $.

A map $L\in{\rm hom}(V_\star,W_\star)$ is \emph{right $A_\star$-linear} if
\begin{eqnarray}
L(v\star f)= {}^{\bar\phi_1}L({}^{\bar\phi_2}v)\star {}^{\bar\phi_3}f = {}^{\bar\phi_1}\big(L({}^{\bar\phi_2}v)\big)\star {}^{\bar\phi_3}f \ .
\end{eqnarray}
We denote the space of all such maps by ${\rm
  hom}_\star(V_\star,W_\star)$; it closes under the
$U\vect^\FF(\Mcal)$-action~\cite{Barnes2014}.
To see this explicitly, we need to show that if $L$ is right $A_\star$-linear, then so is ${}^\xi L$ for all $\xi\in U\vect^\FF(\Mcal)$. This follows from the calculation
\bea
{}^\xi L(v\st f) &=& {}^{\xi_{(1)}}\big(L\big({}^{S(\xi_{(2)_{(2)}})}v\st {}^{S(\xi_{(2)_{(1)}})}f\big)\big) \nn \\[4pt]
&=& {}^{\xi_{(1)}\,\bar\phi_1}\big[\big(L\big({}^{\bar\phi_2\, S(\xi_{(2)_{(2)}})}v\big)\big)\st{}^{\bar\phi_3\,S(\xi_{(2)_{(1)}})}f\big] \nn \\[4pt]
&=& {}^{\phi_1\,\xi_{(1)_{(1)}}}\big[\big(L\big({}^{S(\phi_3\,\xi_{(2)})}v\big)\big)\st{}^{S(\phi_2\,\xi_{(1)_{(2)}})}f\big] \nn \\[4pt]
&=& {}^{\phi_1\,\xi_{(1)_{(1)_{(1)}}}}\big(L\big({}^{S(\phi_3\,\varphi_3\,\xi_{(2)})}v\big)\big)\st {}^{\varphi_1\,\xi_{(1)_{(1)_{(2)}}}\, S(\phi_2\,\varphi_2\,\xi_{(1)_{(2)}})}f \nn \\[4pt]
&=& {}^{\phi_1\,\bar\eta_1\,\xi_{(1)_{(1)}}\, \rho_1}\big(L\big({}^{S(\phi_3\,\varphi_3\,\xi_{(2)})}v\big)\big) \st {}^{\varphi_1\,\bar\eta_2\,\xi_{(1)_{(2)_{(1)}}}\, \rho_2\, S(\phi_2\,\varphi_2\,\bar\eta_3\,\xi_{(1)_{(2)_{(2)}}}\, \rho_3)}f \nn \\[4pt]
&=& {}^{\phi_1\,\bar\eta_1\,\xi_{(1)_{(1)}}}\big(L\big({}^{S(\phi_3\,\varphi_3\, \xi_{(2)})}v\big)\big)\st{}^{\varphi_1\,\bar\eta_2\,\xi_{(1)_{(2)_{(1)}}}\, S(\xi_{(1)_{(2)_{(2)}}})\, S(\bar\eta_3\, \varphi_2\,\phi_2)}f \nn \\[4pt]
&=& {}^{\phi_1\,\xi_{(1)}}\big(L\big({}^{S(\xi_{(2)})\, S(\phi_3)}u\big)\big)\st{}^{S(\phi_2)}f \nn \\[4pt]
&=& {}^{\bar\phi_1}\big({}^\xi L\big)\big({}^{\bar\phi_2}v\big)\st{}^{\bar\phi_3}f \ ,
\eea
where the third equality follows from \eqref{eq:DeltaPhirewrite}, antimultiplicativity of the antipode $S$, and \eqref{Rantsymm}.

For later use, let us explicitly
demonstrate that the composition of $L_1\in {\rm
  hom}_{\star}(W_\star,X_\star)$ and $L_2\in{\rm
  hom}_{\star}(V_\star,W_\star)$ is a right $A_\st$-linear map
$L_1\bullet L_2\in {\rm hom}_{\star}(V_\star,X_\star)$; see
\cite{Barnes2014} for a general proof in the setting of arbitrary quasi-Hopf algebras. For this, we compute
\begin{eqnarray}
(L_1\bullet L_2)(v\star f)= {}^{\phi_1}\big(L_1\big({}^{\phi_2}(L_2({}^{\phi_3}(v\star f))) \big) \big)
\end{eqnarray}
using ${}^{\phi_3}(v\star f)={}^{\phi_{3_{(1)}}}v\star
{}^{\phi_{3_{(2)}}}f$ and the identity \eqref{eq:phicoprodid} to get
\begin{eqnarray}
(L_1\bullet L_2)(v\star f)&=&{}^{\phi_1\,\varphi_1} \big( L_1\big({}^{\phi_2\,\varphi_2}(L_2({}^{\phi_3}v\star {}^{\varphi_3}f)) \big) \big) \nonumber\\[4pt]
&=& {}^{\phi_1\,\varphi_1} \big( L_1\big({}^{\phi_2\,\varphi_2}\big[{}^{\bar\rho_1}( L_2({}^{\bar\rho_2\,\phi_3}v)) \star {}^{\bar\rho_3\,\varphi_3} f\big]\big) \big) \nonumber\\[4pt]
&=& {}^{\phi_1\,\varphi_1} \big( L_1\big({}^{\phi_{2_{(1)}}} \big[ {}^{\varphi_2\,\bar\rho_1} (L_2({}^{\bar\rho_2\,\phi_3}v)) \big] \star {}^{\phi_{2_{(2)}}\,\bar\rho_3\,\varphi_3}f\big) \big) \nonumber\\[4pt]
&=& {}^{\phi_1\,\check\phi_1\,\varphi_1} \big( L_1\big({}^{\phi_2\, \varphi_2\,\bar\rho_1} ( L_2({}^{\bar\rho_2\,\phi_3\,\check\phi_3}v) ) \star {}^{\check\phi_{2}\,\bar\rho_3\,\varphi_3}f\big) \big) \nonumber\\[4pt] 
&=& {}^{\phi_1\,\check\phi_1\,\varphi_1\, \bar\zeta_1} \big( L_1\big({}^{\bar\zeta_2} \big[ {}^{\phi_2\, \varphi_2\,\bar\rho_1} ( L_2({}^{\bar\rho_2\,\phi_3\,\check\phi_3}v) )  \big]\big) \big) \star {}^{\bar\zeta_3\, \check\phi_{2}\,\bar\rho_3\,\varphi_3}f \nonumber\\[4pt] 
&=& {}^{\phi_1\,\check\phi_1\,\varphi_1\, \bar\zeta_1} \big( L_1\big({}^{\bar\zeta_2} \big[ {}^{\phi_2\, \check\varphi_2\, \varphi_2\,\bar\rho_1} ( L_2({}^{\bar\rho_2\,\phi_3\,\check\phi_3\, \check\varphi_3}v) )  \big]\big) \big) \star {}^{\check\varphi_1\, \bar\zeta_3\, \check\phi_{2}\,\bar\rho_3\,\varphi_3}f\nonumber\\[4pt] 
&=& {}^{\phi_1\,\check\phi_1} \big( L_1\big({}^{\phi_2} ( L_2({}^{\phi_3\,\check\phi_3}v) ) \big) \big) \star {}^{\check\phi_2}f \nonumber\\[4pt] 
&=& {}^{\phi_1\,\overline{\check\phi}_1} \big( L_1\big({}^{\phi_2} ( L_2({}^{\phi_3\,\overline{\check\phi}_2}v) ) \big) \big) \star {}^{\overline{\check\phi}_3}f\nonumber\\[4pt] 
&=& {}^{\overline{\check\phi}_1} \big( (L_1\bullet L_2)({}^{\overline{\check\phi}_2}v) \big) \star {}^{\overline{\check\phi}_3}f \ ,
\end{eqnarray}
which establishes that $L_1\bullet L_2$ is right $A_\star$-linear.

For later use in our constructions of connections and curvature, we
will also
prove some properties of tensor products of right $A_\star$-linear
maps. Let $U_\star$, $V_\star$ and $W_\star$ be $A_\star$-bimodules. Then the
lifting of $L\in\hom_\star(U_\star,W_\star)$ to
$L\otimes\Id\in\hom_\st(U_\st\otimes_\st V_\st ,W_\st\otimes_\st V_\st)$ is
defined by
\bea\label{deflid}
(L\otimes\Id)(u\otimes_\star v) :=
({}^{\bar\phi_1}L)({}^{\bar\phi_2}u)\otimes_\star {}^{\bar\phi_3}v = {}^{\bar\phi_1}\big(L({}^{\bar\phi_2}u)\big)\otimes_\star{}^{\bar\phi_3}v
\eea
for $u\in U_\star$ and $v\in V_\star$. Let us first check
equivariance:
\bea
{}^\xi(L\otimes\Id)={}^\xi L\otimes \Id \ .
\eea
For this, we need to check that
\bea
{}^\xi\big((L\otimes\Id)(u\otimes_\st v)\big) =
{}^{\xi_{(1)}}\big(L\otimes \Id \big)\big({}^{\xi_{(2)}}(u\otimes_\st v)\big)
  = \big({}^{\xi_{(1)}}L\otimes
  \Id\big)\big({}^{\xi_{(2)}}(u\otimes_\st v)\big)
\eea
for arbitrary $u,v$ and for any $\xi\in
U\vect^\FF(\Mcal)$. This follows from the calculation
\bea
{}^\xi\big((L\otimes \Id)(u\otimes_\star v) \big) &=&
{}^{\xi_{(1)}}\big(\, {}^{\bar\phi_1}L({}^{\bar\phi_2}u)\big)
\otimes_\star{}^{\xi_{(2)}\,
  \bar\phi_3}v \nonumber \\[4pt] &=&
\big(\, {}^{\xi_{(1)_{(1)}}\,\bar\phi_1}L \big)
\big(\, {}^{\xi_{(1)_{(2)}}\,\bar\phi_2}u \big)
\otimes_\star
{}^{\xi_{(2)}\,\bar\phi_3}v \nonumber \\[4pt] 
&=& \big({}^{\bar\varphi_1\,\varphi_1\,\xi_{(1)_{(1)}}\,
  \bar\phi_1}L\big) \big(  {}^{\bar\varphi_2\, \varphi_2\,
  \xi_{(1)_{(2)}}\, \bar\phi_2}u\big)\otimes_\st {}^{\bar\varphi_3\,
  \varphi_3\, \xi_{(2)}\, \bar\phi_3}v \nn \\[4pt]
&=& \big({}^{\varphi_1\,\xi_{(1)_{(1)}}\,\bar\phi_1}L\otimes\Id\big)
\big( {}^{\varphi_2\,\xi_{(1)_{(2)}}\,\bar\phi_2}u\otimes_\st
{}^{\varphi_3\, \xi_{(2)}\, \bar\phi_3}v\big) \nn \\[4pt]
&=& \big({}^{\varphi_1\,\bar\phi_1\,\xi_{(1)}}L\otimes\Id\big)\big(
{}^{\varphi_2\, \bar\phi_2\, \xi_{(2)_{(1)}}}u\otimes_\st
{}^{\varphi_3\, \bar\phi_3\,\xi_{(2)_{(2)}}}v\big) \nn \\[4pt]
&=& \big({}^{\xi_{(1)}}L\otimes\Id\big)\big({}^{\xi_{(2)}}(u\otimes_\star
v) \big) \ .
\eea
With the definition (\ref{deflid}) the map $L\otimes \Id$ is indeed well-defined on $U_\star\otimes_\star V_\star$:
\bea
(L\otimes\Id)\big((u\star f)\otimes_\star v\big)= (L\otimes\Id)\big({}^{\phi_1}u\otimes_\star({}^{\phi_2}f \star{}^{\phi_3}v)\big) \ .
\eea
For this, we use right $A_\star$-linearity of $L$ to write the left-hand side as
\bea
(L\otimes\Id)\big((u\star f)\otimes_\star v\big)= {}^{\bar\phi_1\,\bar\varphi_1}L({}^{\tau_1\,\bar\phi_2\, \bar\varphi_2}u)\otimes_\star({}^{\bar\varphi_3\, \tau_2}f\star {}^{\tau_3\,\bar\phi_3}v)
\eea
which is indeed equal to the right-hand side
\bea
(L\otimes\Id)\big({}^{\phi_1}u\otimes_\star({}^{\phi_2}f \star{}^{\phi_3}v)\big) = {}^{\bar\varphi_1}L({}^{\bar\varphi_2\,\phi_1}u)\otimes_\star{}^{\bar\varphi_3}({}^{\phi_2}f \star{}^{\phi_3}v) \ .
\eea
Finally, we can show that
$L\otimes\Id$ is right $A_\star$-linear:
\bea\label{eq:Lidlinear}
(L\otimes\Id)\big((u\otimes_\star v)\star f \big) =
\big({}^{\bar\zeta_1}(L\otimes\Id) {}^{\bar\zeta_2}(u\otimes
v)\big) \star{}^{\bar\zeta_3}f \ .
\eea
For this, we note that the left-hand side can be expressed as
\bea
(L\otimes\Id)\big((u\otimes_\star v)\star f \big) =
{}^{\bar\phi_1\,\bar\varphi_1}L({}^{\bar\phi_2\, \bar\varphi_2\,
  \zeta_1}u)\otimes_\star ({}^{\bar\phi_3\,\zeta_2}v\star
{}^{\bar\varphi_3\, \zeta_3}f)
\eea
which is indeed equal to the right-hand side
\bea
\big({}^{\bar\zeta_1}(L\otimes\Id) {}^{\bar\zeta_2}(u\otimes_\st
v)\big) \star{}^{\bar\zeta_3}f = {}^{\varphi_1\,\bar\phi_1\,\bar\zeta_1
  \, \bar\rho_1}L({}^{\tau_1\, \bar\phi_2\,\bar\zeta_2}u)\otimes_\star
({}^{\varphi_2\,\tau_2\, \bar\phi_3\,\bar\rho_2}v\star {}^{\varphi_3\,
  \tau_3\, \bar\rho_3\, \bar\zeta_3}f) \ .
\eea

We also define
\bea
\Id\otimes_\mcR L:=\tau_\mcR\bullet ( L\otimes \Id)\bullet \tau_\mcR \
, 
\label{eq:idL}\eea
with $\tau_\mcR(v\otimes_\st u)= {}^\al u\otimes_\st {}_\al v$ the braiding
operator. This definition is well-posed because $\tau_\mcR$ is compatible
with \eqref{eq:quotientrel}, and
the twisted composition $\bullet$ is associative if one of the maps is
equivariant
(as $\phi_1\otimes \phi_2\, \phi_3=1\otimes1$). Moreover, $\tau_\mcR$ is an equivariant map:
${}^\xi\big(\tau_\mcR(u\otimes_\st v)\big)=
\tau_\mcR\big({}^\xi(u\otimes_\st v)\big)$, and thus the lifting of $L$
to $\Id\otimes_\mcR L$ is equivariant: 
${}^\xi(\Id\otimes_\mcR L)=\tau_\mcR
\bullet({}^\xi L\otimes\Id)\bullet
\tau_\mcR= \Id\otimes_\mcR{}^\xi L$. The lift $\Id\otimes_\mcR L$ is furthermore right $A_\star$-linear:
\bea
(\Id\otimes_\mcR L)\big((u\otimes_\star v)\star f\big) &=& \tau_\mcR\big(\big({}^{\bar\phi_1}(L\otimes\Id)\,\tau_\mcR{}^{\bar\phi_2}(u\otimes_\star v)\big)\star{}^{\bar\phi_3}f\big) \nonumber \\[4pt]
&=& \big(\tau_\mcR\,({}^{\bar\phi_1}L\otimes\Id)\, \tau_\mcR{}^{\bar\phi_2}(u\otimes_\star v)\big)\star{}^{\bar\phi_3}f \nonumber \\[4pt]
&=& \big( {}^{\bar\phi_1}(\Id\otimes_\mcR L){}^{\bar\phi_2}(u\otimes_\star v)\big) \star{}^{\bar\phi_3}f \ ,
\eea
where we used right $A_\star$-linearity of $L\otimes\Id$.

To summarise, if $L:U_\star\to W_\star$ is right $A_\star$-linear, then $L\otimes\Id$ is well-defined on $U_\star\otimes_\star V_\star$ and right $A_\star$-linear, and hence so is $\Id\otimes_\mcR L$. In particular, given another right $A_\star$-linear map $L':V_\st'\to W_\st'$ we obtain a well-defined right $A_\st$-linear map $L\otimes_\mcR L':= (L\otimes\Id)\bullet(\Id\otimes_\mcR L'\,)$, which is compatible with the action of $U\vect^\FF(\Mcal)$ and is quasi-associative~\cite{Barnes2014}:
\bea
(L\otimes_\mcR L'\,)\otimes_\mcR L'' = \Phi^{-1}\bullet\big(L\otimes_\mcR (L'\otimes_\mcR L''\,) \big)\bullet\Phi \ .
\eea

\subsection{Quantum Lie algebra of diffeomorphisms\label{sec:diffeos}}

By applying the twist deformation to the Lie algebra of vector fields
$\vect(\Mcal)$ on phase space $\Mcal$, we obtain the quantum Lie algebra of
nonassociative diffeomorphisms described
in~\cite{Aschieri2015}. Again we deform the usual Lie bracket of
vector fields to the star-bracket
\beq
[u,v]_\star=\big[\, \of\,^\al(u), \of\,_\al(v) \big]~.
\eeq
Defining the star-product between
elements  in $U\vect(\cal M)$ as
$\xi\st\zeta:=\of\,^\al(\xi)\of\,_\al(\zeta)$, the star-bracket equals the deformed commutator
\beq
[u,v]_\star= u\star 
v-{}^\al v\star u{}_\al \ . 
\eeq
This deformed Lie bracket satisfies the star-antisymmetry property
\beq
[u,v]_\star=-[{}^\al v,{}_\al u]_\star
\eeq
and the star-Jacobi identity
\beq
\big[u,[v,z]_\star\big]_\star=\big[[{}^{\bar\phi_1}u,{}^{\bar\phi_2}v]_\star,{}^{\bar\phi_3}z\big]_\star+
\big[{}^\al({}^{\bar\phi_1\,\bar\varphi_1}v)
,[{}_\al({}^{\bar\phi_2\,\bar\varphi_2}u),{}^{\bar\phi_3\,\bar\varphi_3}z]_\star\big]_\star
\ .
\eeq
The star-bracket $[~,~]_\star$ makes $\vect_\star$ into the quantum Lie algebra of vector fields.

To implement the action of nonassociative diffeomorphisms on generic differential forms and tensor fields, we need a suitable definition of star-Lie derivative along a vector $u\in\vect_\star$. From~\cite{Aschieri2015} it is a deformation of the ordinary Lie derivative on phase space $\Mcal$ given by
\bea
{\cal L}^\star_u(T) ={\cal L}_{\, \of{}\,^\al(u)}(\, \of\,_\al(T)) = {\cal L}_{{\rm D}(u)}(T) \ ,
\eea
where we introduced the invertible linear map $\rm D$ on the vector space $U\vect(\Mcal)$ by
\bea
{\rm D} \, : \, U\vect(\Mcal) & \longrightarrow & U\vect(\Mcal) \ , \nn \\
\xi & \longmapsto & {\rm D}(\xi):= \of\,^\al(\xi)\, \of\,_\al \ .
\label{eq:rmDu}\eea
With this definition it follows immediately that
$\Lie_u^\star(v)=[u,v]_\star$ for $u,v\in\vect_\star$. Moreover,
using the inverse of (\ref{cocycle}) shows that
$\Lie_{{\rm D}(\xi)}\bullet \Lie_{{\rm D}(\zeta)}=\Lie_{{\rm D}(\xi\star \zeta)}$,  for all
$\xi,\zeta\in U\vect({\cal M})$, so that
\eq
[\Lie^\star_u,\Lie^\star_v]_{\bullet}=\Lie^\star_{[u,v]_\star} \ .
\en
Thus the star-Lie derivatives provide a representation of the quantum
Lie algebra of vector fields on differential forms and tensor
fields. 

Using \eqref{cocycle} together with $\Delta(u)=u\otimes 1 + 1
\otimes u$, the twisted coproducts of 
$\DD(u)\in U\vect(\Mcal)$ are given by
\bea
\Delta_\FF\big(\DD(u)\big) =
\DD\big({}^{\bar\phi_1}u \big)\, \bar\phi_2\otimes\bar\phi_3+\oR\,^\al\,
\bar\phi_1\, \bar\varphi_1 \otimes\DD\big(\, \oR\,_\al({}^{\bar \phi_2\,
  \bar\varphi_2} u)\big)\, \bar\phi_3\,\bar\varphi_3 \ .
\label{eq:DeltaFFDu}\eea
Using the Leibniz rule for the undeformed Lie derivative ${\cal
  L}_u(\omega\wedge\eta)= {\cal
  L}_u(\omega)\wedge\eta+\omega\wedge{\cal L}_u(\eta)$, it follows
from \eqref{eq:DeltaFFDu} that the star-Lie derivatives satisfy the deformed Leibniz rule~\cite{Aschieri2015}
\bea
\Lie_u^\star(\omega\wedge_\star\eta) =
\Lie^\star_{{}^{\bar\phi_1}u}({}^{\bar\phi_2}\omega)\wedge_\star{}^{\bar\phi_3}\eta+
{}^\al({}^{\bar\phi_1\, \bar\varphi_1}\omega)\wedge_\star \Lie^\star_{{}_\al({}^{\bar\phi_2\, \bar\varphi_2}u)}({}^{\bar\phi_3\, \bar\varphi_3}\eta)
\label{LeibnizLie}\eea
on forms $\omega,\eta\in\Omega_\star^\sharp$. The Leibniz rule
for tensor fields is then obtained by replacing differential forms with tensor fields and
the deformed exterior product $\wedge_{\star}$ with the deformed
tensor product~$\otimes_{\star}$. In particular, since $[u,v\star
f]_\star={\cal L}^\star_u(v\star f)={\cal L}_{\DD(u)}(v\star f)$ for
$f\in A_\star$, we analogously obtain the Leibniz rule for the quantum Lie bracket
of vector fields:
\bea\label{eq:bracketLeibniz}
[u,v\star f]_\star = \big[{}^{\bar\phi_1}u,{}^{\bar\phi_2}v\big]_\star
\star {}^{\bar\phi_3}f +
{}^\al\big({}^{\bar\phi_1\,\bar\varphi_1}v\big) \star {\cal
  L}^\star_{{}_\al({}^{\bar\phi_2\,\bar\varphi_2}
  u)}\big({}^{\bar\phi_3\,\bar\varphi_3} f\big) \ .
\eea

Since the map $\DD$ is invertible, as in the noncommutative and
associative case~\cite{Aschieri2005,Aschieri2017}, the
symmetry properties of the quasi-Hopf algebra of infinitesimal
diffeomorphisms $U\vect^\FF({\cal M})$ are equivalently encoded in the quantum Lie algebra of diffeomorphisms
$\vect_\st$ with bracket $[~,~]_\st$, or in its universal enveloping
algebra generated by sums of star-products of elements in $\vect_\st$.

\newsection{Nonassociative differential geometry\label{sec:connections}}

\subsection{Connections}

A star-connection is a linear map
\begin{eqnarray}
\nabla^\star\,:\, \vect_\star &\longrightarrow& \vect_\star\otimes_\star \Omega^1_\star \nn\\
u&\longmapsto& \nabla^\star u = u^i\otimes_\star \omega_i \ , \label{connection}
\end{eqnarray}
where $u^i\otimes_\star \omega_i\in\vect_\star\otimes_\star
\Omega^1_\star$, which satisfies the right Leibniz rule
\begin{equation}
\nabla^\star (u\star f) = \big({}^{\bar{\phi}_1} \nabla^\star
({}^{\bar{\phi}_2}u) \big) \star {}^{\bar{\phi}_3}f + u\otimes_\star
{\rm d}f \label{LeibnizConn}
\end{equation}
for $u\in\vect_\star$ and $f\in A_\star$. The action of $\phi_{a}$ on
$\nabla^\star$ is the adjoint action (\ref{adjact}), which in the present instance is readily seen to also define a connection. For this, we calculate
\bea
{}^{\phi_a}\nabla^\star(u\star f) &=& {}^{\phi_{a_{(1)}}}\big(\nabla^\star\big({}^{{S(\phi_{a_{(2)}})}_{(1)}}u \star {}^{{S(\phi_{a_{(2)}})}_{(2)}}f \big) \big) \nonumber\\[4pt]
&=& {}^{\phi_{a_{(1)_{(1)}}}}\big({}^{\bar\varphi_1}\nabla^\star\big({}^{\bar\varphi_2\, {S(\phi_{a_{(2)}})}_{(1)}}u\big)\big) \star {}^{\bar\varphi_3\, \phi_{a_{(1)_{(2)}}}\, {S(\phi_{a_{(2)}})}_{(2)}}f \nonumber \\ && +\, {}^{\phi_{a_{(1)_{(1)}}}\, {S(\phi_{a_{(2)}})}_{(1)}} u\otimes_\star {}^{\phi_{a_{(1)_{(2)}}}\, {S(\phi_{a_{(2)}})}_{(2)}}\dd f \nonumber \\[4pt]
&=&
\big({}^{\bar\varphi_1}({}^{\phi_a}\nabla^\star)({}^{\bar\varphi_2}u)\big)\star
{}^{\bar\varphi_3}f +{}^{\phi_{a_{(1)}}\,S(\phi_{a_{(2)}})} (u\otimes_\star \dd f)
\nonumber \\[4pt]
&=&
\big({}^{\bar\varphi_1}({}^{\phi_a}\nabla^\star)({}^{\bar\varphi_2}u)\big)\star
{}^{\bar\varphi_3}f + \epsilon(\phi_a) \, u\otimes_\star \dd f \ ,
\eea
where in the last line we used \eqref{eq:Sepsilon}. Now since 
${}^{\phi_a}\nabla^\star$ will always appear in linear combinations
with the other associator legs $\phi_b$ and $\phi_c$, and since
$\epsilon(\phi_a)\, \phi_b\otimes\phi_c=1\otimes 1$, we
effectively have the Leibniz rule 
\eq\label{phiD*}
{}^{\phi_a}\nabla^\star(u\star
f)=\big({}^{\bar\varphi_1}({}^{\phi_a}\nabla^\star)({}^{\bar\varphi_2}u)\big)\star
{}^{\bar\varphi_3}f+u\otimes_\st \dd
f~.\en
More generally, the adjoint action of an element $\xi\in
U\vect^\FF(\cal M)$ gives the linear map
${}^{\xi}\nabla^\star:  \vect_\st\to \vect_\st\otimes_\st
\Omega_\st^1$ which satisfies
$
{}^\xi\nabla^\star (u\star f) = \big({}^{\bar{\phi}_1\, \xi}\nabla^\star
({}^{\bar{\phi}_2}u) \big) \star {}^{\bar{\phi}_3}f + 
u\otimes_\star
\epsilon(\xi)\, {\rm d}f,
$
i.e., ${}^\xi\nabla^\star$ is a connection with respect to the
rescaled exterior derivative ${}^\xi\dd={\cal L}_{\xi_{(1)}}\, \dd \,{\cal L}_{S(\xi_{(2)})}=\epsilon(\xi)\, {\rm d}$.

The connection on vector fields \eqref{connection} uniquely extends to a covariant
derivative
\bea
\dd_{\nabla^\star} \,:\, \vect_\star^\sharp 
&\longrightarrow& \vect_\star^{\sharp+1} \nn\\
u\otimes_\star\omega &\longmapsto& \big({}^{\bar{\phi}_1} \nabla^\star
({}^{\bar{\phi}_2}u) \big) \wedge_\star {}^{\bar{\phi}_3}\omega + u\otimes_\star
{\rm d}\omega
\label{eq:liftconnection}\eea
on vector fields valued in the exterior algebra $\vect_\star^\sharp=
\vect_\star\otimes_\star\Omega_\star^\sharp$. It satisfies the graded
right Leibniz rule
\bea
\dd_{\nabla^\star} (\psi\wedge_\star \omega) = \big({}^{\bar{\phi}_1} \dd_{\nabla^\star}
({}^{\bar{\phi}_2}\psi) \big) \wedge_\star {}^{\bar{\phi}_3}\omega +
(-1)^{|\psi|}\, \psi \wedge_\star \dd\omega
\eea
for $\psi=u^i\otimes_\star\omega_i\in\vect_\star^\sharp$.

The covariant derivative along a
vector field $v\in\vect_\star$ is defined via the pairing operator as
\begin{eqnarray}
\nabla^\star_v u = \le~ \nabla^\star u \,,\, v~\re_\st = \le~ (u^i\otimes_\star
\omega_i) \,,\,v ~\re_\st = {}^{\phi_1}u^{i}\star \le~ {}^{\phi_2}\omega_i\,,\, {}^{\phi_3}v ~\re_\st \ . \label{CovDerivative}
\end{eqnarray}
From the definition of the pairing (\ref{pairing}), the Leibniz rule for
$\nabla^\star_v$ comes in the somewhat complicated form that we will
need later:
\begin{eqnarray}
\nabla ^\star_v (u\star f) &=& \le~ \nabla^\star (u\star f)\,,\, v~\re_\st \nn\\[4pt]
&=& \le~ ( {}^{\bar{\phi}_1}\nabla^\star( {}^{\bar{\phi}_2}u)\star {}^{\bar{\phi}_3}f)
+ (u\otimes_\star {\rm d}f) \,,\, v~\re_\st  \nn\\[4pt]
&=& \le~ {}^{\phi_1}( {}^{\bar{\phi}_1}\nabla^\star ({}^{\bar{\phi}_2}u)) \star
( {}^{\phi_2\,\bar{\phi}_3}f \,,\, {}^{\phi_3}v )~\re_\st 
+ {}^{\phi_1}u\star \le~ ( {}^{\phi_2}({\rm d}f) \,,\, {}^{\phi_3}v)
    ~\re_\st \label{Q4} \\[4pt]
&=& \le~ {}^{\phi_1}( {}^{\bar{\phi}_1}\nabla^\star
    ({}^{\bar{\phi}_2}u))\,,\,({}^\al({}^{\phi_3} v )~\re_\st \star
{}_\al( {}^{\phi_2\,\bar{\phi}_3}f )) + {}^{\phi_1}u\star \le~ {}^{\phi_2}({\rm d}f) \,,\, {}^{\phi_3}v ~\re_\st \nn\\[4pt]
&=& \big( \le~ {}^{\bar\varphi_1\,\phi_1}(
    {}^{\bar{\phi}_1}\nabla^\star ({}^{\bar{\phi}_2} u
    ))\,,\,{}^{\bar\varphi_2}({}^\al({}^{\phi_3} v )~\re_\st \big) \star
{}^{\bar\varphi_3}({}_\al( {}^{\phi_2\,\bar{\phi}_3}f )) + {}^{\phi_1}u\star \le~ {\rm d}({}^{\phi_2}f) \,,\, {}^{\phi_3}v ~\re_\st
 \ . \nn
\end{eqnarray}
More generally, we define
\bea
\dd_{\nabla^\st_v} \psi := \le~ \dd_{\nabla^\st} \psi \,,\,
  v~\re_\st + \dd_{\nabla^\st}\le~\psi\,,\,v~\re_\st
\eea
for $\psi=u^i\otimes_\star\omega_i\in\vect_\star^\sharp$.

The action of the connection on the basis vectors defines the
connection coefficients $\Gamma^B_{AC}\in A_\star$ through
\begin{equation}
\nabla^\star \partial_A =: \partial_B\otimes_\star \Gamma^B_A=: \partial_B\otimes_\star
(\Gamma^B_{AC}\star {\rm d}x^C) \ . \label{ConnCoeff}
\end{equation} 
Then we have
\begin{eqnarray}
\nabla^\star_A \partial_B &:=& \le~ \nabla^\star \partial_B \,,\, \partial_A ~\re_\st \nn\\[4pt]
&=& \le~ (\partial_C\otimes_\star (\Gamma^C_{BD}\star {\rm d}x^D))\,,\, \partial_A ~\re_\st \nn\\[4pt]
&=&  \le~ (\partial_C\star \Gamma^C_{BD}) \otimes_\star {\rm d}x^D \,,\,
\partial_A ~\re_\st \nn\\[4pt]
&=& {}^{\phi_1}(\partial_C\st\Gamma^C_{BD})\st\le~ {}^{\phi_2}{\rm d}x^D \,,\,
{}^{\phi_3}\partial_A ~\re_\st \nn\\[4pt]
&=& \partial_C\star \Gamma^C_{BA} \ , \label{ConnBasisVectors}
\end{eqnarray}
where we used the definition \eqref{eq:pair1forms}, and the
contributions from nonassociativity vanish because we used 
basis vector fields and basis 1-forms. Using the Leibniz rule (\ref{LeibnizConn}) and
writing an arbitrary vector field $u$ as $u=\partial_A\star u^A$ with
$u^A\in A_\st$ one can calculate
\begin{equation}
\nabla^\star u = \partial_A\otimes_\star ( {\rm
d}u^A+ \Gamma^A_B\star u^B ) \ , \label{ConnArbVector}
\end{equation}
and more generally
\bea
\dd_{\nabla^\star}(\partial_A\otimes_\st\omega^A)=\partial_A\otimes_\st(\dd\omega^A+\Gamma^A_B\wedge_\star
\omega^B) \ ,
\eea
for $\omega^A\in\Omega_\st^\sharp$.

\subsection{Dual connections}

By considering 1-forms as dual to vector fields, we can define
the dual connection ${}^\star\nabla$ on 1-forms in terms of the connection on vector fields and the exterior derivative as
\bea
\le~ {}^\star\nabla\omega\,,\,u~\re_\star=\dd\le~ \omega\,,\,u~\re_\st
-\le~{}^{\phi_1}\omega\,,\, {}^{\phi_2}\nabla^\st({}^{\phi_3}u) ~\re_\st \ .
\label{eq:leftconndef}\eea
Since the pairing is nondegenerate, this defines a connection on the dual bimodule
\bea
{}^\star\nabla\,:\, \Omega_\star^1\ \longrightarrow \ \Omega_\st^1\otimes_\st\Omega_\star^1~.
\eea
This connection acts from the right so that we should more
properly write $(\om){}^\star\nabla$ rather than
${}^\star\nabla(\om)$, but this notation is awkward so we refrain from
using it. That the action is from the right immediately follows by comparing the $U\vect^\FF(\cal
M)$-equivariance property (\ref{eq:evalequiv}) of the evaluation from the left
with the $U\vect^\FF(\cal
M)$-equivariance property of the evaluation of $^\st\nabla$ on $\om$,
${}^\xi({}^\star\nabla\om)={}^{\xi_{(2)}}{{}^\star\nabla}({}^{\,\xi_{(1)}}\om)$,
which shows that evaluation is from the right so that the equivariance reads 
$^\xi\big((\om){}^\star\nabla\big)=( ^{\xi_{(1)}}\om){}^{\xi_{(2)}\star}\nabla$. The proof 
follows from
\bea
{}^\xi\le~ {}^\star\nabla\omega\,,\,u~\re_\star&=&\dd\le~ {}^{\xi_{(1)}}\omega\,,\,{}^{\xi_{(2)}}u~\re_\st
-\le~{}^{\xi_{(1)}\, \phi_1}\omega\,,\,
{}^{\xi_{(2)_{(1)}}\, \phi_2}\nabla^\st( {}^{\xi_{(2)_{(2)}}\, \phi_3}u)
~\re_\st\nn\\[4pt]
 &=&\dd\le~ {}^{\xi_{(1)}}\omega\,,\,{}^{\xi_{(2)}}u~\re_\st
-\le~{}^{\phi_1\, \xi_{{(1)_{(1)}}}}\omega\,,\,
{}^{\phi_2\, \xi_{(1)_{(2)}}}\nabla^\st( {}^{\phi_3\, \xi_{(2)}}u)
~\re_\st\nn\\[4pt]
&=&\le~
{}^{\xi_{{(1)}_{(2)}}}{}^\star\nabla({}^{\xi_{{(1)}_{(1)}}}\omega) \,,\,{}^{\xi_{(2)}}u~\re_\star~,
\label{eq:leftconneqv}\eea
where in the last line we used 
$
\le~ {}^{\xi\, }{}^{\star}\nabla\omega \,,\,u~\re_\star=\epsilon(\xi)\,
\dd\le~ \omega\,,\,u~\re_\st
-\le~{}^{\phi_1}\omega\,,\, {}^{\phi_2\, \xi}\nabla^\st({}^{\phi_3}u) ~\re_\st 
$, which is easily understood by recalling that ${}^\xi\nabla^\st$ is a connection
with respect to the rescaled exterior derivative
${}^\xi\dd = \epsilon(\xi)\, \dd$.

Correspondingly, the connection $^\st\nabla$  satisfies the \emph{left} Leibniz rule
\bea
{}^\st\nabla(f\star \omega)=
{}^{\phi_1}f\star\big({}^{\phi_3}{}^\st\nabla({}^{\phi_2}\omega)\big)
+ \dd f\otimes_\star\omega
\eea
for $f\in A_\st$ and $\omega\in\Omega_\st^1$.
The proof follows from the
definition (\ref{eq:leftconndef}) and the right Leibniz rule for
$\nabla^\st$ after some associator gymnastics.
It uniquely lifts to a
connection
\bea
\dd_{{}^\st\nabla}\,:\, \Omega^\sharp_\st\otimes_\st\Omega^1_\st \
\longrightarrow \ \Omega_\st^{\sharp+1}\otimes_\st\Omega_\st^1 \ .
\eea

Setting $\omega=\dd x^A$ and $u=\partial_B$ in \eqref{eq:leftconndef},
so that $\dd\le~\dd x^A\,,\,\partial_B~\re_\st=\dd\,\delta^A{}_B=0$, we
compute
\bea
\le~{}^\st\nabla(\dd x^A)\,,\,\partial_B~\re_\st &=& -\le~\dd
x^A\,,\,\nabla^\st\partial_B~\re_\st \nn \\[4pt]
&=& -\le~\dd x^A\,,\,\partial_C\otimes_\st\Gamma^C_B~\re_\st \nn
\\[4pt]
&=& -\Gamma^A_B
\eea
so that ${}^\st\nabla(\dd x^A)=-\Gamma^A_B\otimes_\st\dd x^B$. 
Then for $\omega=\omega_A\star\dd x^A\in\Omega_\st^1$ with
$\omega_A\in A_\st$ we have
\begin{eqnarray}
^\star\nabla \omega &=& {}^\star\nabla(\omega_A\star {\rm d}x^A) \nn\\[4pt]
&=& \omega_A\star {}^\star\nabla({\rm d}x^A)+ {\rm d} \omega_A\otimes_\star {\rm
d}x^A\nn\\[4pt]
&=& -\omega_A\star (\Gamma^A_B\otimes_\star {\rm d}x^B) + {\rm d}\omega_A \otimes_\star {\rm d}x^A \nn\\[4pt]
&=& ({\rm d}\omega_B - \omega_A\star\Gamma^A_B )\otimes_\star {\rm
    d}x^B \ .
\label{ConnCoeffOneFormPaolo}
\end{eqnarray}
More generally, for $\omega_A\in\Omega_\st^\sharp$ we
have
\bea
\dd_{{}^\st\nabla}(\omega_A\otimes_\st\dd x^A)=
(\dd\omega_A-\omega_B\wedge_\st\Gamma_A^B)\otimes_\st \dd x^A \ .
\eea
These results are natural nonassociative generalizations of the
usual results in noncommutative differential geometry, since the
associator acts trivially on the basis vector fields and basis 1-forms. 

\subsection{Connections on tensor products}\label{TPC}

Later on we shall need to compute the action of connections on metric
tensors, for which we require a construction of connections on tensor
products of $A_\star$-bimodules. The general construction is an
extension to the nonassociative case
of the noncommutative construction in \cite{AS} and
 is provided in~\cite[Section~4.2]{Barnes2015}. Here we shall give a somewhat
simpler and more explicit treatment. Given $A_\star$-bimodules
$V_\star$ and $W_\star$, together with connections
$\nabla_{V_\star}^\star:V_\star\to V_\star\otimes_\star\Omega_\star^1$
and $\nabla_{W_\star}^\star:W_\star\to
W_\star\otimes_\star\Omega_\star^1$, we wish to construct a connection
$\nabla^\star_{V_\star\otimes_\star
  W_\star}:=\nabla^\star_{V_\star}\oplus_\star\nabla^\star_{W_\star}:(V_\star\otimes_\star
W_\star)\to (V_\star\otimes_\star
W_\star)\otimes_\star\Omega_\star^1$. Using \eqref{eq:idL} we define
\bea\label{eq:conntensordef}
\nabla^\star_{V_\star}\oplus_\star\nabla^\star_{W_\star} =
\nabla_{V_\star}^\star\otimes \Id+\Id\otimes_\mcR\nabla_{W_\star}^\star \ .
\eea
Explicitly, using (\ref{deflid}) and (\ref{eq:idL}) we have
\bea\label{expconnprod}
\nabla^\star_{V_\star\otimes_\star W_\star}(v\otimes_\star w)={}^{\bar\phi_1}\nabla^\star_{V_\star}({}^{\bar\phi_2}v)\otimes_\star{}^{\bar\phi_3}w+{}^{\beta\,\bar\phi_3}{}_\alpha v\otimes_\star{}_\beta{}^{\bar\phi_1}\nabla_{W_\star}^\star({}^{\bar\phi_2\,\alpha}w) \ ,
\eea
where we identify $(V_\star\otimes_\star\Omega_\star^1)\otimes_\star W_\star\cong (V_\star\otimes_\star W_\star)\otimes_\star\Omega_\star^1$ via $(v\otimes_\star\omega)\otimes_\star w=({}^{\bar\phi_1}v\otimes_\star{}^{\bar\phi_2\,\alpha} w)\otimes_\star{}^{\bar\phi_3}{}_\alpha\omega$ for $v\in V_\star$, $w\in W_\star$ and $\omega\in\Omega_\star^1$.

From the general analysis of Section~\ref{sec:modulehom} it follows that this definition is equivariant:
\bea
{}^\xi(\nabla^\star_{V_\star}\oplus_\star\nabla^\star_{W_\star}) = {}^\xi\nabla^\star_{V_\star}\oplus_\star{}^\xi\nabla^\star_{W_\star} 
\eea
for any $\xi\in
U\vect^\FF(\Mcal)$. Next we need to check that this definition is well-defined:
\bea
(\nabla^\star_{V_\star}\oplus_\star\nabla^\star_{W_\star})\big((v\star
f)\otimes_\star w\big) =
(\nabla^\star_{V_\star}\oplus_\star\nabla^\star_{W_\star})\big({}^{\rho_1}
v\otimes_\star ({}^{\rho_2}f\star {}^{\rho_3}w) \big) \ .
\eea
Again by the general analysis of Section~\ref{sec:modulehom}, we know
that this identity holds if the star-connection $\nabla^\st$ is 
substituted by a right $A_\st$-linear map $L$, i.e., it holds
for the terms which come from the right
$A_\star$-linear part of the Leibniz rule for  $\nabla^\st$, so we only need to check
the inhomogeneous terms coming from the exterior derivative: On the left-hand side this comes from the
application of $\nabla_{V_\star}^\star\otimes \Id$ to $(v\star
f)\otimes_\star w$ which gives $(v\otimes_\star \dd f)\otimes_\star w$
on using the fact that ${}^{\phi_a}\nabla_{V_\star}^\star$ is also a
connection, whereas on the right-hand side it comes from applying
$\tau_\mcR\bullet(\nabla_{W_\star}^\star\otimes \Id)$ to $(
{}^{\beta\,\alpha_{(2)}\,
  \rho_3}w\star{}_\beta{}^{\alpha_{(1)}\,\rho_2}f)\otimes_\star
{}_\al{}^{\rho_1} v$ which on using the $\cal R$-matrix identities
\eqref{eq:triangular} and \eqref{eq:DeltaRPhib} yields
\bea\label{eq:sumwelldef}
\tau_\mcR\big({}^\al({}^{\rho_2}\dd f\otimes_\star{}^{\rho_3}w)\otimes_\star {}_\al{}^{\rho_1}v\big) = {}^\gamma{}_\al{}^{\rho_1}v\otimes_\star {}_\gamma{}^\al({}^{\rho_2}\dd f\otimes_\star {}^{\rho_3}w) = (v\otimes_\star \dd f)\otimes_\star w
\eea
as required. Finally, we show that the map $\nabla^\star_{V_\star}\oplus_\star\nabla^\star_{W_\star}$
is a connection because it satisfies the Leibniz rule:
\bea
(\nabla^\star_{V_\star}\oplus_\star\nabla^\star_{W_\star})\big((v\otimes_\star
w)\star f\big) =
\big({}^{\bar\phi_1}(\nabla^\star_{V_\star}\oplus_\star\nabla^\star_{W_\star})\big(
{}^{\bar\phi_2}(v\otimes_\star w)\big)\big) \star {}^{\bar\phi_3} f+ (v\otimes_\star w)\otimes_\star\dd f \ .
\eea
Again it suffices to check the inhomogeneous term, which comes from $(\Id\otimes_\mcR\nabla_{W_\star}^\star)((v\otimes_\star w)\star f)$, and the result follows by a completely analogous calculation to \eqref{eq:sumwelldef}.

We can iterate the twisted sum of
connections to arbitrary numbers of tensors products. The
nonassociativity of $\oplus_\star$ is controlled in the usual way by suitable
insertions of the associator~\cite{Barnes2016}:
\bea
(\nabla_{V_\star}^\star\oplus_\star \nabla_{W_\star}^\star) \oplus_\star
\nabla_{X_\star}^\star = \Phi^{-1} \bullet \big(\nabla_{V_\star}^\star\oplus_\star
(\nabla_{W_\star}^\star \oplus_\star
\nabla_{X_\star}^\star) \big)\bullet \Phi \ .
\eea

\subsection{Torsion}

In order to define the torsion
 ${\sf  T}^\st\in\vect_\st\otimes_\st\Omega_\st^2$ of a connection $\nabla^\st$,
we first observe that the map 
\eq\label{identity}
\le~ \partial_A\otimes_\st \dd
x^A\,,\, ~\re_\st\, :\, \vect_\st \ \longrightarrow \ \vect_\st
\en 
is the identity
map; for this, we simply expand any vector field $u$ as $u=\partial_A\st u^A$, and use
the triviality of the associator when acting on $\partial_A$ and $\dd x^A$. 
Then as in the classical case we define
\begin{eqnarray}
{\sf T}^\st&:=&
          \dd_{\nabla^\st}\big(\partial_A\otimes_\st \dd x^A\big)
  \nn \\[4pt]
&=& \nabla^\st \partial_A\wedge_\star \dd x^A \nonumber\\[4pt]
&=& \partial_B\otimes_{\star}\big( 
    \Gamma^B_{AC}\star(\dd x^C \wedge_\star\dd x^A ) \big) \nn \\[4pt] &=:&
                                                            \partial_A\otimes_{\star} {\sf T}^A \ .
\label{eq:Torsion2form}\end{eqnarray}

We can also regard the torsion in the usual way as a map ${\sf
  T}^\st: \vect_\st\otimes_\st \vect_\st \to \vect_\st$ defined by
\bea
{\sf T}^\st(u,v) =
\le~{\sf T}^\st \,,\, u\otimes_\st 
v ~\re_\st ~,
\label{eq:Torsionuv}\eea
where $\le~{\sf T}^\st \,,\, u\otimes_\st 
v ~\re_\st =\partial_A\star 
\le~{\sf T}^A \,,\, u\otimes_\st 
v ~\re_\st .$ This map is right $A_\star$-linear in its second argument by
\eqref{l-rpairing}, and star-antisymmetric: ${\sf T}^\st(u,v)=-{\sf
  T}^\st({}^\al v,{}_\al u)$. This follows form 
\eq
\le ~ \dd x^A\wedge_\st \dd
x^B\,,\, u\otimes_\st v~ \re_\st= 
\le~ \dd x^A\otimes_\st \dd
x^B\,,\, u\wedge_\st v~ \re_\st~,
\en 
i.e., from 
$\le~ {}^\al\dd x^B\otimes_\st {}_\al\dd
x^A\,,\, u\otimes_\st v~ \re_\st= 
\le~ \dd x^A\otimes_\st \dd
x^B\,,\, {}^\ga v\otimes_\st {}_\ga u~ \re_\st$.
To prove this last equality we recall that the associator is
trivial if it acts on the basis 1-forms and 
apply the definition of the pairing; then reordering we
obtain the equivalent expression
$\le~ {}^\al\dd x^B\,,\,{}^{\be} v~ \re_\st\st \le~ {}_{\be_{(1)}\, \al}\dd
x^A\,,\,{}_{\be_{(2)}}u~ \re_\st=
\le~ {}^{\al_{(1)}} 
\dd x^B\,,\,{}^{\al_{(2)}\, \ga}v~ \re_\st\st\le~ {}_{\al}\dd x^A\,,\,
{}_\ga
u~ \re_\st$, which follows from (\ref{eq:DeltaRPhia}) and
(\ref{eq:DeltaRPhib}). Thus we have shown that the map ${\sf T}^\st$ is the torsion
tensor  ${\sf
  T}^\st\in{\rm hom}_\st(\vect_\st\wedge_\st\vect_\st,\vect_\st)$.

In our good basis one
easily calculates the torsion components from \eqref{eq:Torsion2form} and obtains
\bea
{\sf T}^\st(\partial_A, \partial_B) = \partial_C\star \le~ {\sf T}^C
\,,\, \partial_A\otimes_\st \partial_B ~ \re_\st = \partial_C\star
(\Gamma^C_{AB} - \Gamma^C_{BA}) =: \partial_C\st{\sf T}^C{}_{AB} \ .
\eea
The torsion-free
condition ${\sf T}^\st(\partial_A, \partial_B) =0 $ then results in
the symmetric connection coefficients
\begin{equation}
\Gamma^C_{AB} = \Gamma^C_{BA} \ . \label{TorsionFree}
\end{equation}

We shall now prove the first Cartan structure
equation, which in the present context states that the torsion tensor \eqref{eq:Torsionuv} can be written in terms of
covariant derivatives as
\begin{equation}
\begin{tabular}{|c|}\hline\\
$\displaystyle
{\sf T}^\st(u,v) = {}^{\phi_1}\nabla^\star_{{}^{\phi_2}v}\big({}^{\phi_3}u\big)
- {}^{\phi_1}\nabla^\st_{{}^{\phi_2}{}_\al u}\big({}^{\phi_3\, \al} v\big)
+ [u,v]_\star
$\\\\\hline\end{tabular}
\label{Torsion}\end{equation} 
The expression \eqref{Torsion} agrees with the definition of torsion from~\cite{Blumenhagen2016}. 
To prove 
\eqref{Torsion}, we first check it in our good basis:
we set $u=\partial_A$, $v=\partial_B$ and easily calculate
\begin{eqnarray}
\nabla^\star_B \partial_A - \nabla^\star_{{{}^\al \nn
\partial_A}} {{}_\al} \partial_B +[\partial_A, \partial_B]_\star  
&=& \nabla^\star_B \partial_A - \nabla^\star_{{A}} \partial_B \\[4pt]
&=& \partial_C\star (\Gamma^C_{AB} - \Gamma^C_{BA}) \\[4pt]
&=&{\sf T}^\st(\partial_A, \partial_B) \    , \nn \label{TorsionBasisVectors}
\end{eqnarray}
where we used $[\partial_A, \partial_B]_\star =0$. The equality
\eqref{Torsion} then follows once we establish that the right-hand
side defines a tensor in ${\rm hom}_\st(\vect_\st\wedge_\st\vect_\st,\vect_\st)$.

For this, it is useful to write the right-hand side of \eqref{Torsion} as ${\mbf
  T}^\st(u,v)$, where
\bea\label{defTop}
{\mbf T}^\st := \le \quad
\re_\st \bullet(\nabla^\star\otimes \Id) - \le \quad
\re_\st \bullet(\nabla^\star\otimes \Id) \bullet \tau_\mcR + [ \quad ]_\star \ .
\eea
Here we used the fact that the compositon $\bullet$ is associative since the
pairing and the braiding are $U\vect^\FF(\Mcal)$-equivariant;
this implies that the composition $\bullet$ in this case
reduces to the usual composition of operators. The associators
entering (\ref{Torsion}) are then due to the definition
$(\nabla^\st\otimes\Id)(u\otimes_\st
v)={}^{\bar\phi_1}\nabla^\st({}^{\bar\phi_2}(u))\otimes_\st
{}^{\bar\phi_3} v$ from (\ref{deflid}).

As defined in (\ref{defTop}), the map ${\mbf T}^\st$ is linear
in both of its arguments
because it is a composition of linear maps.  A
first step in showing that ${\mbf T}^\st$ defines a tensor in ${\rm
  hom}_\st(\vect_\st\wedge_\st\vect_\st,\vect_\st)$ is showing that it
is well-defined on
$\vect_\st\otimes_\st\vect_\st$: 
\bea\label{leibT}
{\mbf T}^\st(u\st f,v) = {\mbf
  T}^\st({}^{\phi_1}u,{}^{\phi_2}f \st {}^{\phi_3}v)
\eea
for all $f\in A_\st$ and $ u,v\in \vect_\st$, so that we can write  ${\mbf
  T}^\st(u,v)= {\mbf  T}^\st(u\otimes_\st v)$.
Explicitly, as before with the sum of connections, 
we know that (\ref{leibT}) holds for the terms which come from the
right $A_\st$-linear part of the Leibniz rule, so we only need to
check that the inhomogeneous terms coming from the exterior derivative cancel out. In ${\mbf T}^\star(u\star f,v)$ such
terms come from ${}^{\phi_1}\nabla^\star_{{}^{\phi_2}v}
\big({}^{\phi_3}(u\star f)\big)$, which yields $\le~(u\otimes_\st \dd
f) \,,\,v~\re_\st$, and from
\bea
\big[u\star f\,,\,v\big]_\st &=& - \big[{}^\al v \,,\,{}_\al(u\star f)
\big]_\st \nonumber\\[4pt]
&=& - \big[{}^{\phi_1\,\varphi_1\,\rho_1\,\alpha\,\beta}v\,,\,
{}^{\phi_2\,\varphi_2\,\rho_2}{}_\al u\star
  {}^{\phi_3\,\varphi_3\,\rho_3}{}_\beta f \big]_\st \nonumber\\[4pt]
&=& \big[{}^{\bar\phi_1\,\bar\varphi_1}u \,,\,{}^{\bar\phi_2\,\bar\varphi_2\,\beta}v \big]_\st\st
{}^{\bar\phi_3\,\bar\varphi_3}{}_\beta f - {}^{\bar\phi_1}u\st
\Lie^\st_{{}^{\bar\phi_2\, \beta}v}\big({}^{\bar\phi_3}{}_\beta f \big) \ ,
\label{eq:ustarfbracket}\eea
where we used \eqref{eq:DeltaRPhib}
and \eqref{eq:bracketLeibniz}. By definition of the Lie derivative we
have
\bea\label{eq:Lieufid}
\Lie^\st_u(f) = \Lie_{\of\,^\al(u)}\big(\,\of\,_\al(f)\big) =
\le~\of\,^\al(u) \,,\, \of\,_\al(\dd f)~\re = \le~{}^\beta\dd f\,,\,
{}_\beta u~\re_\st \ ,
\eea
and so $\Lie^\st_{{}^{\bar\phi_2\, \beta}v}({}^{\bar\phi_3}{}_\beta f) =\le~{}^{\bar\phi_3}\dd f\,,\,
{}^{\bar\phi_2} v~\re_\st$. It follows that the inhomogeneous term in
\eqref{eq:ustarfbracket} can be written as
\bea
{}^{\bar\phi_1}u\st
\Lie^\st_{{}^{\bar\phi_2\, \beta}v}({}^{\bar\phi_3}{}_\beta f) = \le~(
{}^{\bar\varphi_1 \,\phi_1}u\otimes_\st {}^{\bar\varphi_2\, \phi_2}\dd
f) \,,\, {}^{\bar\varphi_3\, \phi_3}v~\re_\st = \le~(u\otimes_\st\dd
f)\,,\, v~\re_\st
\eea
and hence cancels the appropriate term. 
Next, it follows immediately that ${\mbf T}^\st$ restricts
from  $\vect_\st\otimes_\st\vect_\st$ to
$\vect_\st\wedge_\st\vect_\st$ because 
it is star-antisymmetric under exchange of its arguments.
Finally, we need to check right
$A_\st$-linearity
\bea
{\mbf T}^\st\big((u\otimes_\st v)\st f\big) = {}^{\bar\phi_1}{\mbf
  T}^\st\big({}^{\bar\phi_2}(u\otimes_\st v) \big) \st {}^{\bar\phi_3}f \ ,
\eea
which is equivalent to
\bea
{\mbf T}^\st({}^{\phi_1}u,{}^{\phi_2}v\st{}^{\phi_3}f) =
{}^{\bar\phi_1\, \bar\varphi_1}{\mbf
  T}^\st({}^{\bar\phi_2}u,{}^{\bar\varphi_2}v) \st
{}^{\bar\phi_3\,\bar\varphi_3}f \ .
\eea
Again we just check that the inhomogeneous terms
coming from the Leibniz rule cancel out: The contribution to the
left-hand side from the first covariant
derivative in \eqref{Torsion} is right $A_\star$-linear by the same
computation that led to \eqref{eq:Lidlinear}, while by \eqref{eq:DeltaRPhia} the star-Lie
derivative term from \eqref{eq:bracketLeibniz} is cancelled by the
inhomogeneous term from
\bea
{}^{\phi_1}\nabla^\st_{{}^{\phi_2}{}_\al u}{}^{\phi_3\, \al}(v\star f) =
\le~ {}^{\phi_1\,
  \rho_1}\nabla^\st({}^{\phi_3\,\bar\varphi_1\,\bar\eta_1\,\zeta_1\,
  \beta}v\st {}^{\bar\varphi_1\,\bar\eta_2\,\bar\zeta_2\, \al\,
  \rho_3}f) \,,\, {}^{\phi_2\,\rho_2\,\bar\varphi_3\,
  \bar\eta_3\,\bar\zeta_3}{}_{\alpha\,\beta}u ~\re_\st
\eea
after using \eqref{eq:Lieufid} and \eqref{Q4}.

\subsection{Curvature}

We proceed by defining the curvature of a connection as in the classical
case, i.e., as the square of the covariant derivative, with the composition
being the $U\vect^\FF({\cal M})$-equivariant $\bullet$-composition of
linear maps:
\bea\label{defRdd}
{\sf R}^\st:=\dd_{\nabla^\st}\bullet \dd_{\nabla^\st}\,:\, \vect_{\st} \
\longrightarrow \ \vect_{\st} \otimes_\st \Omega_{\st}^2 \ .
\eea
This definition is well-posed because the linear map
$\dd_{\nabla^\st}\bullet \dd_{\nabla^\st}$ is right  $A_\st$-linear
and hence defines a tensor 
$\dd_{\nabla^\st}\bullet\dd_{\nabla^\st}\in {\rm
  hom}_{\star}(V_\star,V_\star\otimes_{\star}\Omega_\star^2)$.
Right  $A_\st$-linearity is proven 
 by repeated iteration of the Leibniz rule for $\nabla^\st$, giving
\begin{eqnarray}
(\dd_{\nabla^\st}\bullet\dd_{\nabla^\st})(v\star f) &=& {}^{\phi_1\,\varphi_1}\dd_{\nabla^\st}\big({}^{\phi_2\,\varphi_2}\nabla^\st({}^{\phi_3}v\star {}^{\varphi_3}f)\big) \nonumber \\[4pt]
&=& {}^{\phi_1\,\varphi_1}\dd_{\nabla^\st}\big({}^{\phi_2\,\varphi_2}\big[{}^{\bar\rho_1}\nabla^\st({}^{\bar\rho_2\,\phi_3} v)\star {}^{\bar\rho_3\,\varphi_3} f\big]\big) + {}^{\phi_1\,\varphi_1}\dd_{\nabla^\st}\big( {}^{\phi_2\,\varphi_2} ({}^{\phi_3}v\otimes_{\star}{\rm d}{}^{\varphi_3}f)\big) \nonumber \\[4pt]
&=& {}^{\phi_1\,\check\phi_1\, \varphi_1} \dd_{\nabla^\st}\big( {}^{\phi_2\,\varphi_2\,\bar\rho_1} \nabla^\st({}^{\bar\rho_2\,\phi_3\,\check\phi_3} v)\star {}^{\check\phi_2\, \bar\rho_3\,\varphi_3} f\big) + {}^{\phi_1\,\varphi_1} \dd_{\nabla^\st} \big({}^{\varphi_2\, \phi_3} v\otimes_{\star}{\rm d}{}^{\phi_2\, \varphi_3} f\big)\nonumber \\[4pt]
&=& {}^{\phi_1\,\check\phi_1\, \varphi_1} \dd_{\nabla^\st} \big( {}^{\phi_2\,\varphi_2\,\bar\rho_1} \nabla^\st({}^{\bar\rho_2\,\phi_3\,\check\phi_3} v)\big)\star {}^{\check\phi_2\, \bar\rho_3\,\varphi_3} f+\dd_{\nabla^\st} (v\otimes_{\star}{\rm d}f) \nonumber\\
&& -\, {}^{\phi_1\,\check\phi_1\, \varphi_1} \big({}^{\phi_2\,\varphi_2\,\bar\rho_1} \nabla^\st({}^{\bar\rho_2\,\check\phi_3} v)\otimes_{\star}{}^{\check\phi_2\, \bar\rho_3\,\varphi_3} ({\rm d}f)\big) \nonumber \\[4pt]
&=& {}^{\bar\phi_1} (\dd_{\nabla^\star}\bullet\dd_{\nabla^\st} )({}^{\bar\phi_2} v)\star {}^{\bar\phi_3} f + \dd_{\nabla^\st}(v\otimes_{\star}{\rm d}f) \nonumber \\ 
&& -\, {}^{\check\phi_1\,\varphi_1\,\phi_2\,\varphi_2\, \bar\rho_1} \nabla^\st ({}^{\bar\rho_2\,\phi_3\,\check\phi_3} v)\otimes_{\star} {}^{\phi_1\,\check\phi_2\, \bar\rho_3\,\varphi_3} ({\rm d}f)\nonumber \\[4pt]
&=& {}^{\bar\phi_1} (\dd_{\nabla^\star}\bullet\dd_{\nabla^\st} )({}^{\bar\phi_2} v)\star {}^{\bar\phi_3} f + \dd_{\nabla^\st}(v\otimes_{\star}{\rm d}f) - {}^{\bar\rho_1}\nabla^\st ({}^{\bar\rho_2} v)\otimes_{\star}{}^{\bar\rho_3}({\rm d}f)\nonumber \\[4pt]
&=& {}^{\bar\phi_1} (\dd_{\nabla^\star}\bullet\dd_{\nabla^\st}
    )({}^{\bar\phi_2} v)\star {}^{\bar\phi_3} f \ .
\end{eqnarray}
For trivial associator this definition of curvature
reduces to the noncommutative curvature considered in \cite{AS}, while
the general noncommutative and nonassociative curvature defined in~\cite{Barnes2015}
requires an extra braided commutator in the setting of arbitrary
quasi-Hopf algebras. 

Acting on a basis vector field $\partial_A$ gives
\begin{eqnarray}
{\sf R}^\st(\partial_A) &=& {}^{\phi_1}\dd_{\nabla^\star} \big(
{}^{\phi_2} \nabla^\star ({}^{\phi_3}\partial_A) \big) \nn\\[4pt]
&=& \dd_{\nabla^\star} (\nabla^\star \partial_A ) \nn\\[4pt]
&=& \dd_{\nabla^\star} (\partial_B\otimes_\star \Gamma^B_A)\nn\\[4pt]
&=& {}^{\bar{\phi}_1} \nabla^\star ({}^{\bar{\phi}_2} \partial_B) \wedge_\star
{}^{\bar{\phi}_3} \Gamma^B_A + 
\partial_B\otimes_\star \dd\Gamma^B_A \nn\\[4pt]
&=& (\nabla^\star \partial_B)\wedge_\star \Gamma^B_A 
+ 
\partial_B\otimes_\star {\rm d}\Gamma^B_A \nn\\[4pt]
&=& (\partial_C\otimes_\star \Gamma^C_B)\wedge_\star \Gamma^B_A 
+ 
\partial_C\otimes_\star {\rm d}\Gamma^C_A \nn\\[4pt]
&=& {}^{\phi_1} \partial_C\otimes_\star ({}^{\phi_2}\Gamma^C_B \wedge_\star
{}^{\phi_3} \Gamma^B_A) +
\partial_C\otimes_\star {\rm d}\Gamma^C_A \nn\\[4pt]
&=& \partial_C\otimes_\star ( \Gamma^C_B\wedge_\star \Gamma^B_A ) +
\partial_C \otimes_\star {\rm d}\Gamma^C_A \nn\\[4pt]
&=& \partial_C\otimes_\star ( {\rm d}\Gamma^C_A + \Gamma^C_B\wedge_\star \Gamma^B_A 
) \nn\\[4pt]
&=:& \partial_C\otimes_\star {\sf R}^C_A \ .\label{Curvature2Form}
\end{eqnarray}
We used the fact that the associator acts trivially on basis vector fields and
that the covariant derivative on form-valued vector fields acts on the
form-valued part just as the exterior derivative $\rm d$ (see \eqref{eq:liftconnection}), which commutes
with vector fields (in particular those defining the twist
(\ref{twist})). Taking the exterior derivative of the torsion 2-form from \eqref{eq:Torsion2form} yields the first Bianchi identity
\bea
\dd{\sf T}^A+\Gamma^A_B\wedge_\st {\sf T}^B={\sf R}^A_B\wedge_\st\dd x^B \ ,
\eea
whereas taking the exterior derivative of the curvature 2-form from
\eqref{Curvature2Form} gives the second Bianchi identity which reads
\bea
\dd{\sf R}_A^C+\Gamma_B^C\wedge_\st{\sf R}_A^B-{\sf
  R}^C_B\wedge_\st\Gamma_A^B =
\Gamma_B^C\wedge_\st\big(\Gamma^B_D\wedge_\st\Gamma^D_A\big) -
{}^{\phi_1}\Gamma_B^C\wedge_\st
\big({}^{\phi_2}\Gamma^B_D\wedge_\st {}^{\phi_3}\Gamma^D_A\big) \ .
\eea
We see that the naive expression for the second Bianchi identity is modified by the associator of connection 1-forms. In the associative case, the right-hand side vanishes and one recovers the usual expression of the second Bianchi identity.

Similarly to the torsion, we can also regard the curvature as the
tensor field
${\sf R}^\st\in {\rm hom}_\st\big(\vect_\st\otimes_\st
(\vect_\st\wedge_\st\vect_\st) ,\vect_\st \big)$ given on vectors $u,v,z\in\vect_\st$ by the vector field
\bea\label{defR}
{\sf R}^\st(z,u,v) :=
 \le~{}^{\bar\phi_1}{\sf R}^\st ({}^{\bar\phi_2}z)\,,\, {}^{\bar\phi_3}(u\otimes_\st v)~\re_\st \ . 
\label{eq:Curvzuv}\eea
Indeed from the definition we see that 
${\sf R}^\st(z,u,v)={\sf R}^\st(z,u\otimes_\st v)$, and moreover it is not
difficult to show that it gives the same result when
evaluated on $(z\st f, u\otimes_\st v)$ and on 
$({}^{\phi_1}z,\,{}^{\phi_2}f\st {}^{\phi_3}(u\otimes v))$ so that it is well-defined on $\vect_\st\otimes_\st
(\vect_\st\otimes_\st\vect_\st) $. Hence we can write
\eq\label{Rtens}
{\sf R}^\st(z,u,v)={\sf R}^\st(z\otimes_\st( u\otimes_\st v))~.
\en
This is also consistent with the
$U\vect^\FF({\cal M})$-action:
\eq\label{xiRcomp}
{}^\xi\big({\sf R}^\st(z\otimes_\st( u\otimes_\st v))\big)
={}^{\xi_{(1)}} {\sf R}^\st\big({}^{\xi_{(2)}}(z\otimes_\st( u\otimes_\st v))\big)
~,
\en
for all $\xi\in U\vect^\FF({\cal M})$; this follows by using (\ref{defR}) and ${}^{\xi_{(1)}}\big({}^{\bar\phi_1}{\sf
  R}^\st({}^{\bar\phi_2}z) \big)={}^{\xi_{(1)_{(1)}}\, \bar\phi_1}{\sf
  R}^\st({}^{\xi_{(1)_{(2)}}\, \bar\phi_2}z)$, and then the quasi-associativity
property (\ref{eq:DeltaPhirewrite}) of the coproduct.
Finally, the map ${\sf R^\st}$ is right $A_\star$-linear: 
\bea
{\sf R}^\st\big((z\otimes_\st(u\otimes_\st v))\st f\big) =
{}^{\bar\phi_1}{\sf
R^\st}\big({}^{\bar\phi_2}(z\otimes_\st(u\otimes_\st
v))\big)\st{}^{\bar\phi_3}f \ .
\eea
We have thus shown that ${\sf R}^\st\in {\rm hom}_\st\big(\vect_\st\otimes_\st
(\vect_\st\otimes_\st\vect_\st) ,\vect_\st \big)$; since moreover
${\sf R}^\st(z,u,v)= -{\sf  R}^\st(z,{}^\al v,{}_\al u)$, we conclude
that 
${\sf R}^\st\in {\rm hom}_\st\big(\vect_\st\otimes_\st
(\vect_\st\wedge_\st\vect_\st) ,\vect_\st \big)$.

One readily extracts the explicit expression for the curvature
coefficients with respect to the good basis $\partial_A$. Using
  the star-pairing and \eqref{Curvature2Form} we get
\bea
{\sf R}^\st(\partial_A,\partial_B,\partial_C)
&=& \le~\partial_D\otimes_\st{\sf
  R}^D_A\,,\,\partial_B\wedge_\st\partial_C~\re_\st \nn \\[4pt]
&=& \partial_D\star\le~\dd\Gamma_A^D+\Gamma_{B'}^D\wedge_\st
\Gamma_A^{B'} \,,\, \partial_B\wedge_\st\partial_C~\re_\st \nn \\[4pt] 
&=& \partial_D\star\le~((\partial_{A'}\Gamma_{AE}^D\st\dd
x^{A'})\wedge_\st \dd x^E \nn \\ && \qquad \qquad +\, (\Gamma_{B'E}^D\st\dd
x^E)\wedge_\st(\Gamma_{AF}^{B'}\st\dd
x^F)\,,\, \partial_B\wedge_\st\partial_C~\re_\st \nn \\[4pt] 
&=& \partial_D\st\le~ \partial_{A'}\Gamma^D_{AE}\st(\dd
x^{A'}\wedge_\st \dd x^E) \nn \\ && +\, \Gamma^D_{B'E}\st(\delta^E{}_{E'}\,
\Gamma^{B'}_{AF}+ \ii\kappa\, \RRR^{EG}{}_{E'}\, (\partial_G\Gamma_{AF}^{B'})) \st
(\dd x^{E'}\wedge_\st\dd x^F) \,,\, \partial_B\wedge_\st\partial_C~\re_\st
\nn \\[4pt]
&=& \partial_D
\st\big( \partial_C\Gamma^D_{AB}-\partial_B\Gamma^D_{AC}
-\Gamma^D_{B'E}\st(\delta^E{}_B\, \Gamma^{B'}_{AC} + \ii\kappa\, \RRR^{EG}{}_B\,
(\partial_G\Gamma^{B'}_{AC})) \nn \\ && \hspace{4cm} + \,
\Gamma^D_{B'E}\st (\delta^E{}_C\, \Gamma^{B'}_{AB}+\ii\kappa\, \RRR^{EG}{}_C\,
(\partial_G\Gamma_{AB}^{B'})) \big) \nn \\[4pt]
&=:& \partial_D\st{\sf R}^D{}_{ABC} \ ,
\label{eq:Curvgood}\eea
where once again we used the fact that the associator acts trivially on the
basis vectors and basis 1-forms.

We shall now prove the second Cartan structure equation, which in the present context states that the curvature
tensor \eqref{eq:Curvzuv} can be written in terms of covariant derivatives as 
\bea
\begin{tabular}{|l|}\hline\\
$\displaystyle
{\sf R}^\st(z,u,v)  \ \ = \ \ 
{}^{\kappa_1\,\check\phi_1\,\phi_1'}\mbf\nabla^\st_{{}^{\bar\rho_3\,
    \bar\zeta_3\, \bar\phi_3\, \phi_3'}v}\big({}^{\bar\rho_1\,
  \bar\phi_1\, \kappa_2\, \check\phi_2\,
  \phi_2'}\mbf\nabla^\st_{{}^{\bar\rho_2\, \bar\zeta_2\,
    \check\phi_3}u} {}^{\bar\zeta_1\, \bar\phi_2\, \kappa_3}z \big)
$ \\[2mm] $\qquad\qquad\qquad \ \ \ \ 
 -\, {}^{\kappa_1\,\check\phi_1\,\phi_1'}\mbf\nabla^\st_{{}^{\bar\rho_3\,
    \bar\zeta_3\, \bar\phi_3\, \phi_3'}{}_\al u}\big({}^{\bar\rho_1\,
  \bar\phi_1\, \kappa_2\, \check\phi_2\,
  \phi_2'}\mbf\nabla^\st_{{}^{\bar\rho_2\, \bar\zeta_2\,
    \check\phi_3}{}^\al v} {}^{\bar\zeta_1\, \bar\phi_2\, \kappa_3}z \big) +
\mbf\nabla_{[u,v]_\st}^\st z
$\\\\\hline\end{tabular}
\label{eq:Curvexpl}\eea
where to streamline the notation we introduced the bold-face covariant derivative
\bea
\mbf\nabla_v^\st u:=\le~ {}^{\bar\phi_1}\nabla^\st({}^{\bar\phi_2}u) \,,\,
{}^{\bar\phi_3}v ~\re_\st~.
\label{eq:twistedcovder}\eea
The expression \eqref{eq:Curvexpl} for the curvature agrees with that
of~\cite{Blumenhagen2016} after taking into account their different conventions;\footnote{We are grateful to Michael Fuchs for pointing this out to us.} for trivial associator it reduces to the
general expression in \cite{Aschieri2005}.
To prove \eqref{eq:Curvexpl}  we first check it on our
good basis by setting $z=\partial_A$, $u=\partial_B$ and
$v=\partial_C$. Then the right-hand side reduces to
\bea
&& \nabla^\st_{C}(\nabla^\st_{B}\partial_A)  -
\nabla^\st_{{}_\al\partial_B}(\nabla^\st_{{}^\al \partial_C}\partial_A) +
\nabla^\st_{[\partial_B,\partial_C]_\st}\partial_A \nn 
\nn \\[4pt] && \qquad \qquad \qquad \qquad  \qquad \qquad  \qquad \ \ = \ \ \nabla_C^\st(\nabla^\st_B\partial_A) -
\nabla_B^\st(\nabla^\st_C\partial_A) \nn \\[4pt]
&& \qquad \qquad \qquad \qquad \qquad  \qquad  \qquad \ \ = \ \ \nabla^\st_C(\partial_D\star\Gamma^D_{AB}) -
\nabla^\st_B(\partial_D\st\Gamma^D_{AC}) \nn \\[4pt]
&& \qquad \qquad \qquad \qquad \qquad  \qquad  \qquad \ \ = \ \ \le~ \nabla^\st \partial_D \,,\, {}^\al\partial_C ~\re_\st \st
{}_\al\Gamma^D_{AB} + \partial_D\st\le~ \dd\Gamma_{AB}^D
\,,\, \partial_C ~\re_\st \nn\\
&& \qquad \qquad \qquad \qquad \qquad  \qquad \qquad \ \ \qquad \ \ -\, \le~ \nabla^\st \partial_D \,,\, {}^\al\partial_B ~\re_\st \st
{}_\al\Gamma^D_{AC} - \partial_D\st\le~ \dd\Gamma_{AC}^D
\,,\, \partial_B ~\re_\st \nn\\[4pt]
&& \qquad \qquad \qquad \qquad \qquad  \qquad \qquad \ \ = \ \ {\sf R}^\st(\partial_A,\partial_B,\partial_C) \ ,
\eea
where in the third equality we used the Leibniz rule \eqref{Q4}
while the last equality follows from \eqref{eq:Curvgood}. 
The equality \eqref{eq:Curvexpl} for arbitrary vectors then follows once we establish that the right-hand side defines a tensor in $\hom_\st(\vect_\st\otimes_\st
(\vect_\st \wedge_\st \vect_\st), \vect_\st)$.

For this, as in the case of the torsion, we rewrite the right-hand side of \eqref{eq:Curvexpl} as a
trilinear map $\mbf R^\st$ on vectors $z$, $u$ and $v$, and prove that it is a map in $\hom_\st(\vect_\st\otimes_\st
(\vect_\st \wedge_\st \vect_\st), \vect_\st)$. To arrive at the form of $\mbf
R^\st$, for notational clarity we first consider vectors $z,u,v$ on
which the associator acts trivially (for example basis vectors
$\partial_A,\partial_B,\partial_C$). Then we reproduce
$\nabla^\st_v(\nabla^\st_u\,z)$ as the elementary compositions
\bea
z\otimes_\st (u\otimes_\st v)&\xmapsto{{\nabla^\st\otimes_\st\, \Id^{\otimes_\st
      2}}}&\nabla^\st z\otimes_\st (u\otimes_\st
v)~\xmapsto{{\Phi^{-1}}}~(\nabla^\st z_{\,}\otimes_\st u)\otimes_\st
v\\[7pt]
&\xmapsto{{\le \quad \re_{\!\!\:_\st}\otimes_\st \,\Id}}&\nabla^\st_u\:\!
z\otimes_\st v\nn
~\xmapsto{{\nabla^\st\otimes_\st\, \Id}}~
\nabla^\st(\nabla^\st_u\, z)\otimes_\st
v~\xmapsto{{\le \quad \re_\st}}~\nabla^\st_v(\nabla^\st_u \,z)~.
\eea
This leads to a definition of ${\mbf R}^\st$ written solely in terms of the
connection $\nabla^\st$, the associator $\Phi^{-1}$, and the
equivariant maps studied in Section~\ref{sec:functions}, which reads as
\bea
\mbf R^\st &:=& \le \quad \re_\st \bullet
(\nabla^\st\otimes\Id)\bullet (\le \quad \re_\st
\otimes\Id)\bullet
\Phi^{-1}_{\vect_\st\otimes\Omega_\st^1,\vect_\st,\vect_\st} \bullet
(\nabla^\st\otimes \Id^{\otimes2})\nn
\bullet\, 
(\Id^{\otimes3}- \Id\otimes_{\cal R}\tau_\mcR) \\[5pt]&&\,\,+\,\,\, \le \quad \re_\st \bullet
(\nabla^\st\otimes\Id) \bullet (\Id\otimes_{\cal R} [ \quad ]_\st)~.
\label{eq:Curvoperator}\eea
Even though the composition $\bullet$ is nonassociative, there is no
ambiguity in this definition because of the
equivariance of the maps which are composed and because
${}^{\phi_a}\phi_b=0$ (the associator being generated by an abelian subalgebra).
For these same reasons, there is the more explicit expression
\bea
\mbf R^\st \!&\!\!\!\!=\!\!\!\!& \le \quad \re_\st \circ
({}^{\phi_1}\nabla^\st\otimes\Id)\circ_{\!} (\le \quad \re_\st
\otimes\Id)\circ
\Phi^{-1}_{\vect_\st\otimes\Omega_\st^1,\vect_\st,\vect_\st} \!\!\circ
({}^{\phi_2}\nabla^\st\otimes \Id^{\otimes2})\!\circ {\phi_3}\!
 \circ
(\Id^{\otimes3}\!- \Id\otimes\tau_\mcR)  \nn \\[5pt]
&&\,\,+\,\, \,\le \quad \re_\st \circ
(\nabla^\st\otimes\Id) \circ (\Id\otimes [ \quad ]_\st)~.
\label{eq:Curvoperatorbis}\eea
As sought, explicit evaluation of ${\mbf R}^\st$ on $z\otimes_\st (u\otimes_\st
v)$ gives the right-hand side of (\ref{eq:Curvexpl}):
\bea
{\mbf R}^\st(z,u,v) &=& \le~ {}^{\bar\eta_1\,\phi_1}\nabla^\st 
{\,}^{\bar\eta_2}\le~ {}^{\bar\rho_1}({}^{\bar\varphi_1\,
  \phi_2}\nabla^\st{}^{\bar\varphi_2\,\phi_{3_{(1)}}}z) \,,\, {}^{\bar\rho_2\,
  \bar\varphi_{3_{(1)}}\, \phi_{3_{(2)_{(1)}}}}u~\re_\st \,,\,
{}^{\bar\eta_3\, \bar\rho_3\, \bar\varphi_{3_{(2)}}\, \phi_{3_{(2) _{(2)}}}}v
~\re_\st \nn \\ 
&& -\, \le~ {}^{\bar\eta_1\,\phi_1}\nabla^\st 
{\,}^{\bar\eta_2}\le~ {}^{\bar\rho_1}({}^{\bar\varphi_1\,
  \phi_2}\nabla^\st{}^{\bar\varphi_2\,\phi_{3_{(1)}}}z) \,,\, {}^{\bar\rho_2\,
  \bar\varphi_{3_{(1)}}\, \phi_{3_{(2)_{(1)}}}}{}^\al v ~\re_\st \,,\,
{}^{\bar\eta_3\, \bar\rho_3\, \bar\varphi_{3_{(2)}}\,
  \phi_{3_{(2)_{(2)}}}}{}_\al u
~\re_\st  \nn\\
&& +\, \le~{}^{\bar\varphi_1}\nabla^\st{}^{\bar\varphi_2}z \,,\,
{}^{\bar\varphi_3}[u,v]_\st ~\re_\st \nn \\[4pt]
&=& \le~ {}^{\bar\eta_1\, \kappa_1\, \check\phi_1\, \phi_1'}\nabla^\st
{\,}^{\bar\eta_2}\le~ {}^{\bar\rho_1\, \bar\varphi_1\, \bar\phi_1\,
  \kappa_2\, \check\phi_2\, \phi_2'}\nabla^\st{}^{\bar\zeta_1\,
  \bar\varphi_2\, \bar\phi_2\, \kappa_3}z \,,\, {}^{\bar\rho_2\,
  \bar\zeta_2\, \bar\varphi_3\, \check\phi_3}u ~\re_\st \,,\,
{}^{\bar\eta_3\, \bar\rho_3\, \bar\zeta_3\, \bar\phi_3\, \phi_3'}v
~\re_\st \nn \\
&& -\, \le~ {}^{\bar\eta_1\, \kappa_1\, \check\phi_1\, \phi_1'}\nabla^\st
{\,}^{\bar\eta_2}\le~ {}^{\bar\rho_1\, \bar\varphi_1\, \bar\phi_1\,
  \kappa_2\, \check\phi_2\, \phi_2'}\nabla^\st{}^{\bar\zeta_1\,
  \bar\varphi_2\, \bar\phi_2\, \kappa_3}z \,,\, {}^{\bar\rho_2\,
  \bar\zeta_2\, \bar\varphi_3\, \check\phi_3}{}^\al v ~\re_\st \,,\,
{}^{\bar\eta_3\, \bar\rho_3\, \bar\zeta_3\, \bar\phi_3\, \phi_3'}
{}_\al u ~\re_\st  \nn \\ 
&& + \, \le~{}^{\bar\varphi_1}\nabla^\st{}^{\bar\varphi_2}z \,,\,
{}^{\bar\varphi_3}[u,v]_\st ~\re_\st \ .
\eea

Now the proof that ${\mbf R}^\st$ is a map in 
$\hom_\st(\vect_\st\otimes_\st
(\vect_\st \wedge_\st \vect_\st), \vect_\st)$ requires as a first
step to show that it is a well-defined map on $\vect_\st\otimes_\st(\vect_\st\otimes_\st\vect_\st)$:
\eq\label{middlel}
\mbf R^\st (z,u\st f, v)=\mbf R^\st (z,{}^{\phi_1}u, {}^{\phi_2}f\st
{}^{\phi_3}v)~,
\en 
so that we get a well-defined map $\mbf R^\st
(z,u\otimes_\st v)=\mbf R^\st (z,u, v)$, and 
\eq\label{firsl}
\mbf R^\st (z\st f,u\otimes_\st v)=\mbf R^\st ({}^{\phi_1}z,
{}^{\phi_2}f\st{}^{\phi_3}(u\otimes_\st v))~,
\en
so that we get a well-defined map 
$\mbf R^\st (z\otimes_\st (
u\otimes_\st v))=\mbf R^\st (z,
u,v)$.
The star-antisymmetry of $\mbf R^\st $
under $u\otimes_\st v\to {}^\al v\otimes_\st {}_\al u$ then immediately follows,
and this implies that $\mbf R^\st $  is a linear map from 
 $\vect_\st\otimes_\st(\vect_\st\wedge_\st\vect_\st)$ to $\vect_\st$.
The final step is to show that $\mbf R^\st \in \hom_\st(\vect_\st\otimes_\st
(\vect_\st \wedge_\st, \vect_\st), \vect_\st)$, i.e., that it is  right $A_\st$-linear:
\bea
\mbf R^\st\big((z\otimes_\st(u\otimes_\st v))\st f\big) =
{}^{\bar\phi_1}\mbf
R^\st\big({}^{\bar\phi_2}(z\otimes_\st(u\otimes_\st
v))\big)\st{}^{\bar\phi_3}f \ .
\label{eq:CurvAlin}\eea
In the following we prove right $A_\st$-linearity \eqref{eq:CurvAlin}; the remaining
$A_\st$-linearity properties (\ref{middlel}) and (\ref{firsl})  can be
established with similar techniques.

For this, we note again that if the
star-connection $\nabla^\st$ and the star-Lie derivative
${\cal L}^\st=[\quad]_\st$ were right $A_\st$-linear
maps, then the operator (\ref{eq:Curvoperator}) would also be right
$A_\st$-linear because all composite maps would be right
$A_\st$-linear. Hence as before it suffices to check that the
inhomogeneous terms coming from the Leibniz rule for the connection
and the Lie derivative cancel out. 
We denote by ${\sf Leib}^\st$ the projector onto the inhomogeneous
terms. For example
\bea
{\sf Leib}^\st\big(\nabla^\st(u\st f)\big) = u\otimes_\st\dd f
\ , 
\eea
which induces
\bea 
{\sf  Leib}^\st\big(\nabla^\st_v(u\st f)\big) =
\le~ {\sf
  Leib}^\st(\nabla^\st(u\st f)) \,,\, v ~\re_\st = {}^{\varphi_1}u
\st \le~ {}^{\varphi_2}\dd f \,,\, {}^{\varphi_3}v ~\re_\st =
{\sf  Leib}^\st\big(\mbf\nabla^\st_v(u\st f)\big) \ .
\eea
Here we used the fact that in the inhomogeneous term the
covariant derivative $\mbf\nabla_v^\st $ from (\ref{eq:twistedcovder})
acts as a rescaled exterior
derivative ${}^{\bar\phi_1}\dd = \epsilon(\bar\phi_1)\, \dd$, which is $U\vect^\FF({\cal M})$-equivariant. 
Furthermore, from (\ref{eq:ustarfbracket}) we also have 
\eq
{\sf Leib}^\st\big([u\st f,v]_\st \big) = {{}^{\bar\phi_1}u\st \le~{}^{\bar\phi_3}\dd f \,,\,
  {}^{\bar\phi_2}v ~\re_\st } \ .
\en
The projector ${\sf Leib}^\st$ in these examples is a linear
operator in $u$, $v$ and $f$. We have to show that ${\sf Leib}^\st\big(\mbf
R^\st\big((z\otimes_\st(v\otimes_\st u))\st f\big)\big)=0$ for all
$z,v,u\in \vect_\st$ and $f\in A_\st$.
Since
\bea
\big(z\otimes_\st(v\otimes_\st u)\big)\st
f &=& {}^{\varphi_1\,\rho_1}z\otimes_\st
\big(({}^{\varphi_2}v\otimes_\st {}^{\rho_2}u)\st {}^{\varphi_3\, \rho_3}f\big) \nn \\[4pt]
&=& {}^{\varphi_1\, \rho_1}z\otimes_\st \big(
{}^{\zeta_1\,\varphi_2}v\otimes_\st ( {}^{\zeta_2\,\rho_2}u\st
{}^{\zeta_2\, \varphi_3\, \rho_3}f ) \big) \ ,
\eea
this condition is
equivalent to ${\sf Leib}^\st\big(\mbf
R^\st\big(z\otimes_\st(v\otimes_\st (u\st f) )\big)\big)=0$ because
of linearity in $z,v,u, f$. Hence we
check that ${\sf Leib}^\st\big({\mbf R}^\st(z,v, u\st f)\big) =
0$, or equivalently, using star-antisymmetry and linearity again, that
${\sf Leib}^\st\big({\mbf R}^\st(z, u\st f, v)\big)
= 0$.

From \eqref{eq:Curvexpl} we write
\bea
{\sf Leib}^\st\big({\mbf R}^\st(z, u\st f, v)\big) &=& {\sf
  Leib}^\st\big({}^{\kappa_1\,\check\phi_1\,\phi_1'}\mbf\nabla^\st_{{}^{\bar\rho_3\,
    \bar\zeta_3\, \bar\phi_3\, \phi_3'}v}\big({}^{\bar\rho_1\,
  \bar\phi_1\, \kappa_2\, \check\phi_2\,
  \phi_2'}\mbf\nabla^\st_{{}^{\bar\rho_2\, \bar\zeta_2\,
    \check\phi_3}(u\st f)} {}^{\bar\zeta_1\, \bar\phi_2\, \kappa_3}z \big) \big)
\nn \\
&& +\, {\sf Leib}^\st\big(\mbf\nabla_{[u\st f,v]_\st}^\st z \big)
\label{eq:LeibRAB}\eea
and compare the two contributions. The second contribution in
\eqref{eq:LeibRAB} is equal to
\bea
\mbf\nabla^\st_{{\sf Leib}^\st([u\st f,v]_\st)}z &=&
-\mbf\nabla^\st_{{}^{\bar\phi_1}u\st \le~{}^{\bar\phi_3}\dd f \,,\,
  {}^{\bar\phi_2}v ~\re_\st } z \nn \\[4pt] 
&=& - \big({}^{\bar\kappa_1\,
  \bar\varphi_1}\mbf\nabla^\st_{{}^{\bar\kappa_2\,
    \bar{\check\phi}_2\, \bar\phi_1}u}{}^{\bar{\check\phi}_1\,
    \bar\varphi_2}z \big)\st {}^{\bar\kappa_3\, \bar{\check\phi}_3\,
    \bar\varphi_3}\le~ {}^{\bar\phi_3}\dd f \,,\, {}^{\bar\phi_2}v
  ~\re_\st \ ,
\label{eq:LeibAcontr}\eea
where in the second equality we used the definition \eqref{eq:twistedcovder} to rewrite
\bea
\mbf\nabla^\st_{u\st f}z &=& \le~ {}^{\bar\phi_1\,
  \bar\varphi_1}\nabla^\st{}^{\bar\phi_2\, \bar\varphi_2}z \,,\,
{}^{\bar\phi_3}u\st {}^{\bar\varphi_3}f ~\re_\st \nn \\[4pt]
&=& \le~ {}^{\bar\phi_1'}({}^{\bar\phi_1\,
  \bar\varphi_1}\nabla^\st{}^{\bar\phi_2\, \bar\varphi_2}z) \,,\,
{}^{\bar\phi_2'\, \bar\phi_3}u ~\re_\st \st {}^{\bar\phi_3'\,
  \bar\varphi_3}f \nn \\[4pt]
&=& \big({}^{\bar\kappa_1\,
  \bar\varphi_1}\mbf\nabla^\st_{{}^{\bar\kappa_2\,
    \bar{\check\phi}_2}u}{}^{\bar{\check\phi}_1\, \bar\varphi_2}z
\big) \st {}^{\bar\kappa_3\, \bar{\check\phi}_3\, \bar\varphi_3}f \ ,
\eea
and then replace $u\st f$ with $ {}^{\bar\phi_1}u\st \le~{}^{\bar\phi_3}\dd f \,,\,
  {}^{\bar\phi_2}v ~\re_\st$.
The first contribution in \eqref{eq:LeibRAB} can be rewritten without
the first three associator legs $\kappa_a, \check\phi_a, \phi_a'$, because in the
inhomogeneous term the covariant derivative 
${}^{\kappa_1\,\check\phi_1\,\phi_1'}\mbf\nabla^\st_w$ again acts as a rescaled
exterior derivative
${}^{\kappa_1\,\check\phi_1\,\phi_1'}\dd=\epsilon(\kappa_1)\,
\epsilon(\check\phi_1)\, \epsilon(\phi_1')\, \dd$, which is $U\vect^\FF(\cal M)$-equivariant. Therefore
the first contribution in \eqref{eq:LeibRAB} equals
\bea
&& \!\!\!\!\!\!\!\!\!\!\!\!\!\!\!{\sf Leib}^\st\big( \mbf\nabla^\st_{{}^{\bar\rho_3\, \bar\phi_3'\,
  \bar\zeta_3\, \bar\eta_3\, \bar\phi_3}v} \big( {}^{\bar\rho_1\,
\bar\phi_1'\, \bar\phi_1}\mbf\nabla^\st_{{}^{\bar\rho_2\,
  \bar\zeta_2}u\st {}^{\bar\phi_2'\, \bar\eta_2}f}{}^{\bar\zeta_1\,
\bar\eta_1\, \bar\phi_2}z\big)\big) \nn \\
&& \qquad \ \ = \ \ {\sf Leib}^\st\big( \mbf\nabla^\st_{{}^{\bar\rho_3\,
  \bar\phi_3'\, \bar\zeta_3\, \bar\eta_3\, \bar\phi_3}v}\big( (
{}^{\bar\kappa_1\, \bar\varphi_1\, \bar\rho_1\, \bar\phi_1'\,
  \bar\phi_1}\mbf\nabla^\st_{{}^{\bar\kappa_2\, \bar{\check\phi}_2\,
    \bar\rho_2\, \bar\zeta_2}u}{}^{\bar{\check\phi}_1\,
  \bar\varphi_2\, \bar\zeta_1\, \bar\eta_1\, \bar\phi_2}z) \st
{}^{\bar\kappa_3\, \bar{\check\phi}_3\, \bar\varphi_3\, \bar\phi_2'\,
  \bar\eta_2}f \big) \big) \label{eq:LeibBcalc} \\[4pt]
&& \qquad \ \ = \ \ {}^{\chi_1}\big({}^{\bar\kappa_1\,
  \bar\varphi_1\, \bar\rho_1\, \bar\phi_1'\,
  \bar\phi_1}\mbf\nabla^\st_{{}^{\bar\kappa_2\, \bar{\check\phi}_2\,
    \bar\rho_2\, \bar\zeta_2}u}{}^{\bar{\check\phi}_1\,\bar\varphi_2\,
  \bar\zeta_1\, \bar\eta_1\, \bar\phi_2}z \big) \st \le~ {}^{\chi_2\,
  \bar\kappa_3\, \bar{\check\phi}_3\, \bar\varphi_3\, \bar\phi_2'\,
  \bar\eta_2}\dd f \,,\, {}^{\chi_3\, \bar\rho_3\, \bar\phi_3'\,
  \bar\zeta_3\, \bar\eta_3\, \bar\phi_3}v ~\re_\st \  .\nn
\eea
Replacing $(u,f,v)$ with $
({}^{\bar\eta_1}u,{}^{\bar\eta_2}f,{}^{\bar\eta_3}v)$ in
\eqref{eq:LeibBcalc} gives an action of
$\bar\phi_1'\otimes\bar\eta_1\otimes\bar\zeta_1\otimes \bar\phi_2'\,
\bar\zeta_2\, \bar\eta_2\otimes \bar\phi_3'\, \bar\zeta_3\, \bar\eta_3$
which cancels against that of $\big((\Delta_\FF\otimes\Id)\,
\Delta_\FF(\chi_1) \big) \otimes\chi_2\otimes \chi_3$ and yields
\bea
&& \big( {}^{\bar\kappa_1\, \bar\varphi_1\, \bar\rho_1\,
  \bar\phi_1}\mbf\nabla^\st_{{}^{\bar\kappa_2\, \bar{\check\phi}_2\,
    \bar\rho_2\, \bar\zeta_2}u}{}^{\bar{\check\phi}_1\,
  \bar\varphi_2\, \bar\zeta_1\, \bar\phi_2}z \big) \st \le~
{}^{\bar\kappa_3\, \bar{\check\phi}_3\, \bar\varphi_3}\dd f \,,\,
{}^{\bar\rho_3\, \bar\zeta_3\, \bar\phi_3}v ~\re_\st \\
&& \qquad\qquad\qquad\qquad\qquad\qquad\qquad \qquad \qquad \ \ = \ \ \big({}^{\bar\kappa_1\,
  \bar\varphi_1}\mbf\nabla^\st_{{}^{\bar\kappa_2\,
    \bar{\check\phi}_2}u}{}^{\bar{\check\phi}_1\, \bar\varphi_2}z
\big) \st {}^{\bar\kappa_3\, \bar{\check\phi}_3\, \bar\varphi_3}\le~
\dd f \,,\, v ~\re_\st \ , \nn
\eea
thereby cancelling the contribution \eqref{eq:LeibAcontr} with
the same replacement of $(u,f,v)$.  This shows that
\bea
{\sf Leib}^\st\big({\mbf R}^\st(z,{}^{\bar\eta_1}u\st
{}^{\bar\eta_2}f,{}^{\bar\eta_3}v) \big) = 0
\eea
and hence establishes the right $A_\st$-linearity property \eqref{eq:CurvAlin} as
required. 

\subsection{Ricci tensor\label{sec:Ricci}}

Since the associator acts trivially on the basis $\dd x^A$ and its
dual $\partial_A$, the definition of the Ricci tensor can be given
following the noncommutative case studied in
\cite{Aschieri2005}. We define
\eq\label{defRic}
{\sf Ric}^\st(u,v) := -\le~ {\sf R}^\st(u,v,\partial_A)\,,\,\dd x^A~\re_\st \ ,
\en
for all $u,v\in \vect_\st$, where the pairing between a vector field on the left and a form on the
right is given by 
\eq
\le~ u\,,\,\om
~\re_\st=\le~ \of\,^\al(u)\,,\,\of\,_\al(\om)~\re \ ,
\en
similarly to (\ref{pairing}). The properties of this pairing are completely analogous to those described in Section~\ref{Duality}, by simply interchanging
forms and vector fields in all expressions considered there.

We now show that the map (\ref{defRic}) defines a tensor ${\sf
  Ric}^\st\in \hom_\st(
\vect_\st\otimes_\st\vect_\st, \vect_\st)$. We first prove that it is
a map from $\vect_\st\otimes_\st\vect_\st$ to $\vect_\st$, so that we can
write ${\sf Ric}^\st(u,v)={\sf Ric}^\st(u\otimes_\st v)$; indeed, we have
\bea
{\sf Ric}^\st(u\st f, v)&=&\nn
-\le~ {\sf R}^\st(u\st f,v,\partial_A)\,,\,\dd x^A~\re_\st
\\[4pt]
&=&\nn
-\le~ {\sf R}^\st((u\st f)\otimes_\st(v\otimes_\st\partial_A))\,,\,\dd x^A~\re_\st
\\[4pt]
&=&
-\le~ {\sf R}^\st({}^{\phi_1}u\otimes_\st({}^{\phi_2}f \st {}^{\phi_3}(v\otimes_\st\partial_A)))\,,\,\dd x^A~\re_\st
\nn\\[4pt]
&=&-\le~ {\sf R}^\st({}^{\phi_1}u\otimes_\st(({}^{\phi_2}f \st
{}^{\phi_{3}}v)\otimes_\st\partial_A))\,,\,\dd x^A~\re_\st
\nn\\[4pt]
&=&
{\sf Ric}^\st({}^{\phi_1}u ,{}^{\phi_2}f\st {}^{\phi_3}v)
\eea
where in the second equality we used (\ref{Rtens}), and in the fourth equality
the fact that the associator acts trivially on $\partial_A$.
Next we prove compatibility with the $U\vect^\FF({\cal M})$-action:
\eq
{}^\xi\big({\sf Ric}^\st(u\otimes_\st v)\big)={}^{\xi_{(1)}}{\sf
  Ric}^\st\big({}^{\xi_{(2)}}(u\otimes v)\big)~,
\en
which follows from (\ref{xiRcomp}), iterated use of (\ref{eq:DeltaPhirewrite}) and 
from ${}^{\xi}(\partial_A\otimes_\st \dd x^A)=\epsilon(\xi)\,
\partial_A\otimes_\st \dd x^A$; this latter property follows from writing the
identity map $\Id: \vect_\st\to \vect_\st$ as in (\ref{identity}) and using ${}^\xi\Id=\epsilon(\xi)\, \Id$, i.e.,
${}^\xi\big(\Id(u)\big)={}^{\xi_{(1)}}\Id({}^{\xi_{(2)}}u)=\Id({}^\xi u)$, for
all $\xi\in U\vect^\FF({\cal M})$ and $u\in \vect_\st$.
Finally, we prove right $A_\st$-linearity. For this, we notice that
\beq\label{lemma475}
{}^{\phi_1}u\otimes_\st\big(({}^{\phi_2}v\st{}^{\phi_3}f) \otimes_{\st}\partial_A\big)
&=& {}^{\phi_1}u\otimes_\st\big({}^{\phi_2}v\otimes_\st({}^\al\partial_A\st 
{}_\al{}^{\phi_3}f)\big)\nn\\[4pt] 
 &=&
{}^{\phi_1}u\otimes_\st \big(({}^{\phi_2}v\otimes_\st{}^\al\partial_A)\st 
{}^{\phi_3}{}_\al f\big) \nn\\[4pt]
 &=& \big({}^{\bar\varphi_1\, \phi_1}u\otimes_\st{}^{\bar\varphi_2}({}^{\phi_2}v\otimes_\st{}^\al\partial_A)\big)\st 
{}^{\bar\varphi_3\, \phi_3}{}_\al f
\nn\\[4pt]
 &=& \big(u\otimes_\st(v\otimes_\st{}^\al\partial_A)\big)\st {}_\al f
\eeq
where we used the fact that the associator commutes with each leg of the ${\cal
  R}$-matrix and hence it acts trivially on ${}^\al\partial_A$.
Then the proof follows from (\ref{lemma475}) and the centrality of the tensor $\partial_A\otimes_\st \dd x^A$:
$f\st(\partial_A\otimes_\st \dd x^A)=({}^\al \partial_A\otimes_\st
{}^\be\dd x^A)\st {}_{\be\, \al}f=(\partial_A\otimes_\st \dd x^A)\st f$,
see (\ref{fstardxA}) and (\ref{RRRpartial}), or again (\ref{identity}). Explicitly, we have
\beq
{\sf Ric}^\st\big((u\otimes_\st v)\st f\big)&=
&{\sf
  Ric}^\st\big({}^{\phi_1}u\otimes_\st ({}^{\phi_2}v\st{}^{\phi_3}f)\big)\nn\\[4pt]
&=&\nn
-\le~ {\sf R}^\st\big({}^{\phi_1}u\otimes_\st(({}^{\phi_2}v\st{}^{\phi_3}f)\otimes_{\st} \partial_A)\big)\,,\,\dd x^A~\re_\st 
\\[4pt]
&=&\nn
-\le~ {\sf 
  R}^\st\big(
(u\otimes_\st(v\otimes_\st{}^\al\partial_A))\st {}_\al f\big)\,,\,\dd x^A~\re_\st 
\\[4pt]\nn
&=&- \le~ {}^{\bar\varphi_1}{\sf 
  R}^\st\big(
{}^{\bar\varphi_2} (u\otimes_\st(v\otimes_\st{}^\al\partial_A))\big)\st {}^{\bar\varphi_3}{}_\al f\,,\,\dd x^A~\re_\st 
\\[4pt]\nn
&=&- \le~ {}^{\bar\varphi_1}\big({\sf 
  R}^\st\big(
{}^{\bar\varphi_2} (u\otimes_\st(v\otimes_\st{}^\al\partial_A))\big)\big) \st {}^{\bar\varphi_3}{}_\al f\,,\,\dd x^A~\re_\st 
\\[4pt]\nn
&=&- \le~ {}^{\bar\varphi_1\, \bar\rho_1}\big({\sf 
  R}^\st\big(
{}^{\bar\varphi_2}u\otimes_\st({}^{\bar\rho_2}v\otimes_\st{}^\al \partial_A)\big)\big) \,,\, {}^{\bar\varphi_3\, \bar\rho_3}{}_\al f\st\dd x^A~\re_\st 
\\[4pt]\nn
&=&- \le~ {}^{\bar\varphi_1\, \bar\rho_1}\big({\sf 
  R}^\st\big(
{}^{\bar\varphi_2}u\otimes_\st({}^{\bar\rho_2}v\otimes_\st{}^\al\partial_A)\big) \big)
\,,\,{}^\be\dd x^A~\re_{\st}\st 
{}_{\be\, \al}{}^{\bar\varphi_3\,\bar\rho_3} f 
\\[4pt]\nn
&=&- {}^{\bar\varphi_1\, \bar\rho_{1}}\le~ {\sf 
  R}^\st\big(
{}^{\bar\varphi_2}u\otimes_\st({}^{\bar\rho_2}v\otimes_\st\partial_A)\big) 
\,,\,\dd x^A~\re_{\st}\st 
{}^{\bar\varphi_3\, \bar\rho_3} f 
\\[4pt]\nn
&=&
{}^{\bar\varphi_1\, \bar\rho_1}{\sf Ric}^\st\big({}^{\bar\varphi_2}u\otimes_\st {}^{\bar\rho_2} v\big)\st 
{}^{\bar\varphi_3\, \bar\rho_3}f
\\[4pt]
&=&
{}^{\bar\varphi_1}{\sf Ric}^\st\big({}^{\bar\varphi_2}(u\otimes_\st v) \big)\st 
{}^{\bar\varphi_3}f~.
\eeq

The coefficients of the Ricci tensor in the coordinate basis are given by 
\eq\label{Riccoeff}
{\sf Ric}^\st={\sf Ric}_{AD}\st (\dd x^D\otimes_\st\dd x^A) ~,
\en
where ${\sf Ric}_{BC}:={\sf Ric}^\st(\partial_B,\partial_C)$; indeed we have 
\eq
{\sf Ric}^\st(\partial_B,\partial_C)=\le~{\sf Ric}_{AD}\st (\dd x^D\otimes_\st\dd x^A) \,,\,\partial_B\otimes_\st\partial_C~\re_\st={\sf Ric}_{BC}~.
\en
We use
\eqref{eq:Curvgood} together with the fact that the associator acts trivially on
basis vectors and basis 1-forms to calculate explicitly
\begin{eqnarray} 
{\sf Ric}_{BC} &=& -\le ~ {\sf
  R}^\st(\partial_B,\partial_C,\partial_A) \,,\,\dd x^A ~\re_\st \nn\\[4pt]
&=& -\le~\partial_D\st{\sf R}^D{}_{BCA}
\,,\,\dd x^A ~\re_\st\nn \\[4pt]
&=& -{}^\al{\sf R}^D{}_{BCA} \st\le~{}_\al\partial_D\,,\, \dd
    x^A~\re_\st\nn \\[4pt]
&=& {\sf R}^A{}_{BAC} -\ii\kappa\, \RRR^{EF}{}_D\,\partial_E{\sf
    R}^D{}_{BCA}\st\le~ \partial_F\,,\,\dd x^A~\re_\st \nn\\[4pt]
&=&
 {\sf R}^A{}_{BAC} -\ii\kappa\, \RRR^{EA}{}_D\,\partial_E{\sf
    R}^D{}_{BCA}\nn\\[4pt]
&=& \partial_A\Gamma^A_{BC}-\partial_C\Gamma^A_{BA}
+\Gamma^A_{B'A}\st \Gamma^{B'}_{BC} - \Gamma^A_{B'C}\st
    \Gamma^{B'}_{BA} \nn \\
&& +\, \ii\kappa\, \Gamma^A_{B'E}\st \big( \RRR^{EG}{}_A\,(\partial_G\Gamma^{B'}_{BC}) - \RRR^{EG}{}_C\,(\partial_G\Gamma_{BA}^{B'}) \big) \nn\\ 
&& +\, \ii\kappa\, \RRR^{EG}{}_A\, \partial_G\partial_C \Gamma_{BE}^{A} -\ii\kappa\, \RRR^{EG}{}_A\,\partial_G
\big( \Gamma^A_{B'E}\st \Gamma^{B'}_{BC} - \Gamma^A_{B'C}\st \Gamma^{B'}_{BE} \big) \nn\\
&& +\, \kappa^2\, \RRR^{AF}{}_D\,\big( \RRR^{EG}{}_A\, \partial_F (\Gamma^D_{B'E}\st
\partial_G\Gamma^{B'}_{BC}) - \RRR^{EG}{}_C\, \partial_F (\Gamma^D_{B'E}\st \partial_G\Gamma^{B'}_{BA})\big) \ .
\label{eq:RicBC}\end{eqnarray}
This calculation did not use symmetry of $\Gamma^A_{BC}$, i.e., the torsion-free condition
\eqref{TorsionFree}. Indeed this is the Ricci tensor of an arbitrary affine connection.

\newsection{Nonassociative Riemannian geometry and gravity\label{sec:Riemannian}}

\subsection{Metric and torsion-free connection conditions}

We shall now discuss the metric aspects of nonassociative differential geometry and how it can be used to build a theory of nonassociative gravity. For this, notice first of all that star-symmetric tensors are of the form $ T\otimes_\st T'+{}^\al T'\otimes_\st{}_\al T$. A metric tensor is an element $\g^\st\in\Omega_\st^1\otimes_\st\Omega_\st^1$ which
can be written in the basis $\dd x^A$ as $\g^\st=\g_{AB}\star(\dd
x^A\otimes_\st\dd x^B)$ with real-valued components
$\g_{AB}=\g_{BA}\in A_\st$ (the bracketing here is immaterial due to
the basis 1-forms). We can regard it as a map $\g^\st\in{\rm
  hom}_\st(\vect_\st\otimes_\st\vect_\st, A_\st)$ which on vector fields $u,v\in\vect_\st$ gives the function
\bea
\g^\st(u,v)=\le~\g^\st\,,\, u\otimes_\st v~\re_\st \ .
\eea
It is star-symmetric: $\g^\st(u,v)= \g^\st({}^\al v,{}_\al u)$, and one easily
confirms that $\g^\st(\partial_A,\partial_B)=\g_{AB}$. As usual we assume
that $\g^\st$ is nondegenerate: $\g^\st(u,v)=0$ for all $v\in\vect_\st$ if and
only if $u=0$.

Let us now study the metric compatibility condition for a 
connection $\nabla^\st:\vect_\st\to \vect_\st\otimes_\st\Omega^1_\st$. The connection $\nabla^\st$ gives a connection 
$\nabla^\st :\vect_\st\otimes_\st\vect_\st\to
(\vect_\st\otimes_\st\vect_\st)\otimes_\st\Omega^1_\st$ on the tensor
product $\vect_\st\ot_\st\vect_\st$, defined as in Section~\ref{TPC}. The space
$\Omega_\st^1\otimes_\st\Omega_\st^1$ is dual to
$\vect_\st\otimes_\st\vect_\st$ and hence
$\nabla^\st :\vect_\st\otimes_\st\vect_\st\to
(\vect_\st\otimes_\st\vect_\st)\otimes_\st\Omega^1_\st$
induces a dual connection ${}^\st\nabla:
\Omega_\st^1\otimes_\st\Omega_\st^1\to
(\Omega_\st^1\otimes_\st\Omega_\st^1)\otimes_\st\Omega^1_\st$ as in 
(\ref{eq:leftconndef}), now with $\om\in
\Omega_\st^1\otimes_\st\Omega_\st^1$ and $u\in
\vect_\st\otimes_\st\vect_\st $.
We can then state the metric compatibility condition as
\eq
{}^\st\nabla {\sf g}^\st=0~.
\en
We shall now show that, as in the classical case, this condition
together with the
torsion-free condition uniquely
determine the connection in terms of the metric.

We start by using (\ref{eq:leftconndef}) with $\om=\g^\st\in
\Omega_\st^1\otimes_\st\Omega_\st^1$ and $u=\partial_A \otimes_\st \partial_B\in
\vect_\st\otimes_\st\vect_\st $ to get
\begin{eqnarray}
\dd \g_{AB} &=& \dd \le~ \g^\st \,,\, \partial_A \otimes_\st \partial_B~\re_\st \nn \\[4pt]
&=& \le~ {}^\star\nabla \g^\st \,,\, \partial_A\otimes_\star 
\partial_B ~\re_\star + \le~ {}^{{\phi}_1}\g^\st \,,\,
    {}^{{\phi}_2}\nabla^\star({}^{{\phi}_3}(\partial_A\otimes_\star
\partial_B)) ~\re_\star \nn\\[4pt]
&=& \le~ \g^\st \,,\, \nabla^\star (\partial_A\otimes_\star
\partial_B) ~\re_\star \label{mcc} \ ,
\end{eqnarray}
where we used $\g_{AB} =\g_{BA}$ and the fact that the associators act trivially on the basis
vectors.
We can write this more
explicitly as
\begin{eqnarray}
\dd \g_{AB} &=& \le~\g^\st \,,\, \nabla^\star \partial_A \otimes_\st \partial_B
  +{}^\alpha \partial_A \otimes_\st
  {}_\al \nabla^\star \partial_B ~\re _\star\nn\\[4pt]
&=&\le~\g^\st \,,\, \nabla^\star \partial_A \otimes_\st \partial_B
  +{}^{\alpha\, \gamma} \partial_A \otimes_\st
  {}_\al \nabla^\star{}_\gamma\partial_B \nn~\re _\star\\[4pt]
&=&
\le~\g^\st \,,\, \nabla^\star \partial_A \otimes_\st \partial_B
  +{}^\alpha \partial_A \otimes_\st
  {}_\al (\nabla^\star\partial_B) \nn~\re _\star\\[4pt]
&=& \le ~\g^\st 
    \,,\, (\partial_C\otimes_\star\Gamma_A^C) \otimes_\star \partial_B+{}^\al\partial_A
    \otimes_\star {}_\al(\partial_D\otimes_\star\Gamma^D_B) ~\re _\star 
\nn\\[4pt]
&=&
\le ~\g^\st 
    \,,\, (\partial_C\otimes_\star\Gamma_A^C)
    \otimes_\star \partial_B+{}^\al\partial_A \otimes_\star
    (\partial_D\otimes_\star {}_\al\Gamma^D_B)
    ~\re _\star 
\nn\\[4pt]
&=&\le ~\g^\st 
    \,,\, \partial_C\otimes_\star({}^\al \partial_B \otimes_\star {}_\al
    \Gamma_A^C) + {}^\al\partial_A \otimes_\star
    (\partial_D\otimes_\star {}_\al\Gamma^D_B)
    ~\re _\star 
\nn\\[4pt]
&=&\le ~\g^\st 
    \,,\, \partial_C\otimes_\star({}^\al\partial_B \otimes_\star
    {}_\al\Gamma_A^C) +\partial_D\otimes_\star ({}^\al\partial_A
    \otimes_\star {}_\al\Gamma^D_B)
    ~\re _\star 
\nn\\[4pt]
&=&\le ~\g^\st 
    \,,\, \partial_C\otimes_\star ({}^\al\partial_B \otimes_\star {}_\al\Gamma_A^C
    + {}^\al\partial_A \otimes_\star {}_\al\Gamma_B^C) ~\re _\star
\nn\\[4pt]
&=& \le~ \g_{MC}\star {\rm d}x^M \,,\, {}^\alpha\partial_B \otimes_\star {}_\alpha\Gamma^C_A  +
{}^\alpha\partial_A \otimes_\star {}_\alpha\Gamma^C_B ~\re _\star \nn\\[4pt]
&=& \g_{MN}\star \big( \le~ {\rm d}x^M \,,\, {}^\alpha\partial_B ~\re_\star \star           
{}_\alpha\Gamma^N_A + \le~ {\rm d}x^M \,,\, {}^\alpha\partial_A ~\re_\star \star           
{}_\alpha\Gamma^N_B \big) \ ,
\label{NB5}\end{eqnarray}
where in the first equality we used (\ref{expconnprod}) and the fact that the
${\cal R}$-matrix acts trivially on a pair of
basis vectors, so that $^\beta\partial_A\otimes_\star
{}_\beta\partial_B = \partial_A\otimes_\star \partial_B$. In the second
equality we used 
${}^\xi(\nabla^\st\partial_B) ={}^{\xi_{(1)}}\nabla^\st({}^{\xi_{(2)}}\partial_B)$ (as for all linear maps acting from the left),
the coproduct action (\ref{eq:DeltaRPhib}) on the $\cal R$-matrix, 
and again triviality of the action of $\cal R$ as well as of the associator on
$\partial_A\otimes_\star \partial_B$. 
In the sixth line we used the fact that $^\al\partial_A$ is again a
basis vector and then
star-symmetry of the metric.

We similarly calculate
\begin{eqnarray}
\le~ {\rm d}x^M \,,\, {}^\alpha\partial_B ~\re_\star \star           
{}_\alpha\Gamma^N_A &=& \le~ {\rm d}x^M \,,\, {}^\alpha\partial_B ~\re_\star \star         
{}_\alpha(\Gamma^N_{AC}\star {\rm d}x^C) \nn\\[4pt]
&=& \le~ {\rm d}x^M \,,\, {}^{\beta\, \gamma}\partial_B ~\re_\star \star       
{}_\beta\Gamma^N_{AC} \star {}_\gamma{\rm d}x^C \nn\\[4pt]
&=& \le~ {\rm d}x^M \,,\, {}^\beta\partial_B ~\re_\star\star         
\big({}_\beta\Gamma^N_{AC} \star \dd x^C \big) \nn\\[4pt]
&=& \big(\le~ {\rm d}x^M \,,\, {}^\beta\partial_B ~\re_\star\star         
{}_\beta\Gamma^N_{AC} \big) \star {\rm d}x^C \nn\\[4pt]
&=& \le~ {\rm d}x^M \,,\, \Gamma^N_{AC} \star \partial_B ~\re_\star  \star {\rm
d}x^C \ , \label{NB6}
\end{eqnarray}
where as usual we used again (\ref{eq:DeltaRPhib})
together with ${}^\alpha \partial_A \star {}_\alpha
{\rm d}x^B = \partial_A\star {\rm d}x^B$ for our cochain twist
(\ref{twist}). We can finally write
\begin{eqnarray}
{\rm d} \g_{AB} = \g_{MN}\star \le~ {\rm d}x^M \,,\, \Gamma^N_{AC} \star \partial_B ~\re_\star  \star
{\rm d}x^C + \g_{MN}\star \le~ {\rm d}x^M \,,\, \Gamma^N_{BC} \star \partial_A ~\re_\star 
\star {\rm d}x^C \ .\label{NB7}
\end{eqnarray}
We now contract the expression \eqref{NB7} with $\partial_D$ using the star-pairing to obtain
\begin{eqnarray}
\partial_D \g_{AB} = \le~ {\rm d}\g_{AB} \,,\, \partial_D ~\re_\star = \g_{MN}\star \big( \le~ {\rm d}x^M \,,\, \Gamma^N_{AD} \star
\partial_B ~\re_\star   
+ \le~ {\rm d}x^M \,,\, \Gamma^N_{BD} \star \partial_A ~\re_\star\big)
  \ . \label{NB8}
\end{eqnarray}
We write the expression \eqref{NB8} two more times, with the indices
$A,B,D$ cyclically permuted, and consider the combination $\partial_D
\g_{AB}+\partial_A \g_{BD}-\partial_B \g_{DA}$. Using the fact that the
connection coefficients are symmetric for vanishing torsion (see \eqref{TorsionFree}), we obtain
\begin{equation}
\g_{MN}\star \le~ {\rm d}x^M \,,\, \Gamma^N_{AD} \star
\partial_B ~\re_\star = \mbox{$\frac{1}{2}$}\, (\partial_D \g_{AB} + \partial_A \g_{BD} - \partial_B
\g_{AD}) \ . \label{NB9}
\end{equation}
The left hand side equals 
$\le~ \g_{MN}\star {\rm d}x^M\star
\Gamma^N_{AD}\,,\, \partial_B\re_\st$, so
star-multiplying \eqref{NB9} by
any $v^B\in A_\st$ gives
\begin{eqnarray}
\le~ \g_{MN}\star {\rm d}x^M\star \Gamma^N_{AD}\,,\, v ~\re_\star =
  \mbox{$\frac{1}{2}$}\, (\partial_D \g_{AB} + \partial_A \g_{BD} - \partial_B
\g_{AD})\star v^B \ ,
\end{eqnarray}
with $v=\partial_B\star v^B\in\vect_\st$. Since the vector field $v$ is arbitrary
and the star-pairing is nondegenerate, this shows that 
\begin{eqnarray}\label{gdxG}
\g_{MN}\star {\rm d}x^M\star \Gamma^N_{AD}=\mbox{$\frac{1}{2}$}\, (\partial_D \g_{AB} + \partial_A \g_{BD} - \partial_B
\g_{AD})\star \dd x^B~.
\end{eqnarray}
Since moreover the metric $\g^\st$ is nondegenerate, the expression
\eqref{gdxG} uniquely defines the torsion-free metric compatible
connection in the nonassociative case. 

It remains to explicitly solve for the
connection coefficients $\Gamma^N_{AD}$ from \eqref{gdxG}. For this, we use
(\ref{fstardxA}) to rewrite (\ref{gdxG}) as
\begin{eqnarray}
 {\rm d}x^M\star \G_{MN}\star \Gamma^N_{AD}&=& {\rm d}x^M\star \mbox{$\frac12$}\,\big( \partial_D \g_{AM} +
\partial_A \g_{MD} - \partial_M
\g_{AD} \nn\\
&& \qquad \qquad \qquad +\, \ii\kappa\, \RRR^{EF}{}_M\, (\partial_E\partial_D \g_{AF}
+ \partial_E \partial_A \g_{DF}) \big) ~, 
\label{dxGGamma}\end{eqnarray}
with 
\begin{equation}
\G_{MN} = \g_{MN} +\ii\kappa\, \RRR^{EF}{}_M\, \partial_E \g_{NF} \ ,
\label{GMN} \end{equation}
where the associators act trivially due to the basis vector fields and
basis 1-forms. The tensor $\G_{MN}$ is nondegenerate but not
symmetric; it can be thought of as a realisation the $R$-flux
corrected ``effective metric'' anticipated from the string theory
perspective~\cite{Blumenhagen2016}. We then use the
star-pairing to contract
(\ref{dxGGamma}) from the left with $\partial_C$ and obtain 
\begin{equation}
\G_{CN}\star \Gamma^N_{AD} = \mbox{$ \frac{1}{2}$}\, \big(\partial_D \g_{AC} +
\partial_A \g_{DC} - \partial_C
\g_{AD} +\ii\kappa\, \RRR^{EF}{}_C \, (\partial_E\partial_D \g_{AF}
+ \partial_E \partial_A \g_{DF}) \big) \ .\label{GCN}
\end{equation}
We are now faced with the problem of extracting the connection
coefficients $\Gamma^N_{AD}$ from \eqref{GCN}.

\subsection{Inversion in $A_\st$\label{sec:inversion}}

Before tackling the matrix equation \eqref{GCN}, let us consider a
simpler problem involving ordinary functions: Start from the equation
\begin{equation}
h\star g= w \ ,\label{hgw1}
\end{equation}
where $h,g,w\in A_\st$ are functions,  the star-product is given
by (\ref{fstarg}), and $h$ is invertible with respect to the usual pointwise
product of functions. From \eqref{hgw1} we would like to uniquely determine $g$
in terms of $h$ and $w$. 

Recalling the factorization \eqref{eq:calFFR}
of the cochain twist $\FF$, let $\ast:=\star_F$ be the star-product
induced by the twist $F$ from \eqref{eq:twistF}; this
is of course just the canonical associative Moyal-Weyl star-product on phase space: $f\ast g=\overline{\tt f}\,^\al(f)\cdot
\overline{\tt f}\,_\al(g)$. All the $R$-flux dependence is contained
in the twist $F_{R}$ from \eqref{eq:twistFR}. Let $h^{-1}$ be the usual
pointwise inverse of the function $h$: $h^{-1}\cdot h= h\cdot h^{-1}
=1$. Then $h$ is also $\ast$-invertible~\cite{Aschieri2004}, and
we write $h^{\ast-1}$ for the
$\ast$-inverse function:
\bea
h^{\ast-1}\ast h=h\ast h^{\ast-1}=1 \ .
\eea
It can be expressed explicitly as a power series~\cite{Aschieri2004}
\bea\label{expinvast}
h^{\ast-1} =  h^{-1} + \sum_{n=1}^\infty\, h^{-1}\ast\big(1-h\ast
h^{-1}\big)^{\ast n} \ ,
\eea
where, since $\big(1-h\ast h^{-1}\big)^{\ast n}$ is of order 
${O}(\hbar^n)$, in order to get $h^{\ast-1}$ up to order ${O}(\hbar^n)$ 
we only have to compute the finite sum $h^{-1} + \sum_{k=1}^n\, h^{-1}\ast\big(1-h\ast
h^{-1}\big)^{\ast k}.$
This enables one to uniquely determine the function
$h^{\ast-1}$ order by order in $\hbar$.
If no $R$-flux is present, the equation \eqref{hgw1} becomes $h\ast
g=w$ which has the unique solution
\bea
g=1\ast g= (h^{\ast-1}\ast h)\ast g= h^{\ast-1}\ast(h\ast g)= h^{\ast-1}\ast w \ .
\eea

In the absence of $R$-flux, the problem of determining $g$ from the functions $h$ and $w$ has
been solved in a very specific way: by determining a function $h^{\ast
  -1}$ whose star-product $\ast$ with $w$ gives $g$. 
As similarly observed in~\cite{Blumenhagen2016}, in general we cannot
solve the equation (\ref{hgw1}) in the
nonassociative algebra $A_\st$ in this way.
If a function $h$ is invertible,
and hence $\ast$-invertible, then one can recursively construct a right
$\st$-inverse $h_{\rm r}^{\star-1}\in A_\st$ for $h$: Writing the power series
expansion
\bea
h_{\rm r}^{\st-1}=\sum_{n=0}^\infty\, h_{\rm r}^{\st-1(n)} \ ,
\eea
where $h_{\rm r}^{\st-1(n)}$ is of order $O(\kappa^n)$, by using the fact
that $h^{\ast-1}$ is unique one can solve
the equation $h\star h^{\st-1}_{\rm
  r} =\overline{\tt f}_R^{\,\alpha}( h)\ast \overline{\tt
  f}_{R\,\alpha}( h^{\st-1}) =1$ order by order in $\kappa$ to obtain the iterative solution
\bea
h_{\rm r}^{\st-1(0)} &=& h^{\ast-1} \ , \nonumber \\[4pt] 
h_{\rm
  r}^{\st-1(n)} &=& -h^{\ast-1} \ast \ \sum_{k=0}^{n-1}\, \overline{\tt
  f}_R^{\,n-k}(h)\ast \overline{\tt f}_{R\, n-k}\big(h^{\st-1(k)} \big) \ ,
\quad n\geq1 \ ,
\eea
where the terms of order $O(\kappa^{n-k})$ of the twist $F^{-1}_R$ are
denoted $\overline{\tt f}_R^{\,n-k}\otimes \overline{\tt f}_{R\, n-k}$
(with summation understood).
Similarly, one can construct a left $\st$-inverse
$h_{\rm l}^{\st-1}\in A_\st$ such
that $ h^{\st-1}_{\rm 
  l}\star h = 1$. However, in general 
\bea
h_{\rm r}^{\st-1}=(h_{\rm
  l}^{\st-1}\st h) \st h_{\rm
  r}^{\st-1}= {}^{\phi_1}h_{\rm l}^{\st-1}\st({}^{\phi_2}h\st
{}^{\phi_3}h_{\rm r}^{\st-1}) \neq h_{\rm l}^{\st-1}
\eea
so that the left and right $\st$-inverses do not generally coincide. Moreover,
\eq
g= 1\st g= (h_{\rm
  l}^{\st-1}\st h)\st g = {}^{\phi_1}h_{\rm
  l}^{\st-1}\st({}^{\phi_2}h\st {}^{\phi_3}g)\en 
and so in general the
equation \eqref{hgw1} cannot be solved by using the left
$\st$-inverse. 

In any nonassociative unital algebra
which is \emph{alternative}, i.e., for which $(a\, a)\, b=a\, (a\, b)$ and
$(b\, a)\, a=b\, (a\, a)$, for all algebra elements $a$ and $b$, the theory of inverses is
identical to that in associative algebras (see e.g.~\cite{NA-Book}):
Inverses of elements when they exist are unique, and equations such as
\eqref{hgw1} have unique solutions exactly as in the associative
case. However, the nonassociative algebra $A_\st$ is not alternative,
see e.g.~\cite{Bojowald2016,Kupriyanov2017}, with the violations
always being due to explicit dependence on momentum coordinates $p_\mu$ of phase space. The basic
counterexample to alternativity in this case is the function $\vec
x\,^2:=\sum_{\mu=1}^d \, x^\mu\, x^\mu$ for which
\bea
\vec x\,^2\star(\vec
x\,^2\star\vec x\,^2) = (\vec
x\,^2\star\vec x\,^2)\star \vec x\,^2 + 3\ii\hbar\,\kappa^2 \,
R^{\mu\nu\rho}\, R^{\lambda\nu\rho}\, \delta_{\mu\sigma}\, x^\sigma \,
p_\lambda \ .
\eea

While the equation \eqref{hgw1} cannot be solved in general by taking
$\st$-inverses in $A_\st$, in the spirit of \cite{Aschieri2004}
we can regard the star-product operation $h\st g=\of\,^\al(h)\cdot \of\,_\al(g)$ as the
action of the differential operator $h \st=\of\,^\al(h) \, \of\,_\al$ on the
function $g$, and we can then consider the inverse of the differential
operator $h\st$ with respect to the usual associative composition
product of differential operators. We can actually refine this
procedure by recalling the factorization $\FF^{-1}=F^{-1}\, F^{-1}_R$ and the
$\ast$-inverse expression (\ref{expinvast}) for the Moyal-Weyl twist
$F$. We therefore write
\begin{equation}
h\star g=  \overline{\tt f}_R^{\,\alpha}( h)\ast \overline{\tt f}_{R\,\alpha}( g) \ 
\end{equation}
and star-multiply with $\ast$ from the left by $h^{\ast-1}$. 
This gives
\begin{equation}
h^{\ast-1}\ast (h\star g) =h^{\ast-1}\ast \overline{\tt f}_R^{\,\alpha}( h)\ast \overline{\tt
  f}_{R\,\alpha}( g) \ ,
\end{equation}
so that by defining the differential operator 
\bea
{\rm Y}_{h}:=\overline{\tt f}\,^\be \big(h^{\ast-1}\ast \overline{\tt
  f}_R^{\,\alpha}( h)\big)\, \overline{\tt f}\,_\be\, \overline{\tt
  f}_{R\, \alpha} 
\eea
we have
\begin{equation}
{\rm Y}_{h}(g)=h^{\ast-1}\ast \overline{\tt f}_R^{\,\alpha}( h)\ast \overline{\tt
  f}_{R\,\alpha}( g) = h^{\ast-1}\ast (h\star g) ~.
\end{equation}
The differential operator ${\rm Y}_{h}$
can be regarded as a power series expansion in $\kappa$ (or
equivalently in the
$R$-flux) given by
\begin{eqnarray}
{\rm Y}_{h} &=& 1+\sum_{n=1}^\infty\, \frac{1}{n!}\,
                \Big(\,\frac{\ii\kappa}2\, \Big)^n\, 
                \mbf\varepsilon^{A_1B_1}\cdots\mbf\varepsilon^{A_nB_n}
                \, \overline{\tt
                f}\,^\be \big(h^{\ast-1}\ast (\mbf\partial_{A_1}\cdots
                \mbf\partial_{A_n}h) \big) \, \overline{\tt f}\,_\be\,
                \mbf\partial_{B_1}\cdots\mbf\partial_{B_n}
                \nn \\[4pt]
&=:&
 \sum_{n=0}^\infty \, \frac{(\ii\kappa)^n}{n!} \ {\rm Y}_{h}^{(n)}  \
, \label{Y}
\end{eqnarray}
with
\bea\label{eq:mbfpareps}
\big(\mbf\partial_A\big) = \begin{pmatrix}
R^{\mu\nu\rho}\, p_\nu\, \partial_\rho \\ \partial_\mu 
\end{pmatrix}
\qquad \mbox{and} \qquad \big(\mbf\varepsilon^{AB}\big) =
\begin{pmatrix}
0 & \unit_d \\ -\unit_d & 0
\end{pmatrix} \ .
\eea
The operator ${\rm Y}_h$ is invertible as a formal power series in
$\kappa$ because it starts with the zeroth order term ${\rm
  Y}_h^{(0)}=1$; the first order term is
\begin{eqnarray}
{\rm Y}_{h}^{(1)}  = \mbox{$\frac12$}\, \overline{\tt
                      f}\,^\be\, \big(h^{\ast-1}\ast (
R^{\mu\nu\rho}\, p_\nu\, \partial_\rho h)\big)\, \overline{\tt
                      f}\,_\be\, \partial_\mu - \mbox{$\frac12$}\,
\overline{\tt f}\,^\be\, \big(h^{\ast-1}\ast ( \partial_\mu
h)\big)\, \overline{\tt f}\,_\be\, R^{\mu\nu\rho}\, p_\nu\, \partial_\rho \ .
\end{eqnarray}
We define 
\bea
{\rm X}_{h}= \sum_{n=0}^\infty \, \frac{(\ii\kappa)^n}{n!} \ {\rm X}_{h}^{(n)} 
\label{eq:Xh}\eea
as the formal power series in 
$\kappa$ with coefficients in differential operators that satisfies (with composition $\circ$ of operators understood)
\bea
{\rm X}_h\, {\rm Y}_h = \Id \ . 
\label{F}\eea
Then we can finally invert
\eqref{hgw1} to obtain the unique solution in terms of $h$ and $w$:
\bea\label{IF}
g= ({\rm X}_h\, {\rm Y}_h)(g) ={\rm X}_h\big({\rm Y}_h(g) \big)={\rm X}_h\big(h^{\ast-1}\ast w \big) \ .
\label{g}\eea
Crucially, this inversion cannot be written in general as $g=\tilde h\st w$
where $\tilde h$ is a function depending on $h$; the inversion formula (\ref{IF}) uses differential operators
rather than the nonassociative algebra $A_\st$ of functions on its own.

To explicitly construct ${\rm X}_h$, we expand the left-hand side of
\eqref{F} as a power series in $\kappa$ and equate it order by order
with the right-hand side. The order zero term is $
{\rm X}_h^{(0)} = 1$, while the $n$-th order equation for $n\geq1$ is
\bea
\sum_{k=0}^n\, {n\choose k} \, {\rm X}_h^{(n-k)}\, {\rm Y}_h^{(k)}=0 \ .
\eea 
Since ${\rm Y}_h^{(0)}=1$, this yields a recursion relation for the
differential operators ${\rm X}_h^{(n)}$ given by
\bea\label{eq:Xhnrecursion}
{\rm X}_h^{(n)} = -\sum_{k=1}^n\, {n\choose k} \, {\rm X}_h^{(n-k)}\, {\rm Y}_h^{(k)}
\ .
\eeq
One easily shows by induction on $n$ that the solution to the recursion relation \eqref{eq:Xhnrecursion} with the initial condition ${\rm
  X}_h^{(0)}=1$ is
\bea
{\rm X}_h^{(n)} = \sum_{|\vec \lambda\,|=n} \, (-1)^{l(\vec\lambda\,)}\,
\frac{n!}{\vec\lambda\,!} \  {\rm
  Y}_h^{(\vec\lambda\,)} \ ,
\eea
where the sum runs over all unordered sequences
$\vec\lambda=(\lambda_1,\lambda_2,\dots, \lambda_{l})$ of
positive integers $\lambda_i>0$ with length $l(\vec\lambda\,)=l\leq n$ that
partition $n$, i.e., $|\vec\lambda\,|:= \lambda_1+\lambda_2 + \cdots+ \lambda_l =n$, and we defined
\bea
{\rm Y}_h^{(\vec\lambda\,)} = {\rm Y}_h^{(\lambda_1)}\, {\rm
  Y}_h^{(\lambda_2)}\cdots {\rm Y}_h^{(\lambda_{l})} \qquad \mbox{and}
\qquad \vec\lambda\,! = \lambda_1!\,\lambda_2! \cdots \lambda_l! \ .
\eea
The first few orders are given by
\bea
{\rm X}_h^{(0)} &=& 1 \ , \nn \\[4pt]
{\rm X}_h^{(1)} &=& -{\rm Y}_h^{(1)} \ , \nn \\[4pt]
{\rm X}_h^{(2)} &=& -{\rm Y}_h^{(2)}+ 2\, {\rm Y}_h^{(1)}\, {\rm
  Y}_h^{(1)} \ , \nn \\[4pt]
{\rm X}_h^{(3)} &=& -{\rm Y}_h^{(3)} + 3\, {\rm Y}_h^{(1)}\, {\rm
  Y}_h^{(2)} + 3\, {\rm Y}_h^{(2)}\, {\rm Y}_h^{(1)} - 6\, {\rm
  Y}_h^{(1)}\,{\rm Y}_h^{(1)} \, {\rm Y}_h^{(1)}  \ .
\eea
Altogether, for the power series expansion \eqref{eq:Xh} we obtain
\bea
{\rm X}_h = 1+ \sum_{\vec\lambda} \, \frac{(\ii\kappa)^{|\vec\lambda\,|}}{\vec\lambda\,!} \,
(-1)^{l(\vec\lambda\,)}\ {\rm Y}_h^{(\vec\lambda\,)} \ ,
\eea
where here the sum runs through all unordered finite sequences
$\vec\lambda$ of positive integers.

Let us look at an explicit example. For the tachyon vertex operators
$h(x)= \e^{\ii k_\mu \, x^\mu}$, using antisymmetry of
$R^{\mu\nu\rho}$ we easily obtain from \eqref{Y} the differential operator
\bea
{\rm Y}_{\e^{\ii k_\mu\, x^\mu}} = \exp\big(-\kappa \, R^{\mu\nu\rho}\, k_\mu\, p_\nu\, \partial_\rho\big)
\eea
with inverse
\bea
{\rm X}_{\e^{\ii k_\mu \, x^\mu}} = \exp\big(\kappa \, R^{\mu\nu\rho}\, k_\mu\, p_\nu\, \partial_\rho\big) \ .
\eea
Then the equation $\e^{\ii k_\mu \, x^\mu}\star g=w$ has the unique solution
\bea\label{5qr}
g(x,p) = \big(\e^{-\ii k_\mu\, x^\mu}\ast w\big)\big(x+\kappa \,
R(k,p)\,,\, p\big) =
\e^{-\ii k_\mu\, x^\mu} \, w\big(x+\kappa \,
R(k,p)\,,\, p-\mbox{$\frac\hbar2$}\, k\big)
\eea
where $R(k,p)^\mu:= R^{\mu\nu\rho}\, k_\nu\, p_\rho$. When restricted
to the zero momentum leaf of phase space, this yields $g(x,0) = \e^{-\ii
  k_\mu\, x^\mu}\, w(x, -\frac{\hbar}{2}\, k)$, which for functions $w(x)$
depending only on spacetime coordinates further reduces to $g(x,0) = \e^{-\ii  k_\mu\, x^\mu}\, w(x)$. 

\subsection{Levi-Civita connection\label{sec:LCconnection}}

It is straightforward to extend the analysis of
Section~\ref{sec:inversion} to construct a nonassociative version of
the Levi-Civita connection in Riemannian geometry. If the differential
operator ${\rm Y}_h$ is also matrix-valued, then we just have to
interpret products as the composition of operators together with
matrix multiplication; the algebraic manipulations of
Section~\ref{sec:inversion} are identical because $h$, $g$ and $w$ were
treated there as abstract symbols and not as commuting functions. Hence
all formulas are also valid if $h$, $g$ and $w$ in (\ref{hgw1}) are
matrix-valued functions and matrix multiplication is understood in star-products. Therefore we will only sketch the main steps.

Let $\G^{-1}=\big(\G^{MN}\big)$ be the inverse matrix of the matrix
$\G=(\G_{MN})$ describing the string effective metric:
$
\G^{MC}\cdot \G_{CN}=\G_{NC}\cdot \G^{CM}=\delta^M_N \ .
$
There is the closed expression
\eq
\G^{-1} =\big(\unit_{2d}+\ii\kappa\, \g^{-1}\, \RRR \,\partial\g \big)^{-1}\,  \g^{-1} \ ,
\en
where $\RRR\, \partial\g =\big(\RRR^{EF}{}_M\, \partial_E \g_{NF}\big)$ while
$\g^{-1}=\big(\g^{MC} \big)$ is the inverse of $\g=(\g_{CN})$:
\eq
\g^{MC} \cdot \g_{CN}
=\g_{NC}\cdot \g^{CM}=\delta _N^M~,
\en
and
$(\unit_{2d}+\ii\kappa\, \g^{-1}\, \RRR\, \partial\g )^{-1} $ is understood as a geometric
series, so that
\bea
\G^{MC} = \g^{MC} - \ii\kappa\, \g^{MN}\, \RRR^{AB}{}_N \, (\partial_A \g_{B E}) \,
\g^{EC} \,+\, O(\kappa^2) \ . \label{GInv}
\eea
Let $\G^{\ast-1}=\big(\G^{\ast
  MN}\big)$ be the $\ast$-inverse of the matrix $\G=(\G_{MN})$:
\bea
\G^{\ast MC}\ast \G_{ CN} = \G_{ NC}\ast\G^{\ast CM}= \delta_N^M \ .
\eea
It can be expressed explicitly as a power series in $\hbar$ given by~\cite{Aschieri2004}
\bea\label{astinvclosed}
\G^{\ast MN} = \G^{MN} + \sum_{n=1}^\infty \, \G^{MC}\ast {\big(\unit_{2d}-\G\ast
\G^{-1}\big)^{\ast n}\!}_C{}^N \ .
\eea
Let ${\rm Y}_\G$ be the matrix-valued differential operator defined by
\bea
{\rm Y}_\G:=\big({\rm Y}_\G\,^M_N \big) \ , \quad {\rm Y}_\G\,^M_N := \overline{\tt
  f}\,^\be \big( \G^{\ast MC}\ast \overline{\tt f}_R^{\,\alpha} (\G_{CN}) \big)\, \overline{\tt
  f}\,_\be\, \overline{\tt f}_{R\,\alpha} =: \sum_{n =0}^\infty\,
\frac{(\ii\kappa)^n}{n!}\, {\rm Y}_\G^{(n)}{}^M_N~,
\label{Gamma2}\eea
so that acting on a function $g$ gives ${\rm Y}_\G\,^M_N(g)= \G^{\ast
  MC}\ast \overline{\tt f}_R^{\,\alpha} (\G_{CN}) \ast\overline{\tt
  f}_{R\,\alpha} (g)$. The zeroth and first  order terms are given by 
\bea
{\rm Y}_\G^{(0)}{}^{M}_N&=&\delta_N^M \ , \nn\\[4pt]
{\rm Y}_\G^{(1)}{}^{M}_N&=& \mbox{$\frac12$}\, \overline{\tt
  f}\,^\be \big( \G^{\ast MC} \ast ( R^{\mu\nu\rho}\, p_\nu\,
\partial_\rho \G_{CN})\big) \, \overline{\tt f}\,_\be\, \partial_\mu - \mbox{$\frac12$}\, \overline{\tt
  f}\,^\be \big( \G^{\ast MC} \ast (\partial_\mu \G_{CN}) \big)\,
\overline{\tt f}\,_\be\, R^{\mu\nu\rho}\, p_\nu \, \partial_\rho \nn\\[4pt]
&=&
 \mbox{$\frac12$} \, \mbf\varepsilon^{AB}\, \overline{\tt
  f}\,^\be\big(\G^{\ast
SC(0)}\ast \mbf\partial_{{A}} \g_{CM} \big)\, \overline{\tt
  f}\,_\be\, \mbf\partial_{{B}} \ 
,\label{Y1}
\end{eqnarray}
where in the last equality we used the notation \eqref{eq:mbfpareps}. 

Following the formalism of Section~\ref{sec:inversion}, we determine
from (\ref{GCN}) the connection coefficients
\bea
\begin{tabular}{|l|}\hline\\
$\displaystyle
\Gamma^S_{AD} = \G^{\ast
  SC}\ast \W_{CAD} + \sum_{\vec\lambda} \,
  \frac{(\ii\kappa)^{|\vec\lambda\,|}}{\vec\lambda\,!} \
(-1)^{l(\vec\lambda\,)} \ 
{\rm Y}_{\G}^{(\vec\lambda\,)}{}^S_M
 \big(\G^{\ast
  MC}\ast \W_{CAD} \big) 
$\\\\\hline\end{tabular}
\label{Gamma3}\eea
where
\bea
\W_{CAD} = \mbox{$\frac12$}\, \big( \partial_D \g_{AC} +
\partial_A \g_{DC} - \partial_C
\g_{AD} +\ii\kappa\, \RRR^{EF}{}_C \, (\partial_E\partial_D \g_{AF}
+ \partial_E \partial_A \g_{DF}) \big)
\label{eq:WCAD}\eea
and
\bea
{\rm Y}_\G^{(\vec\lambda\,)}{}^{S}_M
={\rm Y}_\G^{(\lambda_1)}{}^{N_1}_M\circ
{\rm Y}_\G^{(\lambda_2)}{}^{{N_2}}_{N_1} \circ \, \cdots \, \circ
{\rm Y}_\G^{(\lambda_l)}{}^{S}_{N_{l-1}}
\eea
for $\vec\lambda=(\lambda_1,\lambda_2,\dots,\lambda_l)$. We also recall
from Section~\ref{sec:inversion} the notation $\vec\lambda\,! =
\lambda_1!\,\lambda_2! \cdots \lambda_l! $ and
$|\vec\lambda\,|= \lambda_1+\lambda_2 + \cdots+ \lambda_l$, and that
the sum in (\ref{Gamma3}) runs over all finite sequences $\vec\lambda$ of positive
integers. In particular, for $|\vec\lambda\,|=1$ the only contribution
to the sum is given by the term ${\rm Y}_\G^{(1)}{}^{M}_N$ in (\ref{Y1}).

In order to understand better the expansion \eqref{Gamma3} of the Levi-Civita
connection, we will now extract the leading non-trivial terms. For this, we have to expand every 
tensor entering into this expression up to first order in $\kappa$ and first order in $\hbar$; we expect non-trivial
nonassociativity contributions in the $O(\kappa\,\hbar)=O(\ell_s^3)$ terms.
For any tensor $T$ we write
\begin{eqnarray}
T = \sum_{n=0}^\infty\, T^{(n)} \qquad \mbox{with} \quad T^{(n)} =
  \sum_{m=0}^\infty\, T^{(n,m)} \ ,
\end{eqnarray}
where by $T^{(n)}$ we understand the term in the power series
expansion of $T$ which is of $n$-th
order in $\kappa$, and $T^{(n,m)}$ is the term in the double power
series expansion of $T$ which is $n$-th order in $\kappa$ and $m$-th
order in $\hbar$. 
We write \eqref{eq:WCAD} as
\begin{eqnarray}
\W_{CAD} &=& \g_{CM}\, \Gamma^{{\tiny{\sf LC}}\, M}_{AD} + \ii\kappa\,
  \RRR^{EF}{}_C\, \partial_E\big(\g_{FM}\, \Gamma^{{\tiny{\LC}}\,
  M}_{AD}\big) \nn\\[4pt] &=:& \W_{CAD}^{(0)} + \W_{CAD}^{(1)} \  , \label{W} 
\end{eqnarray}
where 
\bea
\Gamma^{{\tiny{\LC}}\,M}_{AD} =
\mbox{$\frac{1}{2} $}\, \g^{MQ}\, (\partial_D \g_{AQ} +
\partial_A \g_{DQ} - \partial_Q
\g_{AD})
\label{eq:GammaLC}\eea
is the usual classical Levi-Civita connection, which is zeroth order in
$\kappa$ and zeroth order in $\hbar$ if the metric $\g_{MN}$ is independent
of $\kappa$ and $\hbar$. Then the definition in \eqref{W} is exact in $\kappa$, i.e., there are no higher
order terms ($\W_{CAD}^{(n)} = 0$ for $n\ge 2$).
The effective metric has the expansion
\begin{eqnarray}
\G_{MN} &=& \g_{MN} +\ii\kappa\, \RRR^{EF}{}_M\, \partial_E \g_{NF}
            \nn\\[4pt]
&=:& \G_{MN}^{(0)}+\G_{MN}^{(1)} \ , \label{GMN01}\\[4pt]
\G^{SC} &=& \g^{SC} - \ii\kappa\, \g^{SM}\, \RRR^{EF}{}_M\,
            (\partial_E \g_{NF})\, \g^{NC} \,+\,{O}(\kappa^2)\nn\\[4pt]
&=:& \G^{SC(0)} +
            \G^{SC(1)} \,+\,{O}(\kappa^2)\ .  \label{GMNInverse}
\end{eqnarray}
Recalling (\ref{Y1}) we can now  write the expansion of $\Gamma_{AD}^S$ in \eqref{Gamma3}
up to first order in $\kappa$ and first order in
$\hbar$ as
\begin{eqnarray}
\Gamma_{AD}^{S(0)} &=& \G^{\ast SC(0)} \ast \W_{CAD}^{(0)} \nn\\[4pt]
&=:& \Gamma_{AD}^{S(0,0)} + \Gamma_{AD}^{S(0,1)} \,+\,{O}(\hbar^2)\ ,\label{Gamma0}\\[4pt]
\Gamma_{AD}^{S(1)} &=& \G^{\ast SC(0)} \ast \W_{CAD}^{(1)} + \G^{\ast SC(1)}
\ast \W_{CAD}^{(0)} \nn\\
&&-\,\mbox{$\frac{\ii\kappa}2$} \, \mbf\varepsilon^{KL}\, \G^{\ast SQ(0)}\ast\mbf\partial_{K}
\g_{QM} \ast \mbf\partial_{L} \big(\G^{\ast MC(0)} \ast
   \W_{CAD}^{(0)}\big)  \nn\\[4pt]
&=:& \Gamma_{AD}^{S(1,0)} +
   \Gamma_{AD}^{S(1,1)} \,+\,{O}(\hbar^2)\ . \label{Gamma1}
\end{eqnarray}

To explicitly calculate these 
terms we observe from (\ref{astinvclosed}) that 
\begin{equation}
\G^{\ast SC} = 2\,\G^{SC} - \G^{SP}\ast \G_{PQ}\ast \G^{QC} \,+\, {O}(\hbar^2)\ , \label{GStarInverse}
\end{equation}
so that
\begin{eqnarray}
\G^{\ast SC(0)} &=& 2\,\g^{SC} - \g^{SP}\ast \g_{PQ}\ast \g^{QC} +
                    O(\hbar^2) \nn\\[4pt]
&=& \g^{SC} -\mbox{$\frac{\ii\hbar}{2}$}\, \big( \partial_\mu \g^{SP} \,\tilde\partial^{{\mu}}\g_{PQ} -
\tilde\partial^{{\mu}}\g^{SP}\, \partial_\mu \g_{PQ} \big)\, \g^{QC}
\,+\,{O}(\hbar^2) \nn\\[4pt]
&=:& \G^{\ast SC(0,0)} + \G^{\ast SC(0,1)} \,+\,{O}(\hbar^2) \
,\label{GStarInverse0}\\[4pt]
\G^{\ast SC(1)} &=& 2\,\G^{SC(1)} - \g^{SP}\ast \g_{PQ}\ast \G^{QC(1)} - \g^{SP}\ast
\G_{PQ}^{(1)}\ast \g^{QC} - \G^{SP(1)}\ast \g_{PQ}\ast \g^{QC} +
                    O(\hbar^2) \nn\\[4pt]
&=& -\ii\kappa\, \RRR^{EF}{}_M \, \g^{SM}\,(\partial_E \g_{NF})\, \g^{NC}\nn\\
&& -\, \mbox{$\frac{\hbar\,\kappa}{2}$}\, \RRR^{EF}{}_M \, \Big(
\partial_E \g_{NF}\, \big( \g^{QM}\, \g^{NC}\, (\partial_\mu \g^{SP}
   \, \tilde\partial^{{\mu}}\g_{PQ} -
\tilde\partial^{{\mu}}\g^{SP}\, \partial_\mu \g_{PQ})\nn\\
&& + \, \g^{SM}\, \g^{QC}\, (\partial_\mu \g^{NP} \, \tilde\partial^{{\mu}}\g_{PQ} -
\tilde\partial^{{\mu}}\g^{NP}\, \partial_\mu\g_{PQ}) + \partial_\mu \g^{SM}\,
\tilde\partial^{{\mu}}\g^{NC} -
\tilde\partial^{{\mu}}\g^{SM}\, \partial_\mu \g^{NC}
\big)\nn\\
&& -\, (\partial_\mu\partial_E \g_{NF})\, (\g^{SM}\, \tilde\partial^{{\mu}}\g^{NC}
-\g^{NC}\, \tilde\partial^{{\mu}}\g^{SM})\nn\\
&& +\, (\tilde\partial^{{\mu}}\partial_E \g_{NF})\, (\g^{SM}\, \partial_\mu\g^{NC}
-\g^{NC}\, \partial_\mu\g^{SM})
\Big) \,+\, {O}(\hbar^2)\nn\\[4pt]
&=:& \G^{\ast SC(1,0)} + \G^{\ast SC(1,1)} \,+\, {O}(\hbar^2)\ 
.\label{GStarInverse1}
\end{eqnarray}

We can now compute the first non-trivial terms of the Levi-Civita
connection as defined in 
(\ref{Gamma0}) and (\ref{Gamma1}). We obtain
\begin{eqnarray}
\Gamma_{AD}^{S(0,0)} &=& \Gamma_{AD}^{\LC\,S} \ ,\label{Gamma00}\\[4pt]
\Gamma_{AD}^{S(0,1)} &=& \mbox{$\frac{\ii\hbar}{2}$}\, \big(
                         -(\partial_\mu \g^{SP} \, \tilde\partial^{{\mu}}\g_{PQ}
- \tilde\partial^{{\mu}}\g^{SP}\, \partial_\mu \g_{PQ})\, \Gamma_{AD}^{\LC\,Q} \nn\\
&& + \, \partial_\mu \g^{SC}\, \tilde\partial^{{\mu}}(\g_{CM}\, \Gamma_{AD}^{\LC\,M})
- \tilde\partial^{{\mu}}\g^{SC}\, \partial_\mu (\g_{CM}\, \Gamma_{AD}^{\LC\,M})
\big)\nonumber \\[4pt]
&=& -\mbox{$\frac{\ii\hbar}{2}$}\, \g^{SP}\, \big( (\partial_\mu
    \g_{PQ})\, \tilde\partial^{{\mu}}\Gamma_{AD}^{\LC\,Q} -
(\tilde\partial^{{\mu}} \g_{PQ})\, \partial_\mu\Gamma_{AD}^{\LC\,Q}
    \big) \ ,\label{Gamma01}\\[4pt]
\Gamma_{AD}^{S(1,0)} &=& \ii\kappa\, R^{\alpha\beta\gamma}\, \Big(
\tilde\g^S_{\ {\gamma}} \, \g_{\beta N}\, \big(
\partial_\alpha \Gamma_{AD}^{\LC\,N} \big) - \g^{SM}\, p_\beta \, (\partial_\gamma
\g_{MN})\, \partial_\alpha \Gamma_{AD}^{\LC\,N}
\Big) \ ,\label{Gamma10} \\[4pt]
\Gamma_{AD}^{S(1,1)} &=& \mbox{$\frac{\hbar\, \kappa}2$}\,
                         R^{\alpha\beta\gamma}\, \Big[
-\partial_\mu \tilde\g^S_{\ {\gamma}}\, 
\tilde\partial^{{\mu}}\, \partial_\alpha(\g_{\beta N}\, \Gamma_{AD}^{\LC\,N})
+ \tilde\partial^{{\mu}}\tilde\g^S_{\ {\gamma}}\, 
\partial_\mu\partial_\alpha(\g_{\beta N}\Gamma_{AD}^{\LC\,N}) \nn\\
&& +\, \big( \partial_\mu \g^{SP}\, \tilde\partial^{{\mu}}\g_{PQ} -
\tilde\partial^{{\mu}}\g^{SP}\, \partial_\mu \g_{PQ} \big)\, \tilde\g^Q_{\
{\gamma}}\, \partial_\alpha(\g_{\beta N}\, \Gamma_{AD}^{\LC\,N}) \nn\\
&& +\, \partial_\mu (\tilde\g^S_{\ {\gamma}}\, \g^{NC}\, \partial_\alpha\g_{N\beta} )
\, \tilde\partial^{{\mu}}( \g_{CT}\, \Gamma_{AD}^{\LC\,T})
- \tilde\partial^{{\mu}}(\tilde\g^S_{\ {\gamma}}\,
\g^{NC}\, \partial_\alpha\g_{N\beta} )\, \partial_\mu( \g_{CT}\, \Gamma_{AD}^{\LC\,T})\nn\\
&& -\, \partial_\alpha \g_{N\beta}\, \Big( \tilde\g^S_{\
   {\gamma}}\, (\partial_\mu \g^{SP}\, 
\tilde\partial^{{\mu}}\g_{PQ} 
-\tilde\partial^{{\mu}}\g^{SP}\, \partial_\mu \g_{PQ}) \, \Gamma_{AD}^{\LC\,N}\nn\\
&& + \, \tilde\g^S_{\ {\gamma}}\, (\partial_\mu \g^{NP}\, 
\tilde\partial^{{\mu}}\g_{PQ} 
-\tilde\partial^{{\mu}}\g^{NP}\, \partial_\mu \g_{PQ}) \, \Gamma_{AD}^{\LC\,Q}\nn\\
&& +\,  (\partial_\mu \tilde\g^S_{\ {\gamma}}\, 
\tilde\partial^{{\mu}}\g^{NC} 
-\tilde\partial^{{\mu}}\tilde\g^S_{\ {\gamma}}\, \partial_\mu
   \g^{NC})\, \g_{CT} \, \Gamma_{AD}^{\LC\,T}
\Big) \nn\\
&& +\, (\partial_\mu\partial_\alpha \g_{N\beta})\, \big( \tilde\g^S_{\ {\gamma}}\,
(\tilde\partial^{{\mu}}\g^{NC})\, \g_{CT} \, \Gamma_{AD}^{\LC\,T}
-(\tilde\partial^{{\mu}}\tilde\g^S_{\ {\gamma}})\, \Gamma_{AD}^{\LC\,N} \big) \nn\\
&& - \, (\tilde\partial^{{\mu}}\partial_\alpha \g_{N\beta})\, \big(
   \tilde\g^S_{\ {\gamma}}\, 
(\partial_\mu\g^{NC})\, \g_{CT} \, \Gamma_{AD}^{\LC\,T}
-(\partial_\mu \tilde\g^S_{\ {\gamma}})\, \Gamma_{AD}^{\LC\,N} \big)\nn\\
&& +\, p_\beta\, \Big( \partial_\mu (\g^{SQ}\, \partial_\gamma\g_{QP})\, 
\tilde\partial^{{\mu}}\partial_\alpha\Gamma_{AD}^{\LC\,P}
- \tilde\partial^{{\mu}}(\g^{SQ}\, \partial_\gamma\g_{QP})\, 
\partial_\mu\partial_\alpha\Gamma_{AD}^{\LC\,P} \Big)\nn\\
&& + \, p_\beta\, \g^{SQ}\, (\partial_\gamma\g_{QP})\, \partial_\alpha \Big(
\partial_\mu \g^{PC} \, \tilde\partial^{{\mu}}(\g_{CT}\, \Gamma_{AD}^{\LC\,T}) -
\tilde\partial^{{\mu}}\g^{PC}\, \partial_\mu (\g_{CT}\, \Gamma_{AD}^{\LC\,T}) \nn\\
&& -\, (\partial_\mu \g^{PX} \, \tilde\partial^{{\mu}}\g_{XY} -
\tilde\partial^{{\mu}}\g^{PX}\, \partial_\mu \g_{XY})\, \Gamma_{AD}^{\LC\,Y}
\Big)\nn\\
&& + \, p_\beta \, \Big( \partial_\mu \g^{SQ} \, \tilde\partial^{{\mu}}\partial_\gamma\g_{QP} -
\tilde\partial^{{\mu}}\g^{SQ}\, \partial_\mu\partial_\gamma \g_{QP}
\Big)\, \partial_\alpha\Gamma_{AD}^{\LC\,P} \nn\\
&& -\, p_\beta \, \Big( \partial_\mu \g^{SM} \, \tilde\partial^{{\mu}}\g_{MN} -
\tilde\partial^{{\mu}}\g^{SM}\, \partial_\mu \g_{MN}
\Big)\, \g^{NQ}\, (\partial_\gamma\g_{QP})\, \partial_\alpha\Gamma_{AD}^{\LC\,P} \nn\\
&& + \, (\partial_\alpha\g^{SQ})\, (\partial_\beta\g_{QP})\, \partial_\gamma\Gamma_{AD}^{\LC\,P}
\Big] \ , \label{Gamma11}
\end{eqnarray}
where $\tilde\g^S_{\ {\gamma}}=\g^{SM}\, \delta_{M,\tilde x_{\gamma}}$ is the part of
the inverse metric tensor $\g^{MN}$ with at least one index in momentum space.

We offer the following remarks on the expanded Levi-Civita connection:
\begin{enumerate}
\item Terms that are of type $(0,1)$ and $(1,0)$, i.e., proportional to
  $\hbar$ or to $\kappa$ alone, are
imaginary; this is analogous to what happens in gravity theories on
Moyal-Weyl spaces~\cite{Aschieri2004}. On the other hand, the term of type $(1,1)$,
i.e., proportional to $\hbar\,\kappa=\frac{\ell_s^3}6$, is real; it represents
the non-trivial nonassociativity contribution.
\item If we restrict ourselves to a metric that does not depend on the
  momenta $p_\mu$, then \eqref{Gamma01} vanishes and all terms
  but $\frac{\ell_s^3}6 \, 
R^{\alpha\beta\gamma}\, (\partial_\alpha\g^{SQ})\,
(\partial_\beta\g_{QP})\, \partial_\gamma
\Gamma_{AD}^{\LC\,P}$ in (\ref{Gamma11}) vanish. This remaining term
is just the associator acting on a
product of classical metric tensors and the classical Levi-Civita
connection \eqref{eq:GammaLC}, as is anticipated from the way in
which we extracted the connection coefficients $\Gamma_{AD}^S$ from
\eqref{GCN}.
\item If we restrict to a metric with no indices in momentum space, i.e., $\tilde\g^S_{\
    {\gamma}}=0$, then many
terms in (\ref{Gamma11}) vanish. The terms that remain are those
linear in momenta $p_\beta$ and
the associator term $\frac{\ell_s^3}6 \, 
R^{\alpha\beta\gamma}\, (\partial_\alpha\g^{SQ})\,
(\partial_\beta\g_{QP})\, \partial_\gamma
\Gamma_{AD}^{\LC\,P}$. If we further restrict to a
momentum-independent metric and constrain it to the zero momentum leaf in phase
space, we obtain a real-valued Levi-Civita connection on 
spacetime which is independent of $\hbar$ and with a non-trivial $R$-flux
dependence due to nonassociativity. However, we must keep the momentum
arbitrary for the time being as such terms will make non-trivial
contributions to the Ricci tensor below.
\end{enumerate}

\subsection{Einstein equations\label{sec:Einsteineqs}}

Given an arbitrary metric tensor $\g$ on
phase space ${\cal M}$ with nonassociative deformation induced by a
constant $R$-flux, we have constructed its unique Levi-Civita
 connection in Section~\ref{sec:LCconnection}. Recalling the
 definition of the Ricci tensor from
 Section~\ref{sec:Ricci}, we can therefore 
consider the vacuum Einstein equations on this nonassociative
deformation of ${\cal M}$. They read ${\sf Ric}^\st=0$, or in
components as
\begin{equation}
{\sf Ric}_{BC} = 0 \ . \label{EinsteinVacuum} 
\end{equation}

This equation is a deformation in $\kappa$ and $\hbar$ of the usual vacuum
Einstein equations for gravity.  It is easy to see that the flat space metric
$\g_{AB}=\eta_{AB}$ gives a
vanishing Levi-Civita connection and hence solves the vacuum equations
(\ref{EinsteinVacuum}). Indeed, in this case $\G_{AB}=\eta_{AB}$ and all
star-products reduce to the usual pointwise products, because there is
no dependence on the phase space coordinates $x$ and $p$ at all. 

A more general solution can be easily obtained by considering
metrics $\g_{AB}(p)$ that depend only on the momentum
coordinates. For these metrics we have  $\G_{AB}=\g_{AB}$ and the
usual inverse $\G^{AB}=\g^{AB}$ is also the $\st$-inverse. Indeed here
too all star-products drop out because the twist $F_R$ always involves
vector fields $\partial_\mu$ and so acts trivially. Moreover, the
Moyal-Weyl twist $F$ also acts trivially on functions that depend only on the momentum
coordinates: Each summand in (\ref{eq:twistF})  contains always at least one 
vector field $\partial_\mu$ that acts trivially in this case.
This implies that if a metric $\g_{AB}(p)$ solves the vacuum Einstein 
equations in the classical case, then it remains a solution of the vacuum
Einstein equations also when the $R$-flux is turned on and
hence it is also a solution of  (\ref{EinsteinVacuum}). See \cite{AC}
for further details in the noncommutative case.

\subsection{Spacetime field equations\label{sec:spacetimeeqs}}

Recall that our original motivation
was to obtain a nonassociative theory of gravity on
\emph{spacetime}. The correct way in which to obtain
a reduction to spacetime dynamics from the nonassociative phase
space formalism was explained in~\cite{Aschieri2015}: We start from
tensors on  $M=\real^d$, lift them to tensors on ${\cal M}=T^\ast
M=\real^d\times (\real^d)^\ast$, construct new composite tensors
using the nonassociative deformation of the geometry of ${\cal M}$, reorder the result
using the associator, and then project back to $M$. The lift from $M$ to
${\cal M}$ for functions and more generally for forms is just the pullback of forms using the
canonical projection $\pi: {\cal M}=T^\ast M\to M$. 
In the opposite direction, using the embedding $\sigma: M\to {\cal M}=\real^d\times (\real^d)^\ast$ given by the zero
section $x\mapsto \sigma(x)=(x,0)$, we pull back forms on ${\cal M}$  to
forms on $M$. For example, the $n$-product of functions on $M$ defined in
\cite[eq.~(3.7)]{Aschieri2015} immediately extends to the $n$-exterior
product of forms on $M$ as
\eq
\wedge_\st^{(n)}(\om_1,\om_2,\ldots,\om_n):=\s^\ast\big[\big(\!\cdot\!\cdot\!\cdot\!((\pi^\ast\om_1\wedge_\st\pi^\ast\om_2)\wedge_\st\pi^\ast\om_3)\wedge_\st\cdots\big)\wedge_\st\pi^\ast\om_n\big]~.
\en
The lifts of
vector fields are obtained by considering a foliation of ${\cal M}$ via constant
momentum leaves, with each leaf being diffeomorphic to $M$. 
Explicitly, the coordinate basis vector field $\partial_\mu$ on $M$ lifts to the
coordinate basis vector field $\partial_\mu$ on ${\cal M}$, and more 
generally $v^\mu(x)\, \partial_\mu\mapsto \pi^\ast(v^\mu)(x, p)\,\partial_\mu$,
where $\pi^\ast(v^\mu)(x, p)=(v^\mu)(\pi(x, p))=v^\mu(x)$. 
In the opposite direction, vector fields on ${\cal M}$ are projected to 
vector fields on $M$ via the zero section $\s: M\to {\cal M}$ as
$v^\mu(x, p)\,\partial_\mu+\tilde v_\mu(x, p)\,\tilde\partial^\mu\mapsto v^\mu(x, 0)\,\partial_\mu$.

The lift of a metric tensor on $M$ to a metric tensor on ${\cal M}$
requires an additional structure: a nondegenerate bilinear form on the
cotangent bundle  ${\cal M}=T^\ast M$, i.e., a bilinear form on each
cotangent space $T_x^\ast M$, which we denote by ${\sf
  h}(x)^{\mu\nu}\, \dd\tilde
 x_\mu\otimes\dd\tilde x_\nu$.
Then a metric
  $\g_{\mu\nu}(x)\, \dd x^\mu\otimes \dd x^\nu$ on $M$ is lifted 
to the metric $\hat \g_{MN} \, \dd x^M\otimes \dd x^N$ on ${\cal M}$ given by
\begin{equation}\label{BDF}
\big(\hat\g_{MN}(x)\big) = \begin{pmatrix}
\g_{\mu\nu}(x) & 0 \\ 
0 & {\sf h}^{\mu\nu}(x)
\end{pmatrix} \ . 
\end{equation}
Next we rewrite $\hat \g_{MN} \, \dd x^M\otimes \dd x^N$ in terms of
the star-tensor product as 
\eq\label{liftedg}
\hat\g_{MN} \, \dd x^M\otimes\dd x^N=\g_{MN}\st (\dd x^M\otimes_\st \dd
x^N) \ .
\en 
We thus obtain metric coefficients that have a linear
correction in the $R$-flux given by
\begin{equation}
\big(\g_{MN}(x)\big) = \begin{pmatrix}
\g_{\mu\nu}(x) & \frac{\ii\kappa}{2}\, R^{\sigma\nu\alpha}\, \partial_\sigma\g_{\mu\alpha}  \\[4pt] 
\frac{\ii\kappa}{2}\, R^{\sigma\mu\alpha}\, \partial_\sigma\g_{\alpha\nu}  & {\sf h}^{\mu\nu}(x)
\end{pmatrix} \ . 
\label{BDF1}\end{equation}

Then the Ricci tensor on spacetime is the pullback
\eq\label{Ric1}
{{\sf Ric}^\st}^\circ:=\s^\ast({\sf Ric}^\st)~.
\en
Recalling the expansion (\ref{Riccoeff}) of the Ricci tensor in the
good basis, we obtain
\eq\label{Ric11}
{{\sf Ric}^\st}^\circ={\sf Ric}_{\mu\nu}^\circ\, \dd
x^\mu\otimes \dd x^\nu 
~,\en
where the products are the usual undeformed products because the 3-tensor product in a good basis is the
usual tensor product:
\eq\otimes_\st^{(3)}(f,\dd x^\mu,\dd x^\nu):=
\s^\ast\big((\pi^\ast f\otimes _\st\pi^\ast\dd
x^\mu)\otimes_\st\pi^\ast\dd x^\nu\big)=
f\, \dd x^\mu\otimes\dd x^\nu~.\en
Comparing (\ref{Ric1}) with (\ref{Riccoeff}) and (\ref{Ric11}) leads to the simple
result that the nonassociative spacetime Ricci tensor is obtained form the phase
space Ricci tensor simply by restricting the components ${\sf Ric}_{MN}$ to
spacetime directions and setting the momentum dependence to zero:
\eq\label{eq:barRic0}
{\sf Ric}^\circ_{\mu\nu}(x)=\s^\ast({\sf Ric}_{\mu\nu})(x,p)=
{\sf Ric}_{\mu\nu}(x,0)~.
\en

The spacetime
vacuum equations for nonassociative gravity then read as
\bea
{\sf Ric}^\circ_{\mu\nu} = 0 \ .
\label{eq:barRic00}\eea
We observe that the flat metric
$\g_{\mu\nu}(x)=\eta_{\mu\nu}$, ${\sf h}^{\mu\nu}(x)=\eta^{\mu\nu}$ is a solution
of (\ref{eq:barRic00}). More generally, every solution of the phase
space Einstein equations (\ref{EinsteinVacuum}) leads to a solution of
(\ref{eq:barRic00}). On the other hand, not all solutions of \eqref{eq:barRic00}
can be lifted to solutions of the phase space vacuum Einstein 
equations \eqref{EinsteinVacuum}. Whether or not such a condition on
solutions should be imposed, i.e., that the dynamics is completely
determined on phase space, is presently unclear and should be ultimately
prescribed by which procedure correctly matches the expectations from
non-geometric string theory. We do not address further this salient
point in the present paper.

Recalling our discussion from Section~\ref{sec:DFT}, it is also
interesting to examine projections of the field equations
\eqref{EinsteinVacuum} with respect to other polarisations of phase
space in the $R$-flux frame. For instance, we could alternatively
choose to foliate phase space with respect to constant position leaves
rather than constant momentum leaves, and hence to reduce the dynamics
from nonassociative phase space onto momentum space. This corresponds
to embedding momentum space $\widetilde M$, with local coordinates $\tilde
x_\mu=p_\mu$, in phase space
via $\tilde \s: \widetilde M\to {\cal M}$, $\tilde x\mapsto \tilde \s(\tilde
x)=(0,\tilde x)$. Correspondingly, we can restrict the
classical metric to the same block diagonal form
\eqref{BDF}, but with the components now dependent only
on momentum. By our general discussion from
Section~\ref{sec:Einsteineqs} it follows that there are no
$R$-flux corrections to the classical Ricci tensor on momentum space,
so that momentum space geometry is uncorrected by stringy contributions; in particular, the
string effective metric \eqref{GMN} coincides with the classical
metric. This would appear to imply the expected result that there are no nonassociative or
noncommutative corrections to the spacetime field equations in a
geometric ($H$-flux or $f$-flux) frame obtained by an $O(d,d)$-rotation of the $R$-flux frame. It would be
interesting to understand how this perspective ties in
precisely with the possibility of Born geometry and
dynamical phase space discussed in~\cite{Freidel2013} using curved
momentum space geometry (see~\cite{Aschieri2015} for further discussion of
this latter point).

\subsection{First order corrections}

We will now study the vacuum Einstein equations \eqref{EinsteinVacuum}
and \eqref{eq:barRic00} in more detail by determining the first non-trivial correction
terms to the classical Einstein equations. For this, we expand the Ricci tensor from \eqref{eq:RicBC} as
\begin{eqnarray}
{\sf Ric}_{BC} &=& \partial_A\Gamma^A_{BC}-\partial_C\Gamma^A_{BA}
+\Gamma^A_{B'A}\st \Gamma^{B'}_{BC} - \Gamma^A_{B'C}\st
    \Gamma^{B'}_{BA} - \ii\kappa\, \RRR^{EG}{}_C\,\Gamma^A_{B'E}\st
\partial_G\Gamma_{BA}^{B'} \nn \\
&&  +\, \ii\kappa\, \RRR^{EG}{}_A\, \big( \partial_G\partial_C \Gamma_{BE}^{A} -
\partial_G\Gamma^A_{B'E}\st
\Gamma^{B'}_{BC} + \partial_G\Gamma^A_{B'C}\st \Gamma^{B'}_{BE} + \Gamma^A_{B'C}\st
\partial_G\Gamma^{B'}_{BE}\big) \nn\\ && + \, O(\kappa^2) \nn\\[4pt]
 &=:& {\sf Ric}_{BC}^{(0,0)} + {\sf Ric}_{BC}^{(0,1)} + {\sf Ric}_{BC}^{(1,0)} + {\sf
Ric}_{BC}^{(1,1)} \, + \, O(\kappa^2,\hbar^2) \label{Ricci} 
\end{eqnarray}
by expanding the star-products $\st$ and using the
expansion of the Levi-Civita connection from Section~\ref{sec:LCconnection}. For the undeformed
contribution we obtain the usual Ricci tensor of the classical
Levi-Civita connection \eqref{eq:GammaLC}:
\begin{equation}
{\sf Ric}_{BC}^{(0,0)} = {\sf Ric}_{BC}^{\sf LC} := \partial_A\Gamma^{{\sf
    LC}\,A}_{BC}-\partial_C\Gamma^{\LC\,A}_{BA} 
+\Gamma^{\LC\, A}_{B'A}\, \Gamma^{\LC\, B'}_{BC} 
-\Gamma^{{\tiny{\sf LC}}\, A}_{B'C}\, \Gamma^{{\tiny{\sf LC}}\, B'}_{BA}
\ . \label{Ricci00} 
\end{equation}
For the order $\hbar$ contribution we have
\begin{eqnarray}
{\sf Ric}_{BC}^{(0,1)} &=& \partial_A\Gamma^{A(0,1)}_{BC}-\partial_C\Gamma^{A(0,1)}_{BA}
+\Gamma^{A(0,1)}_{B'A} \, \Gamma^{\LC\, B'}_{BC} + \Gamma^{\LC\,
                           A}_{B'A} \,
\Gamma^{B'(0,1)}_{BC}\nonumber\\
&& +\, \mbox{$\frac{\ii\hbar}{2}$}\, \big(
\partial_\mu\Gamma^{\LC\, A}_{B'A}\, \tilde\partial^\mu\Gamma^{\LC\, B'}_{BC} 
-\tilde\partial^\mu\Gamma^{\LC\, A}_{B'A}\, \partial_\mu\Gamma^{\LC\,B'}_{BC} \big)
- \Gamma^{A(0,1)}_{B'C} \, \Gamma^{\LC\,B'}_{BA} \nonumber\\ 
&& - \, \Gamma^{\LC\,A}_{B'C} \, \Gamma^{B'(0,1)}_{BA} -
   \mbox{$\frac{\ii\hbar}{2}$}\, \big(
\partial_\mu\Gamma^{\LC\,A}_{B'C}\, \tilde\partial^\mu\Gamma^{\LC\,B'}_{BA} 
-\tilde\partial^\mu\Gamma^{\LC\,A}_{B'C}\, \partial_\mu\Gamma^{\LC\,
   B'}_{BA} \big) \ ,\label{Ricci01} 
\end{eqnarray}
where to obtain the explicit expression in terms of the classical
metric tensor and Levi-Civita connection one has to insert
(\ref{Gamma01}) in (\ref{Ricci01}). Notice that ${\sf Ric}_{BC}^{(0,1)}$
is imaginary. Likewise,
the order $\kappa$ contribution is given by
\begin{eqnarray}
{\sf Ric}_{BC}^{(1,0)} &=& \partial_A\Gamma^{A(1,0)}_{BC} - \partial_C\Gamma^{A(1,0)}_{BA}
+\Gamma^{A(1,0)}_{B'A} \, \Gamma^{\LC\, B'}_{BC} + \Gamma^{\LC\, A}_{B'A}\,
\Gamma^{B'(1,0)}_{BC} \label{Ricci10} \\
&& +\, \ii\kappa \, R^{\alpha\beta\gamma}\, p_\beta\, \big(
\partial_\gamma\Gamma^{\LC\, A}_{B'A}\, \partial_\alpha\Gamma^{\LC\,B'}_{BC}
-\partial_\gamma\Gamma^{\LC\, A}_{B'C}\, \partial_\alpha\Gamma^{\LC\,B'}_{BA} \big) \nonumber\\
&&
- \, \Gamma^{A(1,0)}_{B'C}\, \Gamma^{\LC\, B'}_{BA}  - \Gamma^{\LC\, A}_{B'C} \, \Gamma^{B'(1,0)}_{BA} 
-\ii\kappa \,
   R^{\alpha\beta\gamma}\, \delta_{C,{\tilde x_{\gamma}}}\, \Gamma^{\LC\,A}_{B'\alpha}\, \partial_\beta\Gamma^{\LC\,B'}_{BA}
\nn\\
&& + \, \ii\kappa \,
   R^{\alpha\beta\gamma}\,\delta_{A,{\tilde x_{\gamma}}}\, \big(
 \partial_\beta\partial_C
\Gamma_{B\alpha}^{\LC\, A} - \partial_\beta\Gamma^{\LC\, A}_{B'\alpha}\, \Gamma^{\LC\, B'}_{BC} + \partial_\beta\Gamma^{\LC\, A}_{B'C}\, \Gamma^{\LC\, B'}_{B\alpha} +
\Gamma^{\LC\, A}_{B'C}\, \partial_\beta\Gamma^{\LC\, B'}_{B\alpha}
 \big)  \ , \nonumber
\end{eqnarray}
where here one has to insert (\ref{Gamma10}) to obtain the explicit
expression in terms of classical quantities. Notice that 
${\sf Ric}_{BC}^{(1,0)}$ is also imaginary.

Finally, the order $\kappa\,\hbar=\frac{\ell_s^3}6$ contribution is given by
\small
\begin{eqnarray}
{\sf Ric}_{BC}^{(1,1)} &=& \partial_A\Gamma^{A(1,1)}_{BC} - \partial_C\Gamma^{A(1,1)}_{BA}
+\Gamma^{A(1,0)}_{B'A} \, \Gamma^{B'(0,1)}_{BC} + \Gamma^{A(0,1)}_{B'A}\,
\Gamma^{B'(1,0)}_{BC}\nonumber\\ 
&& +\, \Gamma^{\LC\,A}_{B'A} \, \Gamma^{B'(1,1)}_{BC} + \Gamma^{A(1,1)}_{B'A}\,
\Gamma^{\LC\,B'}_{BC} - \Gamma^{A(1,0)}_{B'C} \, \Gamma^{B'(0,1)}_{BA}\nonumber\\
&& - \, \Gamma^{A(0,1)}_{B'C} \, \Gamma^{B'(1,0)}_{BA} -
   \Gamma^{\LC\,A}_{B'C} \, \Gamma^{B'(1,1)}_{BA}
-\Gamma^{A(1,1)}_{B'C} \, \Gamma^{\LC\,B'}_{BA}\nonumber\\
&& +\, \mbox{$\frac{\ii\hbar}{2}$}\, \big(
\partial_\mu\Gamma^{A(1,0)}_{B'A}\, \tilde\partial^\mu\Gamma^{\LC\,B'}_{BC}
+\partial_\mu\Gamma^{\LC\,A}_{B'A}\, \tilde\partial^\mu\Gamma^{B'(1,0)}_{BC} \nonumber\\
&& -\, \tilde\partial^\mu\Gamma^{A(1,0)}_{B'A}\, \partial_\mu\Gamma^{\LC\,B'}_{BC} 
- \tilde\partial^\mu\Gamma^{\LC\,A}_{B'A}\, \partial_\mu\Gamma^{B'(1,0)}_{BC} 
\big)\nonumber\\
&&- \,\mbox{$\frac{\ii\hbar}{2}$}\, \big(
\partial_\mu\Gamma^{A(1,0)}_{B'C}\, \tilde\partial^\mu\Gamma^{\LC\,B'}_{BA}
+\partial_\mu\Gamma^{\LC\,A}_{B'C}\, \tilde\partial^\mu\Gamma^{B'(1,0)}_{BA}\nonumber\\
&& -\, \tilde\partial^\mu\Gamma^{A(1,0)}_{B'C}\, \partial_\mu\Gamma^{\LC\,B'}_{BA} 
- \tilde\partial^\mu\Gamma^{\LC\,A}_{B'C}\, \partial_\mu\Gamma^{B'(1,0)}_{BA} 
\big)\nonumber\\
&& +\,\ii\kappa\, R^{\alpha\beta\gamma}\, p_\beta\, \big(
\partial_\gamma\Gamma^{A(0,1)}_{B'A}\, \partial_\alpha\Gamma^{\LC\,B'}_{BC}
+ \partial_\gamma\Gamma^{\LC\,A}_{B'A}\, \partial_\alpha\Gamma^{B'(0,1)}_{BC} \nonumber\\
&& -\, \partial_\gamma\Gamma^{A(0,1)}_{B'C}\, \partial_\alpha\Gamma^{\LC\,B'}_{BA} 
-\partial_\gamma\Gamma^{\LC\,A}_{B'C}\, \partial_\alpha\Gamma^{B'(0,1)}_{BA}
\big)\nonumber\\
&& -\,\mbox{$\frac{\kappa\, \hbar}2$}\, R^{\alpha\beta\gamma}\,
   p_\beta \, \big(
\partial_\mu\partial_\gamma\Gamma^{\LC\,A}_{B'A}\,
\tilde\partial^\mu\partial_\alpha\Gamma^{\LC\,B'}_{BC}
- \tilde\partial^\mu\partial_\gamma\Gamma^{\LC\,A}_{B'A}\,
\partial_\mu\partial_\alpha\Gamma^{\LC\,B'}_{BC} \nonumber\\
&& - \, \partial_\mu\partial_\gamma\Gamma^{\LC\,A}_{B'C}\,
\tilde\partial^\mu\partial_\alpha\Gamma^{\LC\,B'}_{BA}       
+ \tilde\partial^\mu\partial_\gamma\Gamma^{\LC\,A}_{B'C}\,
\partial_\mu\partial_\alpha\Gamma^{\LC\,B'}_{BA}
\big)\nonumber\\
&& -\,\ii\kappa \, R^{\alpha\beta\gamma}\, \Big( 
\delta_{A,{\tilde x_{\gamma}}}\, \big( \partial_\beta\Gamma^{\LC\,A}_{B'\alpha}\, \Gamma^{B'(0,1)}_{BC}
+\partial_\beta\Gamma^{A(0,1)}_{B'\alpha}\, \Gamma^{\LC\,B'}_{BC} \big) \nonumber\\
&& +\, \delta_{C,{\tilde x_{\gamma}}}\, \big(
   \Gamma^{\LC\,A}_{B'\alpha}\, \partial_\beta\Gamma^{B'(0,1)}_{BA}
+ \Gamma^{A(0,1)}_{B'\alpha}\, \partial_\beta\Gamma^{\LC\, B'}_{BA}\big) \Big)\nonumber\\
&& +\,\mbox{$\frac{\kappa\, \hbar}2$}\, R^{\alpha\beta\gamma}\, \Big(
   \delta_{A,{\tilde x_{\gamma}}}\, \big(
\partial_\mu\partial_\beta\Gamma^{\LC\,A}_{B'\alpha}\,
\tilde\partial^\mu\Gamma^{\LC\,B'}_{BC} 
-\tilde\partial^\mu\partial_\beta\Gamma^{\LC\,A}_{B'\alpha}\,
\partial_\mu\Gamma^{\LC\,B'}_{BC}
\big)\nonumber\\
&&+ \, \delta_{C,{\tilde x_{\gamma}}}\, \big(
\partial_\mu\Gamma^{\LC\,A}_{B'\alpha}\,
\tilde\partial^\mu\partial_\beta\Gamma^{\LC\,B'}_{BA} 
-\tilde\partial^\mu\Gamma^{\LC\,A}_{B'\alpha}\,
\partial_\mu\partial_\beta\Gamma^{\LC\,B'}_{BA}
\big) \Big) \nonumber\\
&& +\, \ii\kappa\, R^{\alpha\beta\gamma}\,\delta_{A,{\tilde{x}}_\gamma}\,
\big( \partial_\beta\partial_C\Gamma^{A(0,1)}_{B\alpha} \,
+\partial_\beta\Gamma^{\LC\, A}_{B'C}\, \Gamma^{B'(0,1)}_{B\alpha}
+ \partial_\beta\Gamma^{A(0,1)}_{B'C}\, \Gamma^{\LC\, B'}_{B\alpha}\nn\\
&& +\, \Gamma^{\LC\, A}_{B'C}\, \partial_\beta\Gamma^{B'(0,1)}_{B\alpha}
+ \Gamma^{A(0,1)}_{B'C}\, \partial_\beta\Gamma^{\LC\, B'}_{B\alpha} \big) \nn\\
&& -\, \mbox{$\frac{\kappa\, \hbar}{2}$}\, R^{\alpha\beta\gamma}\, \delta_{A,{\tilde{x}}_\gamma}\,
\big( \partial_\mu\partial_\beta\Gamma^{\LC\, A}_{B'C}\,
\tilde\partial^{{\mu}}\Gamma^{\LC\, B'}_{B\alpha} 
-\tilde\partial^{{\mu}}\partial_\beta\Gamma^{\LC\, A}_{B'C}\,
\partial_\mu\Gamma^{\LC\, B'}_{B\alpha}\nn\\
&& +\, \partial_\mu\Gamma^{\LC\, A}_{B'C}\,
\tilde\partial^{{\mu}}\partial_\beta\Gamma^{\LC\, B'}_{B\alpha} 
-\tilde\partial^{{\mu}}\Gamma^{\LC\, A}_{B'C}\,
\partial_\mu\partial_\beta\Gamma^{\LC\, B'}_{B\alpha} \big)
 \ ,\label{Ricci11}
\end{eqnarray}
\normalsize
where again the explicit expression in terms of the classical metric
and connection is obtained after inserting
(\ref{Gamma01})--(\ref{Gamma11}). Like the undeformed contribution
\eqref{Ricci00}, the expression \eqref{Ricci11} is real.

We now consider metrics of the
form (\ref{BDF1}) with the natural choice ${\sf h}^{\mu\nu}(x)=\eta^{\mu\nu}$. The pointwise inverse metric $\g^{MN}$ has
an expansion in $\kappa$, which up to first order is given by
\begin{equation}
\big(\g^{MN}(x)\big) = \begin{pmatrix}
\g^{\mu\nu}(x) &
-\frac{\ii\kappa}{2}\, R^{\alpha\nu\gamma}\, \g^{\mu\rho}\, \partial_\alpha
\g_{\rho\gamma}
\\[4pt] 
-\frac{\ii\kappa}{2}\, R^{\alpha\mu\gamma}\, \partial_\alpha \g_{\gamma\rho}\, \g^{\rho\nu}  & \eta_{\mu\nu}
\end{pmatrix}  \, + \, O(\kappa^2) \ . \label{g(-1)MN}
\end{equation}
One caveat is that the $\kappa$-dependence
of \eqref{BDF1} and \eqref{g(-1)MN} will now reorder the expansion of
the Levi-Civita connection in \eqref{Gamma00}--\eqref{Gamma11}; for example, the
classical contributions $\Gamma_{AD}^{S(0,0)} = \Gamma_{AD}^{\LC\,S} $
in \eqref{Gamma00} will receive both type $(0,0)$ and $(1,0)$
terms. These additional contributions can be easily accounted for by using the fact that there is no momentum dependence in
\eqref{BDF1} and \eqref{g(-1)MN}, and our results below take this reordering into account.

After summing up all expressions, the proper expansion of the
connection coefficients is as follows: For the classical contribution
$\Gamma_{AD}^{S(0,0)}$ the only non-zero components of the classical
Levi-Civita connection in this case are
\begin{equation}
\Gamma_{\mu\nu}^{\LC\, \rho} = \mbox{$\frac{1}{2}$}\,
\g^{\rho\sigma}\, (\partial_\mu
\g_{\sigma\nu} +
\partial_\nu \g_{\mu\sigma} - \partial_\sigma\g_{\mu\nu}) \ . \label{CSpaceGamma00}
\end{equation} 
Using (\ref{Gamma01}) one can check that all contributions $\Gamma_{AD}^{S(0,1)}$ vanish. The only non-zero components
of the corrections $\Gamma_{AD}^{S(1,0)}$ from \eqref{Gamma10} are
\begin{eqnarray}
\Gamma_{\mu\nu}^{\rho(1,0)} &=& -\ii\kappa \,R^{\alpha\beta\gamma}\, p_\beta\,
\g^{\rho\sigma}\, (\partial_\gamma
\g_{\sigma\tau})\, \partial_\alpha \Gamma_{\mu\nu}^{\LC\,\tau} \
                                , \nn \\[4pt]
\Gamma_{\mu\nu}^{{\tilde{x}}_\rho(1,0)} &=& -\mbox{$\frac{\ii\kappa}{2}$} \,
R^{\alpha\rho\sigma}\, \g_{\gamma\sigma} \,
\partial_\alpha\Gamma_{\mu\nu}^{\LC\, \gamma} \ , \nn \\[4pt]
\Gamma_{{\tilde{x}}_\mu,\nu}^{\rho(1,0)} &=& \mbox{$\frac{\ii\kappa}{2}$} \,
R^{\alpha\mu\gamma}\, \g^{\sigma\rho}\, \partial_\alpha\big( \g_{\sigma\tau}\,
\Gamma_{\gamma\nu}^{\LC\, \tau}\big) \ ,\nn \\[4pt]
\Gamma_{\mu, {\tilde{x}}_\nu}^{\rho(1,0)} &=& \mbox{$\frac{\ii\kappa}{2}$} \,
R^{\alpha\nu\gamma}\, \g^{\rho\sigma} \, \partial_\alpha\big( \g_{\sigma\tau}\,
\Gamma_{\mu\gamma}^{\LC\, \tau}\big) \ , 
\end{eqnarray}
while the remaining correction terms $\Gamma_{AD}^{S(1,1)}$ from
\eqref{Gamma11} have non-vanishing contributions
\begin{eqnarray}
\Gamma_{\mu\nu}^{\rho(1,1)} = \mbox{$\frac{\hbar\, \kappa}2$} \, R^{\alpha\beta\gamma}\,
(\partial_\alpha\g^{\rho\sigma})\, (\partial_\beta
\g_{\sigma\tau})\, \partial_\gamma \Gamma_{\mu\nu}^{\LC\, \tau} \ . \label{CSpaceGamma11}
\end{eqnarray}

The non-zero components of the classical Ricci tensor are then
\begin{equation}
{\sf Ric}_{\mu\nu}^{\LC} = \partial_\rho\Gamma^{\LC\, \rho}_{\mu\nu}
-\partial_\nu\Gamma^{\LC\, \rho}_{\mu\rho}
+\Gamma^{\LC\,\rho}_{\sigma\rho}\, \Gamma^{\LC\,\sigma}_{\mu\nu} - \Gamma^{\LC\,\rho}_{\sigma\nu}
\, \Gamma^{\LC\, \sigma}_{\mu\rho} \ .\label{CSpaceRicci00}
\end{equation}
Although the expanded formula for ${\sf Ric}_{BC}$ appears to be unwieldy
and very difficult to analyse in general, in this case all correction terms ${\sf Ric}_{BC}^{(0,1)}$ vanish, while the non-zero components of ${\sf
Ric}_{BC}^{(1,0)}$ are
\begin{eqnarray}
{\sf Ric}_{\mu\nu}^{(1,0)} &=& \ii\kappa\, R^{\alpha\beta\gamma}\,
                               p_\beta \, \Big(
-\partial_\rho\big(\g^{\rho\sigma}\, (\partial_\gamma
\g_{\sigma\tau})\, \partial_\alpha \Gamma_{\mu\nu}^{\LC\,\tau}\big) +
\partial_\nu\big(\g^{\rho\sigma}\, (\partial_\gamma
\g_{\sigma\tau})\, \partial_\alpha \Gamma_{\mu\rho}^{\LC\, \tau}\big)\nonumber\\
&& -\, \Gamma_{\mu\nu}^{\LC\, \omega}\, \g^{\rho\sigma}\, (\partial_\gamma
\g_{\sigma\tau})\, \partial_\alpha \Gamma_{\omega\rho}^{\LC\, \tau}
-\Gamma_{\rho\omega}^{\LC\, \rho}\, \g^{\omega\sigma}\, (\partial_\gamma
\g_{\sigma\tau})\, \partial_\alpha \Gamma_{\mu\nu}^{\LC\, \tau}\nonumber\\
&& + \, \Gamma_{\mu\rho}^{\LC\, \omega}\, \g^{\rho\sigma}\, (\partial_\gamma
\g_{\sigma\tau})\, \partial_\alpha \Gamma_{\omega\nu}^{\LC\, \tau}
+ \Gamma_{\omega\nu}^{\LC\,\rho}\, \g^{\omega\sigma}\, (\partial_\gamma
\g_{\sigma\tau})\, \partial_\alpha \Gamma_{\mu\rho}^{\LC\, \tau} \nonumber\\
&& +\, \big(\partial_\gamma \Gamma_{\sigma\rho}^{\LC\, \rho}\big)\, \partial_\alpha
\Gamma_{\mu\nu}^{\LC\,\sigma}
- \big(\partial_\gamma \Gamma_{\sigma\nu}^{\LC\, \rho}\big)\, \partial_\alpha
\Gamma_{\mu\rho}^{\LC\, \sigma} \Big) \ ,
\label{CSpaceRicci10a}\\[4pt]
{\sf Ric}_{\mu,{\tilde x_{\nu}}}^{(1,0)} &=&
                                             \mbox{$\frac{\ii\kappa}{2}$} \,
R^{\alpha\nu\gamma} \, \Big( 
\partial_\rho\big( \g^{\rho\sigma}\, \partial_\alpha(\g_{\sigma\tau}\,
                                             \Gamma_{\gamma\mu}^{\LC\,
                                             \tau})\big) +
                                             \g^{\rho\sigma}\,
                                             \Gamma_{\rho\omega}^{\LC\,
                                             \omega}\, \partial_\alpha(\g_{\sigma\tau}\,
                                             \Gamma_{\gamma\mu}^{\LC\,
                                             \tau})\nn\\
&& -\, \g^{\rho\sigma}\, \Gamma_{\mu\rho}^{\LC\,
   \omega}\, \partial_\alpha(\g_{\sigma\tau}\,
   \Gamma_{\gamma\omega}^{\LC\, \tau}) + 2\,
   \Gamma_{\sigma\alpha}^{\LC\,
   \rho}\, \partial_\gamma\Gamma_{\mu\rho}^{\LC\, \sigma} \Big) \ , \label{CSpaceRicci10b}\\[4pt]
{\sf Ric}_{{\tilde x_{\mu}},\nu}^{(1,0)} &=&
                                             \mbox{$\frac{\ii\kappa}{2}$} \,
R^{\alpha\mu\gamma} \, \Big( 
\partial_\rho\big( \g^{\rho\sigma}\, \partial_\alpha(\g_{\sigma\tau}\,
                                             \Gamma_{\gamma\nu}^{\LC\,
                                             \tau})\big) - \partial_\nu\big(
\g^{\rho\sigma}\, \partial_\alpha(\g_{\sigma\tau}\,
                                             \Gamma_{\gamma\rho}^{\LC\,
                                             \tau})\big)\nn\\
&& +\, \g^{\rho\sigma}\, \Gamma_{\rho\omega}^{\LC\,
   \omega}\, \partial_\alpha(\g_{\sigma\tau}\,
   \Gamma_{\gamma\nu}^{\LC\, \tau}) - \g^{\rho\sigma}\,
   \Gamma_{\rho\nu}^{\LC\, \omega}\, \partial_\alpha(\g_{\sigma\tau}\,
   \Gamma_{\gamma\omega}^{\LC\, \tau}) \Big) \ . \label{CSpaceRicci10c}
\end{eqnarray}
Notice that all terms of type $(1,0)$ are imaginary.
Finally, the only non-zero components of ${\sf Ric}_{BC}^{(1,1)}$ are
given by
\begin{eqnarray}
{\sf Ric}_{\mu\nu}^{(1,1)} &=& \mbox{$\frac{\hbar\, \kappa}2$}\, R^{\alpha\beta\gamma}\,\Big(
\partial_\rho\big(\partial_\alpha\g^{\rho\sigma}\, (\partial_\beta
\g_{\sigma\tau})\, \partial_\gamma \Gamma_{\mu\nu}^{\LC\, \tau} \big) 
- \partial_\nu\big(\partial_\alpha\g^{\rho\sigma}\, (\partial_\beta
\g_{\sigma\tau})\, \partial_\gamma \Gamma_{\mu\rho}^{\LC\, \tau}\big) \nonumber\\
&& +\,  \partial_\gamma \g_{\tau\omega}\big( \partial_\alpha
(\g^{\sigma\tau}\, \Gamma_{\sigma\nu}^{\LC\,\rho})\, \partial_\beta
\Gamma_{\mu\rho}^{\LC\,\omega} - \partial_\alpha
(\g^{\sigma\tau}\, \Gamma_{\sigma\rho}^{\LC\,\rho})\,\partial_\beta
\Gamma_{\mu\nu}^{\LC\,\omega}\nonumber\\
&& +\, (\Gamma_{\mu\rho}^{\LC\,\sigma}\, \partial_\alpha
\g^{\rho\tau} - \partial_\alpha \Gamma_{\mu\rho}^{\LC\,\sigma} \,\g^{\rho\tau})\,           
\partial_\beta \Gamma_{\sigma\nu}^{\LC\,\omega}\nonumber\\
&& - \, (\Gamma_{\mu\nu}^{\LC\,\sigma}\, \partial_\alpha
\g^{\rho\tau} -\partial_\alpha \Gamma_{\mu\nu}^{\LC\,\sigma}\,\g^{\rho\tau})\,              
\partial_\beta \Gamma_{\sigma\rho}^{\LC\,\omega} \big) \Big) \ .\label{CSpaceRicci11}
\end{eqnarray}

Now we apply the reduction described in Section~\ref{sec:spacetimeeqs}, and altogether we
find for the Ricci tensor ${\sf Ric}^\circ_{\mu\nu}$ on spacetime $M$
up to first order:
\bea
\begin{tabular}{|l|}\hline\\
$\displaystyle
{\sf Ric}^\circ_{\mu\nu} \ \ = \ \ {\sf Ric}_{\mu\nu}^{\LC} +
\mbox{$\frac{\ell_s^3}{12}$} \, R^{\alpha\beta\gamma}\,\Big(
\partial_\rho\big(\partial_\alpha\g^{\rho\sigma}\, (\partial_\beta
\g_{\sigma\tau})\, \partial_\gamma \Gamma_{\mu\nu}^{\LC\, \tau} \big) 
- \partial_\nu\big(\partial_\alpha\g^{\rho\sigma}\, (\partial_\beta
\g_{\sigma\tau})\, \partial_\gamma \Gamma_{\mu\rho}^{\LC\, \tau}\big)
$ \\[2mm] $\qquad\qquad\qquad\qquad \qquad \qquad \quad +\,  \partial_\gamma \g_{\tau\omega}\big( \partial_\alpha
(\g^{\sigma\tau}\, \Gamma_{\sigma\nu}^{\LC\,\rho})\, \partial_\beta
\Gamma_{\mu\rho}^{\LC\,\omega} - \partial_\alpha
(\g^{\sigma\tau}\, \Gamma_{\sigma\rho}^{\LC\,\rho})\,\partial_\beta
\Gamma_{\mu\nu}^{\LC\,\omega}
$ \\[2mm] $\qquad\qquad\qquad\qquad\qquad \qquad \quad +\, (\Gamma_{\mu\rho}^{\LC\,\sigma}\, \partial_\alpha
\g^{\rho\tau} - \partial_\alpha \Gamma_{\mu\rho}^{\LC\,\sigma} \,\g^{\rho\tau})\,           
\partial_\beta \Gamma_{\sigma\nu}^{\LC\,\omega}
$ \\[2mm] $\qquad\qquad\qquad\qquad\qquad \qquad \quad - \, (\Gamma_{\mu\nu}^{\LC\,\sigma}\, \partial_\alpha
\g^{\rho\tau} -\partial_\alpha \Gamma_{\mu\nu}^{\LC\,\sigma}\,\g^{\rho\tau})\,              
\partial_\beta \Gamma_{\sigma\rho}^{\LC\,\omega} \big) \Big)
$\\\\\hline\end{tabular}
\label{eq:Ricpolarised}\eea
for $\mu,\nu=1,\dots,d$. One readily checks that the linear $R$-flux
correction to the classical Ricci tensor is not a total derivative: While the first two lines of the correction in \eqref{eq:Ricpolarised} are total derivatives, the last two lines are not. The consistent reduction to
spacetime has thus achieved \emph{two} remarkable and desirable
features: Not only do we find that the nonassociative $R$-flux gravitational corrections
lead to non-trivial dynamical consequences on spacetime, but they are also
independent of $\hbar$ and
real-valued, in contrast to what happens in the usual metric
formulations of noncommutative gravity~\cite{Aschieri2004}. Notice
that this latter feature in itself singles out the zero momentum leaf among
all constant momentum leaves: Pulling back to a leaf of constant
momentum $p=p^\circ$ generally gives a non-vanishing imaginary contribution
${\sf Ric}_{\mu\nu}^{(1,0)}\big|_{p= p^\circ}$ from \eqref{CSpaceRicci10a} to the
spacetime Ricci tensor \eqref{eq:Ricpolarised}; indeed, in
that case the pullback via the phase space star-product $\star$ yields associative
Moyal-Weyl star-product deformations of the usual closed string scattering
amplitudes with constant bivector
$\theta^{\alpha\beta}=2\, \kappa\, R^{\alpha\beta\gamma}\, 
p^\circ_\gamma$~\cite{Aschieri2015}, and ${\sf
  Ric}_{\mu\nu}^{(1,0)}\big|_{p= p^\circ}$ coincides with the first
order contribution to the noncommutative Ricci tensor
from~\cite{Aschieri2004}. The potential physical significance of the
$ p^\circ \neq0$ leaves is discussed in~\cite{Aschieri2015}.

It would be
interesting to confirm explicitly that these features all persist to
higher orders, and to find explicit solutions of the spacetime
vacuum equations \eqref{eq:barRic00}
for nonassociative gravity. Note that these equations are linear in the $R$-flux, whereas the $H$-flux modified Einstein equations at leading order in string worldsheet perturbation theory and for constant dilaton read ${\sf Ric}^{\LC}_{\mu\nu} = \frac14\, H_{\mu\alpha\beta}\,H_\nu{}^{\alpha\beta}$ which by T-duality would naively imply that the leading corrections should be of quadratic order in the $R$-flux. Here the first non-trivial contribution to the spacetime curvature tensor
is of order $O(\kappa \,\hbar)$ which is the order in which the first
nonassociative contributions appear. As this is a second order
contribution when one expands the twist element \eqref{eq:calFFR}, it is natural that the curvature (and torsion) have corrections at this order.

\newsection{Conclusions\label{sec:Conclusions}}

In this paper we have provided and developed a formalism leading to a
consistent approach to nonassociative gravity induced by
locally non-geometric constant $R$-flux backgrounds of string theory
in the parabolic phase space model of~\cite{Lust2010}. The construction relied on the
proper characterization of tensor fields in nonassociative geometry
as well as their covariance under the quasi-Hopf algebra generated by
infinitesimal diffeomorphisms on twisted nonassociative
phase space. The unique Levi-Civita connection of any metric $\g$ has
been determined at all orders in the nonassociative deformation parameters. 
The vacuum Einstein equations have been obtained also at all orders, and the first order
corrections to the classical equations explicitly calculated, which is
the order at which the corresponding string theory calculations
are reliable.

We have then pulled back the vacuum Einstein equations on
phase space ${\cal M}$  to spacetime $M$ via the zero momentum
section $\s: M\to {\cal M}$.
General covariance of these latter equations is on the one hand guaranteed
by the geometric pullback operation. On the other hand, it could be
studied explicitly by considering the projection of the quantum Lie algebra of
nonassociative diffeomorphisms from Section~\ref{sec:diffeos} to the
zero momentum leaf, as pursued in \cite{Aschieri2015}, where it was
illustrated how nonassociativity survives in the action of
diffeomorphisms on spacetime. Ultimately, these symmetries should be compared to the classical diffeomorphism symmetries of closed string theory and to the generalised diffeomorphism symmetries of double field theory.

Further insights into this nonassociative
theory of gravity on spacetime should be obtained by studying the
pullbacks to spacetime also of the torsion and the Riemann curvature tensors. Additional investigations relating the curved phase
space geometry to the curved spacetime geometry, and in particular the other
possible spacetime geometries obtained by considering different
foliations of the manifold ${\cal M}$, and not only those defined by
constant momentum leaves and constant position leaves, are left for future work.
These investigations, and the construction of a dynamical action
principle for nonassociative gravity, should clarify the expected
relevance in the contexts of closed
string theory and double field theory of the field
equations we have obtained, and
in particular their interpretations as low-energy effective field equations of closed string theory.

\subsection*{Acknowledgments}

We thank Ralph Blumenhagen, Leonardo Castellani, Michael Fuchs, Chris Hull, Dieter L\"ust, Emanuel
Malek, Eric Plauschinn, Alexander Schenkel and Peter Schupp for
helpful discussions. This work was supported in part by the Action MP1405 QSPACE from 
the European Cooperation in Science and Technology (COST). The work of P.A. and R.J.S. was supported in
part by the Research Support Fund of the Edinburgh Mathematical
Society, and by the Consolidated Grant ST/L000334/1 from the UK Science and
Technology Facilities Council. The work of P.A. is partially supported
by INFN, CSN4, Iniziativa Specifica GSS. P.A. is also affiliated to INdAM,
GNFM (Istituto Nazionale di Alta Matematica, Gruppo Nazionale di
Fisica Matematica). The work of M.D.C. is 
supported by Project~ON171031 of the Serbian Ministry of
Education and Science.

\end{document}